\documentclass[10pt]{article}
\usepackage[utf8]{inputenc}
 \pdfoutput=1

\usepackage{tikz, adjustbox, float} 
\usepackage[authoryear]{natbib} 
\usepackage{amsmath, amsfonts, amssymb, amsthm, dsfont} 
\usepackage{mathrsfs} 
\usepackage{soul} 
\usepackage{bm} 
\usepackage{booktabs}  
\usepackage{enumitem} 
\usepackage{mathtools}
\usepackage{mathabx}
\usepackage{nicefrac}
\usepackage{tabularx}
\usepackage{algorithm}
\usepackage{algpseudocode}

\usepackage{float}

\usepackage{titlesec}
\usepackage{cancel}
\titleformat{\paragraph}
{\normalfont\normalsize\scshape}{\theparagraph}{1em}{}
\titlespacing*{\paragraph}
{0pt}{3.25ex plus 1ex minus .2ex}{1.5ex plus .2ex}

\usepackage{bbm}

\usepackage{wrapfig}

\DeclareMathOperator*{\argmin}{argmin}
\DeclareMathOperator*{\tr}{tr}

\DeclareMathOperator*{\cov}{cov}
\def\0{{\bf 0}}

\DeclareMathOperator*{\var}{var}

\DeclareMathOperator*{\diag}{diag}

\DeclareMathOperator*{\op}{op}

\usepackage{latexsym}
\usepackage{amsthm}
\usepackage{amsmath,amssymb,amsfonts,bm,amsbsy,bbm}
\usepackage{epsfig}
\usepackage{graphicx,rotating,lscape}
\usepackage[colorlinks=true,  urlcolor=blue, citecolor=blue, psdextra, pdfencoding=auto]{hyperref}
\usepackage{verbatim}
\usepackage{multirow,color}
\usepackage{geometry}
\usepackage{setspace}
\usepackage{subfigure}
\usepackage{bbm}
\usepackage{ulem} 

\usepackage{cleveref}

\def\0{{\bf 0}}

\def\bse{\begin{eqnarray*}}
\def\ese{\end{eqnarray*}}
\def\be{\begin{eqnarray}}
\def\ee{\end{eqnarray}}
\def\bsq{\begin{equation*}}
\def\esq{\end{equation*}}
\def\bq{\begin{equation}}
\def\eq{\end{equation}}

\def\boxit#1{\vbox{\hrule\hbox{\vrule\kern6pt  \vbox{\kern6pt#1\kern6pt}\kern6pt\vrule}\hrule}}
\def\bse{\begin{eqnarray*}}
\def\ese{\end{eqnarray*}}
\def\be{\begin{eqnarray}}
\def\ee{\end{eqnarray}}
\def\bsq{\begin{equation*}}
\def\esq{\end{equation*}}
\def\bq{\begin{equation}}
\def\eq{\end{equation}}

\def\diag{\hbox{diag}}

\def\diag{\hbox{diag}}
\def\log{\hbox{log}}

\def\squarebox#1{\hbox to #1{\hfill\vbox to #1{\vfill}}}

\def\0{{\bf 0}}

\def\diag{\hbox{diag}}
\def\log{\hbox{log}}

\def\diag{\hbox{diag}}

\newtheoremstyle{mytheoremstyle} 
    {0.3cm}                      
    {0cm}                        
    {\itshape}                   
    {}                           
    {\bf}                   
    {: }                          
    {0em}                       
    {}  

\theoremstyle{mytheoremstyle}
\newtheorem{Theorem}{Theorem}

\newtheorem{Lemma}{Lemma}
\newtheorem*{Lemma*}{Lemma}

\newtheorem{Proposition}{Proposition}

\newtheoremstyle{myExampleRemarkstyle} 
    {0.3cm}                    
    {0cm}                           
    {\itshape}                   
    {}                           
    {\bf}                   
    {: }                          
    {0em}                       
    {}  

\theoremstyle{myExampleRemarkstyle}
 
\newtheorem{Remark}{Remark}
\newtheorem{Assumption}{Assumption}

\newtheorem{innercustomgeneric}{\customgenericname}
\providecommand{\customgenericname}{}
\newcommand{\newcustomtheorem}[2]{%
  \newenvironment{#1}[1]
  {%
   \renewcommand\customgenericname{#2}%
   \renewcommand\theinnercustomgeneric{##1}%
   \innercustomgeneric
  }
  {\endinnercustomgeneric}
}

\newcustomtheorem{customthm}{Theorem}
\newcustomtheorem{customlemma}{Lemma}
\usepackage{mathalfa}
\usepackage{oubraces}

\let\refBKP\ref
\renewcommand{\ref}[1]{{\upshape\refBKP{#1}}}

\usepackage{titlesec}
\titleformat{\section}{\normalfont\Large\scshape}{\thesection.}{1em}{}
\titleformat{\subsection}{\normalfont\large\scshape}{\thesubsection.}{1em}{}
\titleformat{\subsubsection}{\normalfont\scshape}{\thesubsubsection.}{1em}{}

\titlespacing*{\section}{0pt}{1em}{0em}
\titlespacing*{\subsection}{0pt}{1em}{0em}
\titlespacing*{\subsubsection}{0pt}{1em}{0em}

\usepackage[numbib]{tocbibind}

\usepackage[section]{placeins}

\usepackage{titlesec}

\titleformat*{\section}{\large \scshape}

\usepackage{setspace}
\usepackage{graphicx}
\usepackage{subcaption}
\begin{document}

\begin{center}
{	\centering


\LARGE{\textsc{Inference  for Large Scale Regression Models\\ with Dependent Errors}}

	}
    \vspace{0.5cm} \normalsize
	{\textsc{Lionel~Voirol}$^{1}$, \textsc{Haotian~Xu}$^{2}$,  \textsc{Yuming~Zhang}$^{3}$, \textsc{Luca~Insolia}$^{1}$\\
 \textsc{Roberto~Molinari}$^{4}$
 \& \textsc{St\'ephane~Guerrier}$^{1,5}$}
	\vspace{0.5cm} 

	{\footnotesize $^{1}$Geneva School of Economics and Management, University of Geneva, Switzerland;  $^{2}$Department of Statistics, University of Warwick, United Kingdom; $^{3}$Department of Biostatistics, Harvard T.H. Chan School of Public Health, Harvard University $^{4}$Department of Mathematics and Statistics, Auburn University, United States; $^{3}$Faculty of Science, University of Geneva, Switzerland 
	}
\end{center}

\begin{abstract}

  The exponential growth in terms of data sizes and consequent storage costs has brought considerable challenges to the data science community who have had to find solutions to run learning methods on such data. While approaches from machine learning and artificial intelligence have scaled to achieve predictive accuracy also in these big data settings, the availability of statistical inference and uncertainty quantification tools is still lagging under many aspects in the face of this challenge. Indeed, areas of priority scientific and social relevance collect vast amounts of data to understand (interpret) and test the significance of different phenomena which is a task usually associated with statistical learning methods such as regression. In this setting, the estimation of the regression parameters can benefit from efficient computational procedures but the main challenge lies in the computation of the parameters of the error process when this has complex covariance (kernel) structures. The identification and estimation of these structures are essential for inference on the regression parameters (needed for interpretation) and, aside from non-parametric modelling, they are often used for uncertainty quantification in machine learning through the use of Gaussian Processes, for example. However the estimation of these structures remain burdensome (or even impossible) to estimate as data scales, thereby requiring various levels of approximations that consequently affect the reliability of their outputs. These approximations become even more unreliable when complexities such as long-range dependencies, missing and/or contaminated observations are present (which are common to find in big data settings). In this work we define and prove the statistical properties of the Generalized Method of Wavelet Moments with Exogenous variables (GMWMX) which provides a highly scalable, numerically stable and statistically valid method to estimate and deliver inference for linear models (and non-linear adaptations thereof) using stochastic processes in the presence of data complexities such as latent dependence structures and missing data. On top of the efficiency of wavelet convolutions, all this is achieved through new efficient approaches to compute theoretical and empirical quantities for the wavelet variance (which are of interest in their own right), allowing to model complex features in big data settings with high computational efficiency. Applied examples from the field of Earth Sciences are used to highlight the important advantages of the GMWMX, whose theoretical, numerical and computational properties are supported by extensive simulation studies.

\end{abstract}


\section{Introduction}

The performance of standard statistical procedures and the delivery of their corresponding inferential tools is currently undergoing a major challenge as the size of data scales increasingly faster. In particular, the estimation of parameters linked to observed predictors (i.e. fixed-effect parameters) can benefit from approaches and algorithmic procedures that allow to compute these parameters on large-scale data within reasonable timeframes, but the estimation of the stochastic components (or error processes) that are used to perform inference on the fixed-effect parameters is often challenging even in moderate sample sizes and can become infeasible as data scales. Hence there is a strong need for scalable methods to estimate the parameters of stochastic processes with complex dependence structures. Indeed, while in this work we will put forward a scalable method to perform inference on the parameters of the linear model through an efficient estimation of its error process,  in general the use and estimation of stochastic processes has become of increased interest over the past years in its own right. The latter is due to their natural capacity of providing interpretable inference as well as delivering measures needed for uncertainty quantification in machine-learning domains, such as Gaussian Process Regression (GPR) (e.g. \citealp{banerjee2013efficient}). In particular, stochastic processes are of increasing relevance in many areas where growing amounts of data are expensive to record/simulate and where complex physical, multi-agent or deep-learning models become computationally burdensome to predict from and are approximated through so-called ``surrogate models'' in the form of (Gaussian) stochastic processes \citep[see e.g.][and references therein]{gramacy2020surrogates}. With respect to applications, the vast field of natural sciences heavily relies on stochastic processes to interpret and monitor natural phenomena going from climate, ocean and earth sciences to the study of huge amounts of high-dimensional satellite and telescope images \citep[e.g.][]{muyskens2022star}. An important example comes from tracking and understanding Earth's systems, including changes and irregularities caused by global warming (see e.g. \citealp{montillet2024big}). The tools created to perform these tasks generate large amounts of data, necessitating new techniques to manage and analyse them. In the majority of these applications the models used to analyze these data are linear for the mean-function (deterministic component), with possible transformations of the predictors to accommodate for non-linear relations. However this does not represent the main challenge when performing inference and uncertainty quantification for these models: the covariance-function characterizing the stochastic-error component can suffer massively from complex dependence structures in the data which are problematic to estimate in general. Common examples of such complex dependence structures are those represented by the sum of different underlying stochastic error processes (hereinafter \textit{latent processes}) which in many cases have physical justifications and interpretations, such as a process representing the accumulated error of the measurement device and another representing the correlated error resulting from, for instance, missing predictors. Focusing on Earth Sciences as the main example in this work where such processes are widely applied, these models are used to study concentrations of ethane or carbon dioxide \citep{proietti2022modelling, maddanu2023trends}, water storage \citep{schmidt2008periodic}, solar irradiance \citep{montillet2022data} and global surface temperature \citep{mudelsee2019trend}, among others. Of particular interest in this context are the large networks of time-stamped signals recorded by the Global Navigation Satellite Systems (GNSS) which collect information on many geophysical phenomena. More specifically, the time series recorded by these GNSS receivers consist in sequences of position measurements reflecting dynamic changes such as, for example, post-glacial rebound, hydrological loading, tectonic rates and crustal deformations. Networks of permanent GNSS stations worldwide track these positional changes using satellite constellations like GPS, GLONASS, Galileo, and BeiDou to pinpoint their locations on Earth, typically recording daily over decades. In this context, the focus often lies in accurately estimating and understanding the trends in these signals, which are crucial for monitoring long-term tectonic movements, fault strain accumulation, and Glacial Isostatic Adjustment among many others.

\subsection{Related work}

The task of inference for stochastic processes can become extremely problematic, especially when dealing with natural phenomena and measurement devices for example. In fact, many real-world processes are often the result of a combination of different underlying latent processes (i.e. composite processes) each with their distinct covariance structure which can include long-memory and non-stationary components (see e.g. \citealp{amiri2009noise}). Discarding the issues in estimating the latter components, composite processes are particularly complicated to address due to the need to distinguish the underlying processes before retrieving their respective parameters, especially for likelihood-based approaches which have to heavily rely on a Kalman Filter (or variants thereof) to estimate the underlying states, thereby introducing an additional step in the overall estimation procedure. Indeed, likelihood-based approaches often use the Expectation-Maximization algorithm to obtain such estimates and incur in many computational and numerical issues which often lead to algorithmic convergence problems (see e.g. \citealp{stebler2014generalized}). Alternative approaches, mainly from the field of engineering, use informative quantities such as the Power Spectral Density (PSD) or the Allan Variance (AV) to detect the different underlying processes as well as to estimate their parameters. While these approaches have practical advantages over likelihood-based methods, they do not possess appropriate statistical properties making them overall unreliable in terms of inference \citep{guerrier2016theoretical}.

However, the challenge of performing inference and uncertainty quantification does not solely reside in the complexity of the dependence structure of stochastic processes, but also in the size of the data they are estimated from. Indeed, this feature of the data often leads to unmanageable storage and computational bottlenecks which severely hinder the estimation of the parameters of these processes. To address this issue, various strategies have been devised for efficient stochastic process estimation. Some methods focus on sparsifying the correlation matrix, either by partitioning the domain into blocks or by altering the covariance matrix to induce independence in near-zero entries \citep{fuentes2002spectral, sang2011covariance}. Others assume sparsity in the precision matrix, achieved by modeling only a subset of conditioning sets or employing spectral methods \citep{stein2013stochastic, guinness2019spectral}. Variational low-rank approximations and predictive processes are popular in machine learning and spatial statistics, respectively, where a reduced set of knot locations or basis functions enables faster computation \citep{gibbs2000variational, lazaro2011variational}. Moreover, methods like Gapfill use quantile regression without relying on statistical distributions \citep{gerber2018predicting}. A benchmark study by \cite{heaton2019case} compared most of these methods on predicting missing measurements in land surface temperature datasets highlighting how the approximations achieve good computation times but often at the cost of accuracy. A particularly fast implementation of the Maximum-Likelihood Estimator (MLE) for these problems was put forward by \cite{bos2013fast} which, as mentioned earlier, delivers accurate estimations and inference tools in reduced computation times. Nevertheless, while these approximations are able to (partly) address the large sample sizes for which stochastic processes are often employed, they preserve a variety of limitations which make their use in practice narrow. Among these we can find common scenarios in applied settings, such as missing observations (e.g. due to energy-cuts in the measurement devices), contaminated data points (e.g. due to abrupt bursts in natural phenomena) and long-range dependence structures in the errors.

\subsection{Problem Formulation and Motivation}

We formalize the definition of stochastic process $\boldsymbol{Y} = \{Y_i\}_{i=1,\hdots,n}$ considered in this work, with $Y_i \in \mathbb{R}$, which takes on the standard linear regression form as follows:

\begin{equation}
    \boldsymbol{Y} = \boldsymbol{X} \boldsymbol{\beta} + \boldsymbol{\varepsilon}, \quad \boldsymbol{\varepsilon} \sim \mathcal{F}\left\{\boldsymbol{\Sigma}\left(\boldsymbol{\gamma}\right)\right\},
    \label{eq.reg_complete}
\end{equation}

where $\boldsymbol{X} \in \mathbb{R}^{n \times p}$ is a design matrix of observed predictors, $\boldsymbol{\beta} \in \mathbb{R}^p$ is the regression parameter vector and $\boldsymbol{\varepsilon} = \{\varepsilon_{i}\}_{i=1,\hdots,n}$ is a zero-mean process following an unspecified joint distribution $\mathcal{F}$ with positive-definite covariance function $\boldsymbol{\Sigma}(\boldsymbol{\gamma}) \in \mathbb{R}^{n\times n}$ characterizing the second-order dependence structure of the process and parameterized by the vector $\boldsymbol{\gamma} \in \boldsymbol{\Gamma} \subset \mathbb{R}^q$. While the design matrix $\boldsymbol{X}$ can accommodate for non-linear relationships in the deterministic mean-function (i.e. $\boldsymbol{X} \boldsymbol{\beta}$) through appropriate transformations of the predictors, the error term $\boldsymbol{\varepsilon}$ captures the stochastic component of the process $\boldsymbol{Y}$ whose (often complex) structure must be learned to perform adequate inference on the parameter vector of interest $\boldsymbol{\beta}$. As mentioned earlier, for this work we assume that the process $\boldsymbol{Y}$ is indexed over time and that the covariance matrix therefore characterizes the dependence between observations at different points in time. However the results of this work can be extended to processes indexed over different domains in a relatively straightforward manner.

We now extend the above definition by equipping the stochastic process $\boldsymbol{Y}$ with a random variable $\boldsymbol{Z} =\{Z_{i}\}_{i=1,\hdots,n}$ which describes the missing observation mechanism. Indeed, in this work we assume that we only observe the stochastic process $\widetilde{\boldsymbol{Y}} = \boldsymbol{Z} \odot \boldsymbol{Y}$, where $\odot$ denotes the Hadamard product. More specifically, the vector $\boldsymbol{Z} \in \{0, 1\}^n$ is a binary-valued stationary process independent of $\boldsymbol{Y}$ with expectation $\mu(\boldsymbol{\vartheta}) = \mathbb{E}[Z_i] \in [0, \, 1)$, $\forall \, i$, and covariance matrix $\boldsymbol{\Lambda}(\boldsymbol{\vartheta}) = \mathbb{V}[\bm{Z}] \in \mathbb{R}^{n\times n}$ whose structure is assumed known up to the parameter vector $\boldsymbol{\vartheta} \in \boldsymbol{\Upsilon} \subset \mathbb{R}^k$. Hence, $\boldsymbol{Z}$ determines which elements of $\boldsymbol{Y}$ are observed (i.e. when $Z_i = 1$). As a result of the missing observations mechanism, the (overall) observed process $\widetilde{\boldsymbol{Y}}$ is parametrized by the vector:

$$\boldsymbol{\theta} := \left[\boldsymbol{\beta}^T, \;\;\; \boldsymbol{\gamma}^T, \;\;\;\boldsymbol{\vartheta}^T\right]^T \in \boldsymbol{\Theta},$$

where $\boldsymbol{\Theta} := \mathbb{R}^p \times \boldsymbol{\Gamma} \times \boldsymbol{\Upsilon}$. As described above, estimating the parameter $\boldsymbol{\theta}$ by directly maximising the likelihood function is usually computationally demanding (see e.g. \citealp{guerrier2013wavelet,proietti2022modelling}). For example, in the analysis of GNSS time series, most likelihood-based methods have a computational complexity of $\mathcal{O}(n^\delta)$, where $\delta \in [2, 3]$, depending on the matrix inversion algorithm used (see e.g. \citealp{bos2013fast}). Although there have been improvements, such as the restricted MLE in \cite{tehranchi2021fast} or faster approximations in \cite{Bos2008} and \cite{bos2013fast} which can reduce computation times by factors of 10 to 100 compared to traditional methods \citep{williams2008cats}, the challenges of parameter estimation remain significant with large datasets and complex noise structures which are very common in these applications. To highlight the computational burden of such analyses,  \cite{tunini2024global} recently analysed a small network of relatively short GNSS time series (up to two decades of daily observations) that they managed to process in one week through sophisticated high-performance computing techniques. Based on this, the issue of scalability of the existing approaches is clearly still problematic, even without considering the additional limitations of these methods in presence of long-range dependence structures and missing observations. Moreover, if one were to add the task of model selection in this paradigm, there are no available solutions to do so in a computationally efficient manner.



\subsection{Main Contributions}

This work delivers new statistical and computational properties of the Generalized Method of Wavelet Moments with Exogenous Variables (GMWMX) derived from the method in \cite{cucci2023generalized}, itself being an extension of the Generalized Method of Wavelet Moments (GMWM) put forward in \cite{guerrier2013wavelet}. In particular, with respect to \cite{cucci2023generalized}, this work delivers theoretical guarantees and computational improvements which were not available before and extends them to model selection as well as the missing observation and long-range dependence paradigms which were not adequately addressed in previous work. More specifically, under broad dependence conditions and weak distributional assumptions, we show the following \textit{statistical properties} for the GMWMX:

\begin{enumerate}
    \item It delivers consistent estimators $\hat{\boldsymbol{\theta}}$ for the true parameter vector $\boldsymbol{\theta}_0$, i.e. $\hat{\boldsymbol{\theta}} \overset{p}{\to} \boldsymbol{\theta}_0$;
    \item Under common short-memory processes, its estimate of the true parameter $\boldsymbol{\beta}_0$, which we denote as $\hat{\boldsymbol{\beta}}$, achieves asymptotic normality, i.e. $\hat{\boldsymbol{\Phi}}^{-1/2} (\hat{\boldsymbol{\beta}} - \boldsymbol{\beta}_0) \overset{d}{\to} \mathcal{N}(\mathbf{0}, \mathbf{I})$, where $\hat{\boldsymbol{\Phi}}$ is the estimated covariance matrix of $\hat{\boldsymbol{\beta}}$ defined further on.
\end{enumerate}

As mentioned earlier, these statistical properties are achieved with high computational efficiency via important intermediate results and under very general and weak conditions as listed below:

\begin{enumerate}
    \item The moment-based nature of the GMWMX generally allows it to make fewer assumptions on the distribution of the stochastic error $\boldsymbol{\varepsilon}$, making the proposed method more flexible compared to likelihood-based approaches with respect to which the GMWMX does not suffer from significant loss in precision in the large sample settings it is proposed for;
    \item The GMWMX is applicable to a wide range of covariance (kernel) structures for the stochastic error $\boldsymbol{\varepsilon}$, including sums of different processes (i.e. \textit{latent processes}) composed of long-memory and/or non-stationary processes for which standard methods often fail to provide satisfactory results;
    \item To compute the quantities needed for the GMWMX, this work provides new and general forms of the theoretical Wavelet Variance (WV) which can now be computed for any (positive-definite) covariance function and any wavelet filter. This result is of great interest in its own right since these quantities are of extreme relevance in the natural sciences (e.g. metrology), finance and other fields. Based on the results in \cite{zhang2008allan}, these forms were only known for certain classes of covariance functions and only for the Haar wavelet filter. Moreover, if adding the assumption that $\boldsymbol{\varepsilon}$ follows a GP, this work also delivers a recursive formula to efficiently compute the variance of the WV in linear time. This result generalizes the results of \cite{zhang2008allan} to higher moments and opens to the possibility to perform computationally efficient model selection (among others).
\end{enumerate}

The above properties and advantages are highlighted and confirmed in simulation studies under different settings. The relevance of the GMWMX is then put forward through an application involving large scale GNSS signals.

\section{Estimation Framework}

To introduce the proposed inferential framework, let us firstly define it without considering the missing observation process $\boldsymbol{Z}$ which leaves us with the process defined in \eqref{eq.reg_complete} parametrized by $\boldsymbol{\theta}^* = [\boldsymbol{\beta}^T , \;\;\; \boldsymbol{\gamma}^T]^T$ (i.e. without the missing observation parameter $\boldsymbol{\vartheta}$). In this setting, a step-wise procedure to estimate $\boldsymbol{\theta}^*$ and perform inference on the parameter $\boldsymbol{\beta}$ is given by:

\begin{enumerate}[label=\textbf{Step \arabic*}, , leftmargin=1.5cm, itemsep=0.5em]
    \item Compute $\hat{\boldsymbol{\beta}}^*$ through least-squares to obtain residuals $\hat{\boldsymbol{\varepsilon}}^* = \boldsymbol{Y} - \boldsymbol{X}\hat{\boldsymbol{\beta}}^*$;
    \item Estimate $\boldsymbol{\gamma}$ by running the GMWM on the residual process $\hat{\boldsymbol{\varepsilon}}^*$ to obtain $\hat{\boldsymbol{\gamma}}^*$;
    \item Use $\hat{\boldsymbol{\gamma}}^*$ to compute $\boldsymbol{\Sigma}(\hat{\bm{\gamma}}^*)$ and obtain the covariance matrix of $\hat{\boldsymbol{\beta}}^*$ for inference.
\end{enumerate}

In particular, the GMWM used in \textbf{Step 2} was proposed in \cite{guerrier2013wavelet} as a computationally efficient approach to estimate complex (latent) stochastic models, parametrized by $\bm{\gamma}$, and is defined as follows:
\begin{equation}
    \hat{\bm{\gamma}}^* = \argmin_{\bm{\gamma} \in \bm{\Gamma}} \,[\hat{\bm{\nu}}^* - \bm{\nu}(\bm{\gamma})]^T \, \bm{\Omega} \, [\hat{\bm{\nu}}^* - \bm{\nu}(\bm{\gamma})],
    \label{eq:gmwm}
\end{equation}
where $\hat{\bm{\nu}}^* \in \mathbb{R}^J$ (with $J < \log_2(n)$) is the estimated WV from the process (i.e. the residuals $\hat{\boldsymbol{\varepsilon}}^*$ in the above step-wise procedure), $\bm{\nu}(\bm{\gamma})$ is the theoretical WV implied by the assumed stochastic model, and $\bm{\Omega}$ is any positive-definite weighting matrix whose choice is made to achieve statistical efficiency of the estimator $\hat{\bm{\gamma}}^*$. Although in finite-samples this may not always be the best choice, {the optimal choice for this weighting matrix is given by $\bm{\Omega}^{\star \star} := \bm{V}^{-1}$, where $\bm{V} = \mathbb{V}[\hat{\bm{\nu}}^*]$ is the covariance matrix of the WV estimator} \citep[see][]{hansen1982large, guerrier2013wavelet}.  

The above step-wise procedure is a common one in the Generalized Method of Moments (GMM) literature which also considers iterative versions in the form of a Generalized Least Squares (GLS) solution:
\begin{equation}
   \hat{\boldsymbol{\beta}}^{(k)^*} = (\boldsymbol{X}^T\boldsymbol{\Psi}^{(k)^*}\boldsymbol{X})\boldsymbol{X}^T\boldsymbol{\Psi}^{(k)^*}\boldsymbol{Y}, 
   \label{eq:gls}
\end{equation}
where, with $k = 1, \hdots, K$ representing the iteration step, $\boldsymbol{\Psi}^{(k)^*} = [\boldsymbol{\Sigma}(\hat{\boldsymbol{\gamma}}^{(k)^*})]^{-1}$ is the inverse of the estimated covariance matrix of $\bm{Y}$ based on the residuals $\hat{\boldsymbol{\varepsilon}}^{(k)^*}$ at the $k^{th}$ iteration. This is, for example, the strategy used in the GMWMX put forward in \cite{cucci2023generalized} where the GMWM was indeed proposed for \textbf{Step 2} of the above step-wise procedure and where the final $\hat{\boldsymbol{\beta}}^{(K)^*}$ and $\boldsymbol{\Sigma}(\hat{\boldsymbol{\gamma}}^{(K)^*})$ were used for inference on $\boldsymbol{\beta}_0$. In the latter work, the one-step ($K=1$) and two-step ($K=2$) strategies were compared under specific settings confirming that there can be an improved statistical performance when increasing $K$ at the cost of an increased computational burden. In this work we only consider the case of $K=1$, allowing however our results to hold when increasing the number of iterations indefinitely. More specifically, with respect to the GMWMX procedure discussed in \cite{cucci2023generalized}, we adequately address the problem of missing observations by delivering theoretical guarantees of the resulting estimator of $\boldsymbol{\beta}$. Moreover we provide an appropriate estimator of the covariance matrix of the WV (which we denote as $\hat{\bm{V}}$) that can be computed efficiently and deliver more precise estimates of the parameter vector $\bm{\gamma}$ in addition to providing a wide range of tools for inference on the stochastic component (and consequently on the mean-function parametrized by $\bm{\beta}$). 

To formalize our contribution, recalling that $\widetilde{\boldsymbol{Y}} \in \mathbb{R}^n = \bm{Y} \odot \bm{Z}$ represents the observed process vector with null elements in the positions where observations are missing, and using $\otimes$ to denote the Kronecker product, we similarly define $\widetilde{\boldsymbol{X}} = \bm{Z} \otimes \bm{1}^T \odot \bm{X} \in \mathbb{R}^{n \times p}$ as the design matrix $\bm{X}$ with zero-valued vectors for the rows where observations are missing in $\boldsymbol{Y}$ (here $\bm{1}$ represents a vector of ones of dimension $p$). It is important to note that, throughout this work we will assume that all rows in $\bm{X}$ are observed (or can be deterministically derived/computed) implying that, as defined earlier, the missing observation process $\bm{Z}$ is solely related to $\bm{\varepsilon}$ and hence $\bm{Y}$ (therefore not to $\bm{X}$). Since we set $K = 1$, we have that the matrix $\boldsymbol{\Psi}$ in \eqref{eq:gls} is simply the identity matrix which, for \textbf{Step 1}, gives the estimator: 
\begin{equation}
    \hat{\boldsymbol{\beta}} = \left(\widetilde{\boldsymbol{X}}^T \widetilde{\boldsymbol{X}}\right)^{-1} \widetilde{\boldsymbol{X}}^T \widetilde{\boldsymbol{Y}}.
    \label{eq:beta_observed}
\end{equation}
Assuming that $\widetilde{\boldsymbol{X}}^T \widetilde{\boldsymbol{X}}$ is invertible (see \Cref{ass:missing} further on), the estimator $\hat{\bm{\beta}}$ is affected by a loss of statistical efficiency, compared to $\hat{\bm{\beta}}^*$, due to the impact of missing observations. Indeed, the main challenge for inference on $\hat{\bm{\beta}}$ is the estimation of its covariance matrix, i.e. $\bm{\Phi}$, given that it depends on the error model which is itself also affected by the missing observation process. As a consequence, the covariance matrix $\bm{\Phi}$ ultimately depends on the predictor design matrix $\bm{X}$ and the covariance matrix of the error process $\bm{\varepsilon}$ where missingness introduces irregular lags between observations that need to be adequately handled to correctly identify its dependence structure. The paragraphs below describe how we propose to deal with this setting. 

\subsection{Missingness in the Error Process}

In the absence of missing values, the covariance matrix of the estimator $\hat{\bm{\beta}}^*$ is defined as follows:
$$\bm{\Phi}^* = (\bm{X}^T\bm{X})^{-1} \bm{X}^T \bm{\Sigma}(\bm{\gamma}) \bm{X} (\bm{X}^T\bm{X})^{-1}.$$
It must be noted, at this stage, that $\bm{\Phi}^* = \bm{\Phi}^*_n$ is a sequence in $n$, since $\bm{X} = \bm{X}_n$ depends on $n$. Given the setting of this work, we know that the missing observations are only present in the response $\bm{Y}$ and we therefore have the entire predictor design matrix $\bm{X}$ available to us. As mentioned above, the main issue however consists in the form of the covariance matrix of the error process under missingness since it becomes a function also of the parameter $\bm{\vartheta}$ (and not only $\bm{\gamma}$).

To address the issue of missing observations in the error process, we firstly define the error process with missing observations as $\widetilde{\boldsymbol{\varepsilon}} = {\widetilde{\boldsymbol{Y}}} - \widetilde{{\boldsymbol{X}}} \boldsymbol{\beta}_0$ with corresponding residual process $\hat{\boldsymbol{\varepsilon}} = {\widetilde{\boldsymbol{Y}}} - \widetilde{{\boldsymbol{X}}} \hat{\boldsymbol{\beta}}$ (i.e. the estimated error process). Following these definitions, we propose to compute the empirical WV $\hat{\bm{\nu}}$ on the estimated error process $\hat{\boldsymbol{\varepsilon}}$, which does not correspond to the approach that was put forward for the GMWMX in \cite{cucci2023generalized}. In the latter the authors propose to perform the wavelet decomposition only on the available observations and to then compute the WV on the wavelet coefficients resulting from this procedure. However this approach heavily relies on the availability of sets of consecutive observations which does not necessarily correspond to more general missing observation settings where the mechanism can generate random missing observations in consecutive and non-consecutive patterns.

Having defined the estimated error process $\hat{\boldsymbol{\varepsilon}}$, let $\lambda_{\min}(\bm{X}^T \bm{X})$ represent the minimum eigenvalue of the matrix $\bm{X}^T \bm{X}$, with $\bm{X}_{i \cdot} \in \mathbb{R}^p$ representing the $i^{th}$ row of the design matrix $\bm{X}$. With these definitions we can deliver Lemma \ref{lemma:beta_cov} which provides the form of the asymptotic covariance matrix of $\hat{\bm{\beta}}$ in the presence of the missing observation process. Before giving this result, we need to specify a key assumption needed for this lemma for which we consider the following Markov chain-based model for missingness patterns:
\begin{equation}\label{eq:markov_missinigness}
    Z_i \begin{cases}
        = Z_{i-1}, & \text{with probability $\rho$},\\
        \sim \mathrm{Bernoulli}(\mu(\bm{\bm{\vartheta}})), & \text{with probability $1-\rho$},
    \end{cases}
\end{equation}
where $\rho \in [0,1)$ and $\mathrm{Bernoulli}(\mu(\bm{\bm{\vartheta}}))$ is a Bernoulli random variable with parameter $\mu(\bm{\bm{\vartheta}}) \in (0,1]$.  Note that $\rho$ governs the strength of temporal dependence (for example, in \Cref{lemma:ergodic} in the appendix we prove that $\bm{Z} = \{Z_i\}$ is a geometric ergodic sequence for any $\rho \in [0,1)$) and it can also be verified that the parameter $\mathbb{E}[Z_i] = \mu(\bm{\bm{\vartheta}})$. Similar models have been used to model dependent missingness patterns, for example in the context of dynamic networks \citep{xu2021online}.


\begin{Assumption}[Missingness]\label{ass:missing}
    Let the missinig observation process $\bm{Z}$ be a stationary Markov chain in the form of \eqref{eq:markov_missinigness} with $\rho \in [0, 1)$ and $\mu(\bm{\bm{\vartheta}}) \in (0,1]$. Then the expected value $\mu(\bm{\bm{\vartheta}}) = \mathbb E[Z_i]$ is such that
    $$\mu(\bm{\bm{\vartheta}})\,\lambda_{\min}(\bm{X}^T \bm{X}) \geq C_{x}\, \alpha_n \, \max_i (\|X_{i \cdot}\|_{2}^2)\,n^{1/2}\,\log (n),$$
    where $C_{x} > 0$ is some constant and $\alpha_n > 0$ is any slowly diverging sequence.
\end{Assumption}

The implication of this assumption is essentially that the average proportion of missing observations does not affect the invertibility of the matrix $\widetilde{\bm{X}}^T \widetilde{\bm{X}}$ which is needed to compute $\hat{\bm{\beta}}$ and all quantities derived from it (including its covariance matrix). As highlighted in the proof of Lemma \ref{lemma:beta_cov}, this assumption guarantees (asymptotically) that this matrix is indeed invertibile. Based on this, we deliver the following lemma.


\begin{Lemma}\label{lemma:beta_cov}
    Suppose \Cref{ass:missing} holds. Then, as $n \rightarrow \infty$, we have that

\begin{align}
    (\hat{\bm{\beta}} - \bm{\beta}_0) - \mu(\bm{\vartheta})^{-1}(\bm{X}^T\bm{X} )^{-1} \bm{X}^T \widetilde{\bm{\varepsilon}} = o_p(1),
\end{align}
and hence we have that
    $$\bm{\Phi} = \mathbb{V}[\mu(\bm{\vartheta})^{-1}(\bm{X}^T\bm{X} )^{-1} \bm{X}^T \widetilde{\bm{\varepsilon}}] = \mu(\bm{\vartheta})^{-2}(\bm{X}^T\bm{X})^{-1} \bm{X}^T \bm{\Sigma}(\bm{\gamma}, \bm{\vartheta}) \bm{X} (\bm{X}^T\bm{X})^{-1} , 
    $$
    where $\bm{\Sigma}(\bm{\gamma}, \bm{\vartheta}) = [\bm{\Lambda}(\bm{\vartheta}) + \mu(\bm{\vartheta})^2 \bm{1}\bm{1}^T] \odot \bm{\Sigma}(\bm{\gamma})$.
\end{Lemma}

The proof of this lemma is given in Appendix \ref{app:lemma_beta_cov}. It must be noted that, if we assume additionally that the parameter space of $\bm{\beta}_0$ is compact, we could also state that asymptotically $\mathbb{V}[\hat{\bm{\beta}} - \bm{\beta}_0] - \bm{\Phi} \to 0$. Having said this, in practice we use $\bm{\Phi}$ to estimate $\bm{\Phi}^*$ where we note that $\bm{\Phi} = \bm{\Phi}_n$ is a degenerate sequence in $n$ whose limit can be obtained after proper rescaling (the asymptotic variance after rescaling is given in \Cref{sec:asy_property} under two dependence regimes). Hence, this result is very helpful since, given also the independence between $\bm{Z}$ and $\bm{\varepsilon}$, we can separately estimate the parameters of these two processes to obtain an estimator of the covariance matrix $\bm{\Sigma}(\bm{\gamma}, \bm{\vartheta})$ and consequently of the covariance matrix $\bm{\Phi}$. In this perspective, we consider a two-step strategy in the estimation of $\bm{\Sigma}(\bm{\gamma}, \bm{\vartheta})$ where we firstly estimate the missing observation process parameter $\bm{\vartheta}$, for which many approaches exist \citep[see e.g.][and references therein]{chib2001markov}, and then use the GMWM in \eqref{eq:gmwm} to obtain an estimate of the error process parameter $\bm{\gamma}$. Nevertheless, to run the GMWM we need the corresponding form of the theoretical WV which has this far only been defined for specific wavelet filters and certain classes of stochastic process \citep[see e.g.][]{zhang2008allan, guerrier2013wavelet, guerrier2021robust}. For this reason we put forward a general form of the theoretical WV for a zero-mean stochastic process (as is the case of the error process $\bm{\varepsilon}$). In this perspective, let $\bm{h}_{j} = [h_{j,L_j - 1}, \hdots, h_{j, 0}]^T \in \mathbb{R}^{L_j}$ be a generic wavelet filter vector at scale $j \in \mathbb{N}^+$ of length $L_j$ and let $M_j$ represent the number of wavelet coefficients issued from the wavelet decomposition at scale $j$. This allows us to define 
\begin{equation}
    \mathbf{A}_j = \sum_{l=0}^{L_j-1}\frac{h_{j,l}}{M_j} \bm{H}_{j,l},
    \label{eq:filt_mat}
\end{equation}

where $\bm{H}_{j,l}$ is an $n \times n$ matrix such that its $n \times (n-L_j+1)$ partitioned matrix with its $1^{st}$ to $n^{th}$ rows and its $(L_j-l)^{th}$ to $(n-l)^{th}$ columns given by
$$\begin{pmatrix} \bm{h}_{j} & 0           & \ldots & 0    \\ 
0 & \bm{h}_{j} & 0 & \vdots     \\  
\vdots & 0 & \ldots & \vdots       \\ 
\vdots      & \vdots      & \vdots & 0     \\  \ldots      & \ldots      & 0 & \bm{h}_{j} \end{pmatrix}.$$

With the above definitions and using $\tr[\cdot]$ to denote the matrix trace operator, we can deliver the general form of the theoretical WV for a zero-mean stochastic process in the following proposition.
\begin{Proposition}
    Assume that $\bm{\varepsilon}$ is a zero-mean stochastic process with covariance matrix $\bm{\Sigma}(\bm{\eta})$ parametrized by $\bm{\eta}$, then the theoretical WV at scale $j$ is given by
    $$\nu_j(\bm{\eta}) = \tr[\bm{A}_j \bm{\Sigma}(\bm{\eta})].$$
    \label{prop:theo_wv}
\end{Proposition}
The proof of this proposition is a result of the expectation of a quadratic form of the WV for a zero-mean error process such as $\bm{\varepsilon}$. Indeed, letting $W_{j,i} = \bm{\varepsilon}_j^T \bm{h}_{j}$ represent the wavelet coefficients (where $\bm{\varepsilon}_j = [\varepsilon_{i}, \hdots, \varepsilon_{i+L_j-1}]^T \in \mathbb{R}^{L_j}$), the WV for a zero-mean process is given by $\mathbb{E}[W_{j,i}^2] = \mathbb{E}[(\bm{\varepsilon}_j^T \bm{h}_{j})^2] = \mathbb{E}[\bm{\varepsilon}_j^T \bm{h}_{j} \bm{h}_{j}^T \bm{\varepsilon}_j] = \tr[A_j \bm{\Sigma}_j(\bm{\eta})]$, where $A_j = \bm{h}_{j} \bm{h}_{j}^T$ is the outer product of the wavelet filter vector and $\bm{\Sigma}_j(\bm{\eta}) = \mathbb{V}[\bm{\varepsilon}_j]$. It is then straightforward to extend this form to the entire covariance matrix $\bm{\Sigma}(\bm{\eta})$ by defining $\bm{A}_j$ as in \eqref{eq:filt_mat}. Following this proposition, the theoretical WV for the process $\tilde{\bm{\varepsilon}}$ is given by:
\begin{equation}
    \nu_j\left(\boldsymbol{\gamma},{\boldsymbol{\vartheta}} \right) = \text{tr}\left\{\mathbf{A}_j \left[\boldsymbol{\Sigma}\left(\boldsymbol{\gamma}\right) \odot \left\{\boldsymbol{\Lambda}(\boldsymbol{\vartheta}) + \mu(\boldsymbol{\vartheta})^2 \mathbf{1} \mathbf{1}^T \right\}\right] \right\}.
    \label{eq:wv_missing}
\end{equation}

\begin{Remark}
\label{remark:compute_wvtheo}
    Proposition \ref{prop:theo_wv} generalizes the form of the theoretical WV for all wavelet filters and covariance structures as well as provide a useful result to study the properties of the GMWM. However the matrix multiplication in \eqref{eq:wv_missing} can be computationally-demanding to evaluate. For this purpose, the results in \cite{xu2017study} provide a computationally efficient form of the theoretical Allan variance (equivalent to the Haar WV up to a constant) for zero-mean stochastic processes such as $\bm{\varepsilon}$. Indeed in \cite{xu2017study} they generalize the results in \cite{zhang2008allan} to zero-mean non-stationary processes by using averages of the diagonals and sub-diagonals of the covariance matrix of $\bm{\varepsilon}$. What this implies is that the GMWM, which uses this form, does not require the storage of the $n \times n$ covariance \textbf{matrix} of $\bm{\varepsilon}$ (hence a storage cost of order $\mathcal{O}(n^2)$), but only of a \textbf{vector} of dimension $L_J = \mathcal{O}(n)$ which is then plugged into an explicit formula consisting in a linear combination of the elements of this vector (these elements being averages of the diagonals and sub-diagonals of the covariance matrix). This delivers an important computational advantage to the GMWM compared to other approaches which require the storage of the entire matrix and the use of all its elements to obtain the required estimators. We postulate that such a strategy can be extended to other filters based on the form in Proposition \ref{prop:theo_wv} and we leave this for future research.
\end{Remark}

With $\bm{\nu}(\bm{\gamma}, \bm{\vartheta}) = [\nu_j(\bm{\gamma}, \bm{\vartheta})]_{j = 1, \hdots, J}$ and using an appropriate estimator for $\bm{\vartheta}$, we can consequently define the GMWM with missing observations as follows:
\begin{equation}
    \hat{\bm{\gamma}} = \argmin_{\bm{\gamma} \in \bm{\Gamma}} \,[\hat{\bm{\nu}} - \bm{\nu}(\bm{\gamma}, \hat{\bm{\vartheta}})]^T \, \bm{\Omega} \, [\hat{\bm{\nu}} - \bm{\nu}(\bm{\gamma}, \hat{\bm{\vartheta}})],
    \label{eq:gmwm_new}
\end{equation}
where $\hat{\bm{\nu}}$ is the estimated WV computed on the residuals $\hat{\bm{\varepsilon}}$ and $\hat{\bm{\vartheta}}$ is the pre-computed estimator of the parameter $\bm{\vartheta}$. The estimators $\hat{\bm{\gamma}}$ and $\hat{\bm{\vartheta}}$ can then be plugged into the covariance function $\bm{\Sigma}(\bm{\gamma}, \bm{\vartheta})$ to estimate the matrix $\bm{\Phi}$ and perform inference on $\bm{\beta}_0$.

\begin{Remark}
\label{remark:sigma}
    It must be underlined that $\bm{\Sigma}(\bm{\gamma})$ represented in Lemma \ref{lemma:beta_cov} and in \eqref{eq:wv_missing} represents the covariance matrix of the true error process $\bm{\varepsilon}$ disregarding the missing observation process $\bm{Z}$. However, the empirical WV $\hat{\bm{\nu}}$ is computed on the estimated error (residual) process where the exact covariance matrix of the residuals would be $\bm{\Sigma}_{\hat{\bm{\varepsilon}}}(\bm{\gamma}) = (\bm{I} - \bm{P})\,\bm{\Sigma}(\bm{\gamma})\,(\bm{I} - \bm{P})$, where $\bm{P} = \bm{X}(\bm{X}^T\bm{X})^{-1}\bm{X}^T$ is the projection (or ``hat'') matrix. Therefore, a first option would consist in ignoring the estimated nature of this error process and make a large sample approximation (which is the setting for this work) and simply state that $\bm{\Sigma}_{\hat{\bm{\varepsilon}}}(\bm{\gamma}) \approx \bm{\Sigma}(\bm{\gamma})$. Otherwise, the second option would consist in using the exact form given by $\bm{\Sigma}_{\hat{\bm{\varepsilon}}}(\bm{\gamma})$. The latter form however can be computationally burdensome as data scales, therefore we propose an approximation of this form in Appendix \ref{app:IminusH} and study its performance in Section \ref{sec:simulations}.
\end{Remark}

Letting $\tilde{\bm{y}}$ and $\bm{z}$ denote the realizations of the random variables $\widetilde{\bm{Y}}$ and $\bm{Z}$ respectively, the complete proposed procedure is summarized in Algorithm~\ref{algo:gmwmx} which delivers the estimators $\hat{\bm{\beta}}$ and its covariance matrix $\bm{\Phi}$ (also under missingness) based on (i) the approximation of $\bm{\Sigma}_{\hat{\bm{\varepsilon}}}(\bm{\gamma})$ given in Appendix \ref{app:IminusH} and on (ii) a two-step GMWM approach which uses a first estimate $\hat{\bm{\gamma}}$ to obtain an ``optimal'' weighting matrix $\bm{\Omega}$ for the second step GMWM. Algorithm \ref{algo:gmwmx} therefore corresponds to the most computationally expensive procedure within the proposed framework and is used within the simulation and case-study sections further on. Considering that the statistical properties of the GMWMX would generally remain asymptotically (approximately) valid even without these additional steps, the results in Sections \ref{sec:simulations} and \ref{sec:case_study} should be interpreted as being the most computationally expensive (one-step) version of the proposed procedure (with the two or more steps delivering optimal statistical properties).


\begin{algorithm}
\caption{One-Step GMWMX}
\begin{algorithmic}[1]

\vspace{0.4em}
\Require Complete design matrix $\boldsymbol{X}$; response vector with zero-values in positions with missing observations $\widetilde{\boldsymbol{y}}$; index vector of missing values
$\boldsymbol{z}$.
\vspace{1em}
\State  Compute $ \hat{\boldsymbol{\beta}} = (\widetilde{\boldsymbol{X}}^T \widetilde{\boldsymbol{X}})^{-1} \widetilde{\boldsymbol{X}}^T \widetilde{\boldsymbol{y}}$, where $\widetilde{\boldsymbol{X}}= \boldsymbol{z} \otimes \mathbf{1}^T \odot \boldsymbol{X}$.

\vspace{0.4em}
\State Compute  $\hat{\boldsymbol{\varepsilon}} = {\widetilde{\boldsymbol{y}}} - \widetilde{{\boldsymbol{X}}} \hat{\boldsymbol{\beta}}$.

\vspace{0.4em}
\State Compute $\hat{\boldsymbol{\vartheta}}$ based on $\boldsymbol{z}$ and obtain $\mu(\hat{\boldsymbol{\vartheta}})$ and $\boldsymbol{\Lambda}(\boldsymbol{\hat{\vartheta}})$.

\vspace{0.4em}
\State Following Lemma \ref{lemma:beta_cov} and Remark \ref{remark:sigma}, define $$\bm{\Sigma}_{\hat{\bm{\varepsilon}}}(\bm{\gamma}) = [\boldsymbol{\Lambda}(\hat{\boldsymbol{\vartheta}}) + \mu(\hat{\boldsymbol{\vartheta}})^2 \mathbf{1} \mathbf{1}^T ] \odot ( \boldsymbol{I} - \bm{P})\boldsymbol{\Sigma}(\boldsymbol{\gamma}) (\boldsymbol{I} - \bm{P})$$ using the approximation in Appendix \ref{app:IminusH}.


\vspace{0.4em}
\State Compute $\hat{\bm{\gamma}}$ through \eqref{eq:gmwm_new} by using $\bm{\Sigma}_{\hat{\bm{\varepsilon}}}(\bm{\gamma})$ instead of $\bm{\Sigma}(\bm{\gamma})$ in Proposition \ref{prop:theo_wv}; using a diagonal matrix $\bm{\Omega}$ with entries proportional to estimated variances of $\{\hat{\nu}_j\}_{j = 1, \hdots, J}$; and employing computational approaches in Remark \ref{remark:compute_wvtheo}.

\vspace{0.4em}
\State Compute an estimator for $\hat{\bm{V}}$ (the covariance matrix of $\hat{\bm{\nu}}$) based on the form in Proposition \ref{prop:V} and the computational approaches in Appendix \ref{app:compute_v} (see Section \ref{sec:covariance_wv} further on).


\vspace{0.4em}
\State Define $\boldsymbol{\Omega}^\star = \boldsymbol{\hat{V}}^{-1}$ and use this in \eqref{eq:gmwm_new} to re-compute $\hat{\bm{\gamma}}$ (denoted as $\hat{\bm{\gamma}}^\diamond$).



\vspace{0.4em}
\State Compute $$\hat{ \boldsymbol{\Phi}}  = \mu(\hat{\bm{\vartheta}})^{-2}(\bm{X}^T\bm{X})^{-1} \bm{X}^T \left\{[\bm{\Lambda}(\hat{\bm{\vartheta}}) + \mu(\hat{\bm{\vartheta}})^2 \bm{1}\bm{1}^T] \odot \hat{\bm{\Sigma}}(\hat{\bm{\gamma}}^\diamond)\right\} \bm{X} (\bm{X}^T\bm{X})^{-1},$$ based on Lemma \ref{lemma:beta_cov}.

\vspace{1em}
\hspace{-1.2cm}\textbf{Output}: Parameter vector $\hat{\bm{\beta}}$ and its estimated covariance matrix $\hat{\bm{\Phi}}$

\end{algorithmic}
\label{algo:gmwmx}
\end{algorithm}

Notice that the algorithm is called ``One-Step GMWMX'' since the matrix $\hat{\bm{\Sigma}}(\hat{\bm{\gamma}}^\diamond)$ used in point 8 of Algorithm~\ref{algo:gmwmx} could be used as $\Psi = [\hat{\bm{\Sigma}}(\hat{\bm{\gamma}}^\diamond)]^{-1}$ from \eqref{eq:gls} to perform an iteration from point 1 of the same algorithm.

\subsection{Asymptotic Properties}\label{sec:asy_property}

In this section we describe assumptions and relative results regarding the asymptotic properties of the estimator of interest $\hat{\bm{\beta}}$. In particular, we provide two results for this estimator under two different regimes for the rate of decay of the dependence structure of the error process $\bm{\varepsilon}$, more specifically the \textit{short-memory} and \textit{long-memory} regimes. For both regimes, we prove the asymptotic properties of the proposed GMWMX procedure in the presence of missing observations.

\subsubsection{Short-Memory Regime}\label{sec:short-memory}

For the short-memory regime, we consider that the error process $\bm{\varepsilon} = \{\varepsilon_{i}\}_{i \in \mathbb{Z}}$ satisfies the following representation
\begin{equation}\label{eq:functional_wold}
    \varepsilon_{i} = G(\mathcal{F}_i),
\end{equation}
where $G$ is an $\mathbb{R}$-valued measurable function, $\mathcal{F}_i = \sigma(\ldots, e_{i-1}, e_i)$ is the natural filtration of the innovations process $\{e_i\}_{i \in \mathbb{Z}}$ which consists in a sequence of independent and identically distributed (i.i.d.) random variables.  Note that the representation in \eqref{eq:functional_wold} implies stationarity and is general, which covers a large class of linear and nonlinear processes. For instance, the linear process \eqref{eq:error_lp} considered in \Cref{sec:long-memory} is a special case of this representation.

We allow $\{\varepsilon_{i}\}_{i \in \mathbb{Z}}$ to possess temporal dependence whose strength can be measured by the functional dependence measure \citep{wu2005nonlinear} which we proceed to define and adapt to the context of this work. We start by defining the coupled random variables which, for any $i \in \mathbb{Z}$ and any integer $s$, is represented as $$\varepsilon_{i, \{s\}} = G_i(\mathcal{F}_{i,\{s\}}),$$ where the filtration is defined as
\begin{align*}
    \mathcal{F}_{i,\{s\}} = 
    \begin{cases}
        \sigma(\dots, e_{s-1}, e_{s}^*, e_{s+1}, \dots, e_i),  \quad &s < i,\\
        \sigma(\dots, e_{i}),  \quad &i < s,\\
    \end{cases}
\end{align*}
with $e_i^*$ representing an independent copy of $e_i$. Generally speaking, the random variable defined by this filtration is a ``copy'' of the original random variable $\varepsilon_i$ in which some elements of the independently distributed random variable $\{e_i\}$ are replaced by other independent copies of this variable. In particular, since these sequences are input into the function $G$, we expect the resulting process $\bm{\varepsilon}$ to be affected by these modifications (i.e. the copy is a ``perturbed'' version of the original). The reason for defining these is to then be able to measure to what extent a process is affected by changes in the past values of the innovation process $\{e_i\}$, thereby quantifying the range of dependence of the process $\bm{\varepsilon}$ on its past. With this in mind, for some integer $q > 0$ such that 
$\sup_{i \in \mathbb{Z}}\Vert \varepsilon_{i} \Vert_q < \infty$, we define the functional dependence measure of order $q$ and its tail cumulative version as
\begin{equation}\label{FDM}
        \delta_{s,q} = \sup_{i \in \mathbb{Z}}\Vert \varepsilon_{i} - \varepsilon_{i,\{i-s\}} \Vert_{q} \quad \text{and} \quad \Delta_{m, q} = \sum_{s = m}^{\infty} \delta_{s,q}.
    \end{equation}

Intuitively, $\delta_{s,q}$ quantifies how much the $q^{th}$ moment of $\bm{\varepsilon}$ differs from its copy and $\Delta_{m, q}$ quantifies how much this difference accumulates over lags in the past, where the latter being large indicates a significant dependence of ${\varepsilon}_i$ on its past values. Having these definitions, we can now deliver the required results on the asymptotic properties of $\hat{\bm{\beta}}$ defined in \eqref{eq:beta_observed}. In this perspective, we also provide the following assumptions for which we also define $\bm{D}_n \in \mathbb R^{p \times p}$ as being a diagonal matrix given by
\begin{align*}
    \bm{D}_n = \diag(\bm{X}^T\bm{X}) = \diag(\|\bm{X}_{\cdot 1}\|_2^2, \dots, \|\bm{X}_{\cdot p}\|_2^2),
\end{align*}
where $\bm{X}_{\cdot b} \in \mathbb R^n$ is the $b^{th}$ column of $\bm{X}$. Moreover, we define the following quantities that will also be used throughout them:
\begin{itemize}
    \item $\bm{\Pi}_n = \bm{X}^T\bm{X}$;
    \item $\widetilde{\bm{\Pi}}_n = \widetilde{\bm{X}}^{\top}\widetilde{\bm{X}}$;
    \item $\bm{C}_n = \bm{D}_n^{-1/2}\bm{X}^T\bm{X}\bm{D}_n^{-1/2}$;
    \item $\bm{Q}_{i \cdot} = \bm{\Pi}_n^{-1/2}\bm{X}_{i \cdot}$ and $\widetilde{\bm{Q}}_{i \cdot} = \bm{Q}_{i \cdot} \odot \bm{Z}$.
\end{itemize}

We can now list the assumptions needed, starting from the first two assumptions which place conditions on the behaviour of the design matrix $\bm{X}$ and its derived quantities.

\begin{Assumption}[Fixed Design]\label{ass:X}
    Let $\{\bm{X}_{i \cdot}\}_{i = 1, \hdots, n} \subset \mathbb{R}^p$ be a deterministic covariate process. Then we have that $$\max_i(\bm{X}_{i \cdot}^T \bm{D}_{n}^{-1} \bm{X}_{i \cdot}) = o(1),$$
    and there exists a nondegenerate deterministic matrix $\bm{C} \in \mathbb R^{p \times p}$ such that
$$\bm{C} = \lim_{n \to \infty} \bm{C}_n.$$
\end{Assumption}

\begin{Assumption}\label{ass:lagk}
There exists, for each $k \in \mathbb{Z}$, $\bm{\Phi}_k \in \mathbb{R}$, such that
    \begin{align*}
        \lim_{n \to \infty}\sum_{i = 1}^{n - |k|}\bm{Q}_{i}\bm{Q}_{i+k}^{\top} = \bm{\Phi}_k.
    \end{align*}
\end{Assumption}

The above assumptions are relatively weak and basically ensure that the design matrix is well defined in order to obtain stable quantities to derive the asymptotic distribution of $\hat{\bm{\beta}}$. The following assumption places one final condition related to the error process in order to deliver asymptotic properties in the short-memory regime.

\begin{Assumption}[Short-Memory Random Errors]\label{ass:epsilon}
    $\{\varepsilon_{i}\}_{i = 1, \hdots, n} \subset \mathbb{R}$ is a stationary process in the form of \eqref{eq:functional_wold} such that $\mathbb{E}(\varepsilon_{i}) = 0$, $\|\varepsilon_{i}\|_2 < \infty$ and there exists an absolute constant $ C_{\mathrm{FDM}} > 0$ such that
    \begin{equation*}
        \Delta_{0,2} \leq C_{\mathrm{FDM}}.
    \end{equation*}
\end{Assumption}

The latter assumption is common in the analysis of stochastic processes and places conditions on the variance of the error process $\bm{\varepsilon}$ and its covariance structure. Generally speaking, it requires the variance of $\bm{\varepsilon}$ to be finite and its covariance structure to be summable, in the sense that the covariance goes to zero as the lag $k$ grows. Using these assumptions, we now state the asymptotic properties of $\hat{\bm{\beta}}$.

\begin{Theorem}[Asymptotic Normality -- Short-Memory]\label{theorem:short_memory}
Under Assumptions \ref{ass:X} and \ref{ass:epsilon} we have that
    \begin{align*}
         \widetilde{\bm{\Pi}}_n^{1/2}(\hat{\bm{\beta}} - \bm{\beta}_0) = \sum_{i = 1}^n \widetilde{\varepsilon}_{i}\widetilde{\bm{Q}}_{i} = O_p(1).
    \end{align*}
In addition, if \Cref{ass:lagk} holds, then we have that
\begin{align}\label{eq:apply_CLT}
    \widetilde{\bm{\Pi}}_n^{1/2}(\hat{\bm{\beta}} - \bm{\beta}_0) \overset{\mathcal{D}}{\to} \mathcal{N}(\bm{0}, \bar{\bm{\Phi}}),
\end{align}
where $\bar{\bm{\Phi} }= \sum_{k \in \mathbb{Z}}\mathbb{E}[Z_0Z_k]\mathbb{E}[\varepsilon_{0}\varepsilon_{k}]\bm{\Phi}_k$. Consequently,
\begin{align*}
    \bm{D}_n^{1/2}(\hat{\bm{\beta}} - \bm{\beta}_0) 
    \xrightarrow{\mathcal{D}} \mu(\bm{\vartheta})^{-1}\bm{C}^{-1/2}\mathcal{N}(\bm{0}, \bar{\bm{\Phi}}).
\end{align*}
\end{Theorem}

The proof of this theorem can be found in Appendix \ref{proof:theorem_short_memory}. We underline that as $n \to \infty$, it can easily be shown that $D_n^{1/2}\bm{\Phi}D_n^{1/2} \to \mu(\bm{\vartheta})^{-2}\bm{C}^{-1/2} \bar{\bm{\Phi}}\bm{C}^{-1/2}$, where $\bm{\Phi}$ is given in \Cref{lemma:beta_cov}.

\subsubsection{Long-Memory Regime}\label{sec:long-memory}

For the long-memory regime we will only consider the linear process model for the error process $\bm{\varepsilon}$ which is defined as follows:
\begin{align}\label{eq:error_lp}
    \varepsilon_i = \sum_{k = 0}^{\infty}a_k e_{i-k},
\end{align}
where $e_h$, for $h \in \mathbb Z$, are i.i.d. innovations such as those defined in Section \ref{sec:short-memory} with $\mathbb E(e_h) = 0$ and $\var(e_h) = \sigma_{e}^2 < \infty$, and $a_h$, $h \in \mathbb N$, are real coefficients of the form
\begin{align*}
    a_h = c_a\,h^{d-1},
\end{align*}
where $c_a$ is some constant and $0 < d < 1/2$. Then, for $k \in \mathbb Z$, the autocovariance with lag-$k$ is given by:
\begin{align*}
    \cov(\varepsilon_i, \varepsilon_{i+k}) = \sigma_{e}^2\sum_{i = 0}^{\infty}a_i\,a_{i+k} \asymp C_{a}\frac{\sigma_{e}^2}{k^{1-2d}},
\end{align*}
where $C_{a} > 0$ is a constant. Note that when $0 < d < 1/2$ we have that $\cov(\varepsilon_i, \varepsilon_{i+k})$ is not summable and $\bm{\varepsilon}$ is a long-memory process. In contrast, when $d < 0$, we have that $\cov(\varepsilon_i, e_{i+k})$ is summable and $\bm{\varepsilon}$ is a short-memory process. We refer to \cite{beran2013long} for a comprehensive review of long-memory processes. In this regime, we define an additional assumption.

\begin{Assumption}[Long-Memory Random Errors]\label{ass:epsilon_long-memory}
    $\{\varepsilon_{i}\}_{i = 1, \hdots, n} \subset \mathbb{R}$ is a stationary process in the form of \eqref{eq:error_lp} such that $\mathbb{E}(\varepsilon_{i}) = 0$, $\|\varepsilon_{i}\|_2 < \infty$ and with $0 < d < 1/2$.
\end{Assumption}

With this assumption, which basically formalizes the conditions of the long-memory process we use for this section, we can now deliver the following result.

\begin{Theorem}[Asymptotic Distribution -- Long-Memory]\label{theorem:long_memory}
Under Assumptions \ref{ass:X} and \ref{ass:epsilon_long-memory}, we have that
\begin{align*}
    n^{-d}\bm{D}_n^{1/2}(\hat{\bm{\beta}} - \bm{\beta}_0) \xrightarrow{\mathcal{D}} \mu(\bm{\vartheta})^{-1}\, \bm{C}^{-1}\int_{0}^1 \bm{G}(u) \, \mathrm{d}B^{d}(u),
\end{align*}
where $\{B^d(u)\}_{u \geq 0} \subset \mathbb{R}$ is a standard fractional Brownian motion, the matrix $\bm{C}$ is given in \Cref{ass:X}, and $\bm{G}: \mathbb{R} \mapsto \mathbb{R}^p$ defined in the proof.
\end{Theorem}

The proof of this theorem can be found in Appendix \ref{proof:theorem_long_memory}.  While the limiting distribution given in \Cref{theorem:long_memory} is irregular, we could simulate this stochastic integral using the Monte Carlo method and obtain the numerical quantiles in order to perform inference on $\bm{\beta}_0$.

\subsection{Variance of the WV}
\label{sec:covariance_wv}

The results presented this far regarding the estimation of $\bm{\gamma}$ through the GMWM in \eqref{eq:gmwm} (and consequently in \eqref{eq:gmwm_new}) all implicitly rely on the matrix $\bm{\Omega}$ which can be any positive-definite weighting matrix that contributes to the asymptotic variance of the estimator $\hat{\bm{\gamma}}$. Indeed, the main purpose of this matrix is to guarantee efficiency of the resulting estimator $\hat{\bm{\gamma}}$ where, as stated earlier, the optimal choice consists in $\bm{\Omega} = \bm{V}^{-1}$ where $\bm{V} = \mathbb{V}[\hat{\bm{\nu}}]$ \citep[see][]{hansen1982large}. While in practice it is often more effective to use a simple matrix for $\bm{\Omega}$ (such as an identity matrix or a diagonal matrix whose entries are proportional to the variances of the elements of $\hat{\bm{\nu}}$), the derivation of the matrix $\bm{V}$ can be extremely useful under many aspects beyond the GMWM setting, for example in approaches that employ the WV for Portmanteau or isotropy tests (among others). To date there has not been an explicit form, even of approximate nature, that can be used to define and directly estimate $\bm{V}$. Among the few solutions we can find the form based on the asymptotic variance of M-estimators that was put forward in \cite{guerrier2021robust} which however can be computationally cumbersome as data scales.

Here we provide an explicit form to the matrix $\bm{V}$ under the assumption that $\bm{\varepsilon}$ is a strongly stationary process. More specifically, we make the assumption that the wavelet coefficients issued from the wavelet decomposition of $\bm{\varepsilon}$ follow a zero-mean Gaussian process with stationary covariance matrix $\bm{\Sigma}_W(\bm{\gamma})$. This is exactly the case when $\bm{\varepsilon} \sim \mathcal{N}(\bm{0}, \bm{\Sigma}(\bm{\gamma}))$ or, using a central-limit theorem argument, can be a reasonable approximation given that the wavelet coefficients are the result of a linear combination of elements of $\bm{\varepsilon}$ at different scales. Under this assumption, the following proposition provides the form of the matrix $\bm{V}$.

\begin{Proposition}
\label{prop:V}
    Assuming that the wavelet coefficients issued from the wavelet decomposition of the error process $\bm{\varepsilon}$ follow a stationary Gaussian distribution, then the $(j,l)^{th}$ element of the covariance matrix $\bm{V} \in \mathbb{R}^{J \times J}$ is given by

    $$v_{j,l} = 2 \tr\left[\mathbf{A}_j \boldsymbol{\Sigma}({\boldsymbol{\gamma}} )\mathbf{A}_l \boldsymbol{\Sigma}({\boldsymbol{\gamma}} ) \right],$$

    where $\bm{A}_j$ is the wavelet filter matrix for scale $j$ defined in \eqref{eq:filt_mat}.
\end{Proposition}

The proof of this proposition is a straightforward consequence of the variance of quadratic forms of Gaussian variables (similarly to the proof of Proposition \ref{prop:theo_wv}). From the definition in this proposition, we consequently have that $v_{j,j} = \mathbb{V}[\hat{\nu}_j^2]$ and $v_{j,l} = \cov(\hat{\nu}_j^2, \hat{\nu}_l^2)$. While this form is extremely useful to study the properties of the resulting estimators, also as a reasonable approximation even if the error process is non-Gaussian, the direct use of this form can be computationally demanding as (and even more than) the computation of the theoretical form of the WV given in Lemma \ref{prop:theo_wv}. Therefore, as for Lemma \ref{prop:theo_wv}, we put forward a recursive algorithm (inspired by the pyramid algorithm used for the wavelet decomposition) that allows us to compute the elements $v_{l,j}$ of this matrix in linear time. Using the Haar wavelet filter to illustrate the procedure, Appendix \ref{app:compute_v} describes these recursive steps and defines the explicit quantities computed in these recursions. More specifically, the quantities defining the elements $v_{j,l}$ are different based on whether the error process $\bm{\varepsilon}$ is stationary and non-stationary (with the stationary version being a special case of the non-stationary version).


\section{Simulation Study}
\label{sec:simulations}

In this section, we evaluate the performance of the proposed GMWMX framework compared to the MLE implemented in the software \texttt{Hector} v2.1 of \cite{Bos2008}. The reason for this comparison is two-fold in nature because we intend to compare the statistical performance of the GMWMX to the method (MLE) which benefits from optimal statistical properties and also benefits from low algorithmic complexity ($\mathcal{O}(n^2)$) among the methods which guarantee similar statistical performance (see \citealp{bos2013fast, he2017review}). To compare these two approaches we study their performance on the so-called \textit{functional} model commonly used in the field of Earth Sciences for the analysis of GNSS signals. More specifically, we consider a trajectory model based on the Equation 4 of \cite{he2017review} which represents a linear combination of seasonal signals and tectonic rate. Therefore, with $t_i$ representing the $i^{th}$ ordered time point (epoch) and $t_0$ representing the reference (initial) epoch, the deterministic component can be represented as follows:
\begin{equation}
\label{eq:func_model}
    \begin{aligned}
& \boldsymbol{X}_{i \cdot} \boldsymbol{\beta}=\beta_0+\beta_1\left(t_i-t_0\right) +\sum_{h=1}^2\left[\beta_{h+1} \cos \left(2 \pi f_h t_i\right) + \beta_{h+2} \sin \left(2 \pi f_h t_i\right)\right], 
\end{aligned}
\end{equation}
where $\beta_0$ is the initial position at the reference epoch $t_0$,  $\beta_1$ is the velocity parameter and $\beta_k$ ($k = 2, 3, 4, 5$) are the periodic motion parameters (where $h=1$ and $h=2$ represent the annual and semiannual seasonal terms respectively with $f_1 = \nicefrac{1}{365.25}$ and $f_2 = \nicefrac{2}{365.25}$). Hence, we have that $\boldsymbol{\beta} = [\beta_0, \ldots, \beta_5]^T$. Section \ref{sec:case_study} presents a case study where this  is of interest for many applications in the field of Earth Sciences, in particular for inference on the parameter $\beta_1$ (i.e. the velocity parameter).

To perform inference on these parameters, it is necessary to correctly estimate the parameters of the stochastic component, i.e.  the error process $\bm{\varepsilon}$,  since the latter is affected by complex forms of dependence which can considerably bias inferential conclusions. For this section we consider a general model that is widely used for the errors of large GNSS signals which consists in a combination of white noise (WN) and power-law (PL) processes which is also considered for our following case study section.  In fact, it is known that most GNSS times series exhibit a combination of WN and PL noise \citep{bos2013fast, he2017review, amiri2007assessment}, where  the WN generally characterizes the noise at high spectral frequencies while the PL model does the same for the lower frequencies. PL processes were originally discussed in \cite{mandelbrot1968fractional} (and in \citealp{gardner1978mathematical})  and are commonly observed in a wide variety of geophysical processes \citep{williams2004error}. Various studies highlight that a combination of WN and PL noise seems to best describe the noise of GNSS time series \citep{zhang1997southern, calais1999continuous, bock2000instantaneous}. More specifically, the PL noise is defined by its power spectrum $P(f) = \nicefrac{P_0}{f^{\alpha}}$ where $f$ is the frequency, $P_0$ is a constant and $\alpha$ is the spectral index.  Alternatively, the PL noise can be represented as a linear process like the one in \eqref{eq:error_lp}, where the spectral index $\alpha$ determines the coefficients $a_k$ while the variance of the i.i.d. innovations is represented by $\sigma^2_{\text{PL}}$ (see Appendix \ref{app:IminusH} for a more detailed discussion on these processes). In particular, if the spectral index is such that $\alpha<1$, then the PL noise is stationary whereas it becomes non-stationary for $\alpha \geq 1$. Therefore, for our simulations we consider three general settings for the error process: 
\begin{enumerate}[label=\textbf{Setting \Alph*}, , leftmargin=2.5cm, itemsep=0.5em]
    \item WN and a PL process with spectral index $\alpha < 1$ (stationary model);
    \item WN and a \textit{flicker} noise which is a PL process with spectral index $\alpha = 1$ (non-stationary model);
    \item WN and a Matérn process (stationary model).
\end{enumerate}

In this section however, for ease of reading, we only highlight the results for \textbf{Setting A} and, while briefly discussing the results of the other settings in this section, we refer the reader to Appendix \ref{app:extended_simu_results} for the complete results. This being said, our simulations take inspiration from the models and data structures coming from Earth Sciences (on which Section \ref{sec:case_study} is based) and therefore consider different lengths of synthetic GNSS daily position time series, i.e.  $10$, $20$, $30$ and $40$ years. Moreover, for the missingness process $\boldsymbol{Z}$, we consider a Markov model with transition probabilities defined as follows:

\begin{equation}
    \label{eq:markov_model_def}
    \begin{aligned}
& P\left( Z_2=1 \mid Z_1=1\right)=1 - p_1 \\
& P\left(Z_2=1 \mid Z_1=0\right) = p_2 \\
& P\left(Z_2=0 \mid Z_1=1\right)=p_1 \\
& P\left(Z_2=0 \mid Z_1=0\right)=1-p_2.
\end{aligned}
\end{equation}

It can be shown that the general Markov chain represented in \eqref{eq:markov_model_def} is a special example of \eqref{eq:markov_missinigness} and hence respects this requirement for the GMWMX properties to hold. Proof of this is given in Appendix \ref{app:beta_mixing}. With this definition, we can interpret $p_1 \in (0,1)$ as the probability of an observation becoming missing while $p_2 \in (0,1)$ is the opposite (hence, for example, a higher the value of $p_1$ implies a higher probability of missing observations, all other things remaining equal). Based on this, we have $ \mathbb{E}[Z]= \nicefrac{p_2}{p_1 + p_2}$ which can be interpreted as the general probability of a observing a value. Modeling the missingness in GNSS time series using such a process seems appropriate since missing observations can occur due to environmental or human factors, like receiver antenna replacement, poor observation conditions, signal interruptions, or intense crustal movement \citep{bao2021filling}. These factors generally entail a missingness process where there is a low probability of not observing the next data point, but when a point is not observed, it is likely that several consecutive observations will also be missing. Based on this we consider 5 settings of missingness which are summarized in Table \ref{tab:parameters:missingness}.

\begin{table}[h!]
\centering
\begin{tabular}{c c c c}
\toprule
Setting & $p_1$ & $p_2$ & $\mathbb{E}[Z]$\\ 
\midrule
1 & 0.00 & - & 1 \\ 
2 & 0.05 & 0.45 & 0.9 \\ 
3 & 0.05 & 0.20 & 0.8 \\ 
4 & 0.05 & 0.17$^*$ & 0.70$^*$  \\ 
5 & 0.10 & 0.15 & 0.6 \\ 
6 & 0.10 & 0.10 & 0.5 \\ 
\bottomrule
\end{tabular}
\caption{Parameter values for the missigness process $\boldsymbol{Z}$ for each setting. Entries marked with $^*$ have been rounded to two decimal places.}
\label{tab:parameters:missingness}
\end{table}

While the parameters of the other settings are described in Appendix \ref{app:extended_simu_results}, in \textbf{Setting A} we fix the parameter values to be $\sigma^2_{\text{WN}} = 10$ (WN variance), $\sigma^2_{\text{PL}} = 6$ (PL variance), $\alpha_{\text{PL}} = 0.9$ (PL spectral index). The choice of the values is in line with those that are often estimated on real-world GNSS data. For all settings, $M=2000$ replications are performed with the goal of accurately measuring coverage of the elements of the parameter vector $\bm{\beta}$ (among others). Given the different combination of sample sizes and missing observation scenarios, we split each setting into version 1 and 2: so, for example, for \textbf{Setting A} we have that \textbf{Setting A1} fixes the proportion of missing observations to 10\% and varies the sample size, while in \textbf{Setting A2} we fix the sample size to 20 years of daily measurements (i.e. $n = 7300$) and let the proportion of missing observations vary.

\begin{figure}
    \centering
 \includegraphics[width=.5\linewidth]{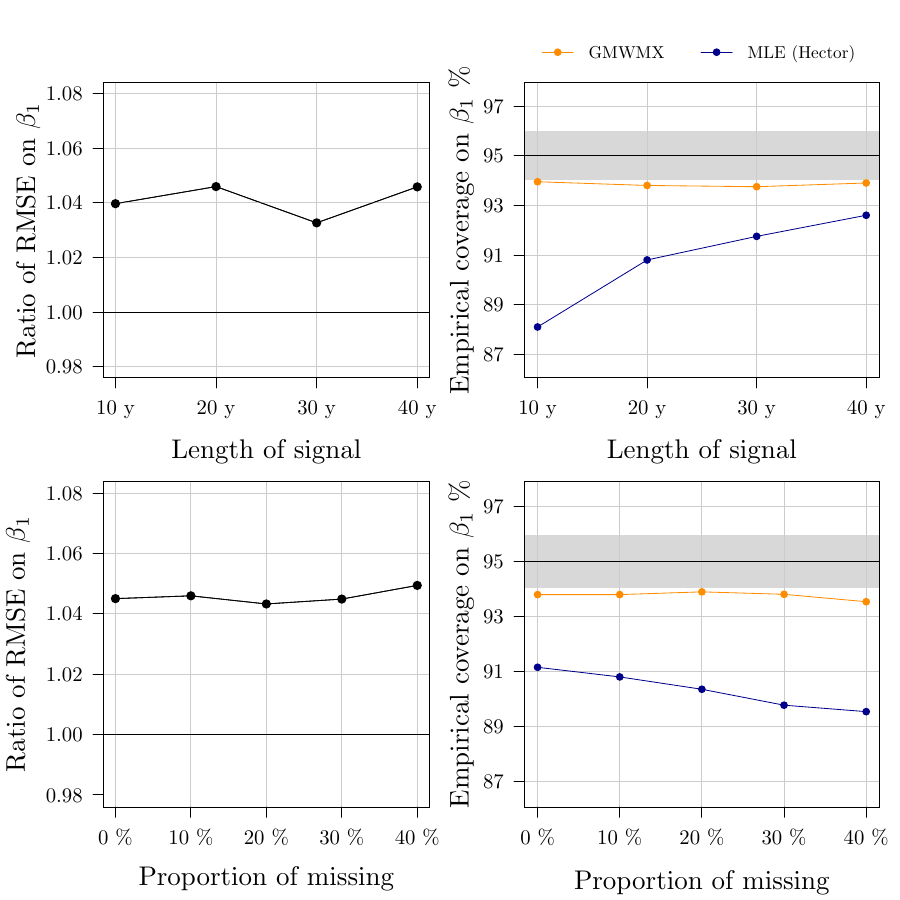}
     \caption{Ratio of RMSE and empirical coverage for parameter $\beta_1$ for Setting A1 and A2.}
      \label{fig:beta_setting_a}
\end{figure}
\begin{figure}
    \centering
    \includegraphics[width=0.6\linewidth]{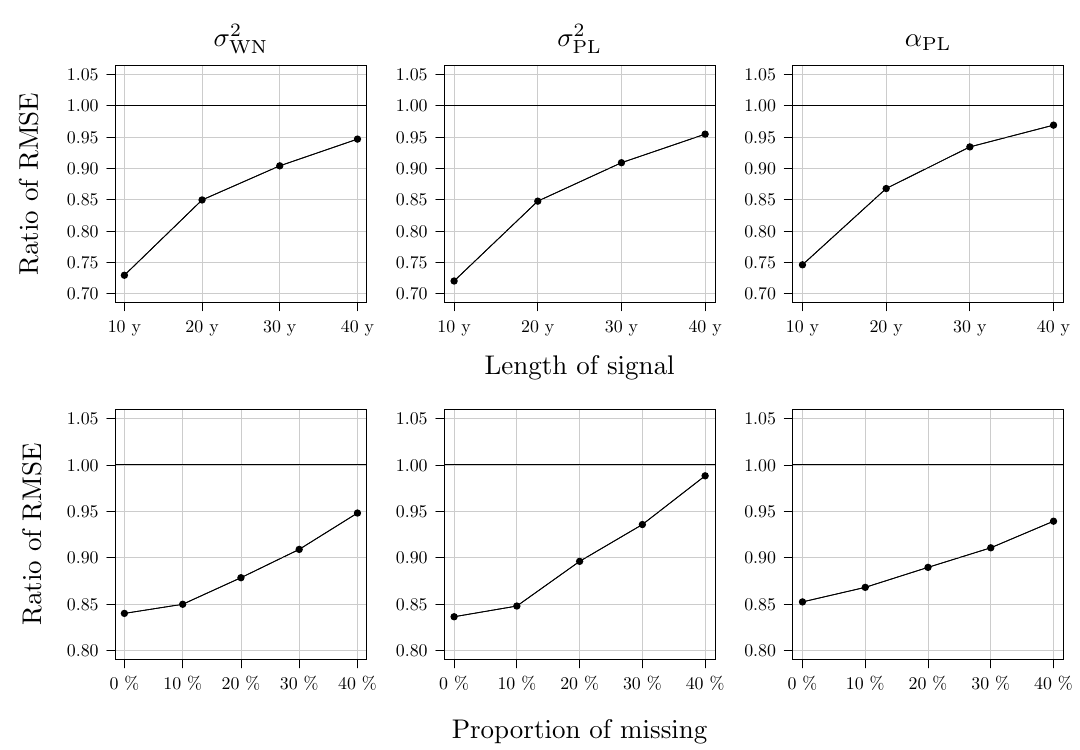}
    \caption{Ratio of RMSE of the stochastic parameters for Setting A1 and A2.}
    \label{fig:gamma_setting_a}
\end{figure}

Since we have various components to the parameter vector $\bm{\beta}$, we focus on the element that is commonly of more interest to geoscientists, i.e. $\beta_1$, which measures the velocity (or trend) of crustal movements (many geophysical applications focus on trend estimation in GNSS signals, see e.g. \citealp{mudelsee2019trend, maddanu2023trends, kermarrec2024modeling}). For this parameter, in Figure \ref{fig:beta_setting_a} we present the ratio of the Root Mean Squared Error (RMSE) of the GMWMX over the RMSE of the MLE obtained by Hector. For all of the considered sample sizes, this ratio remains below $1.05$ which indicates an acceptable loss in efficiency of the GMWMX (considering that the finite-sample bias is low and in line with the MLE) given the computational gains highlighted further on. What we must underline however is the empirical coverage of the true underlying parameter which, as part of the inference process, is the main objective of the proposed method in this work. In fact, we can observe how the coverage of the GMWMX lies close to the nominal 95\% level for all sample sizes and missing observation scenarios, whereas the MLE only starts approaching similar levels for the largest sample size considered (and worsens faster than the GMWMX as the proportion of missing observations increases). The results on coverage are possibly a consequence of the better performance of the GMWMX, compared to the MLE, when estimating the parameters of the error process. Indeed, as can be seen in Figure \ref{fig:gamma_setting_a}, the ratio of the RMSE for these parameters remains below 1 in all settings, with the MLE improving and approaching the RMSE of the GMWMX only as the sample sizes and proportion of missing observations increase. A more detailed description of the empirical distributions of the parameter estimates shows that the MLE often presents a higher bias than the GMWMX but reports a smaller variance. An extended overview of these results can be found in Appendix \ref{app:extended_simu_results}.

We can make the same conclusions for the other elements of $\bm{\beta}$ as those made for $\beta_1$ (see Appendix \ref{app:extended_simu_results}). For the other settings, we generally observe fluctuating RMSE ratios for the parameter $\beta_1$ (always close to 1), with this ratio being favorable to the GMWMX mainly for \textbf{Setting C}, while the RMSE of the error processes tend to vary showing how, on average, the GMWMX and MLE tend to give similar performances according to the sample size and proportion of missing observations. The main result that we are interested in, however, is the inference derived from the proposed GMWMX framework: as for \textbf{Setting A}, also in all other settings the GMWMX is consistently close to the desired nominal 95\% coverage level whereas the MLE either overcovers or only approaches an acceptable empirical level with larger sample sizes and worsens faster than the GMWMX as the proportion of missing observations increases.

\begin{Remark}
    Based on our experimental findings, ignoring that the estimated WV $\hat{\bm{\nu}}$ is computed on the residuals (and not on the true error process) can introduce a bias in the last scales of the WV. This, in turn, can lead to biased estimates of the stochastic parameters, particularly if a stochastic process is primarily distinguishable in the last scales of the WV. Nevertheless, since we know that the last scales of the WV are also those that generally suffer from a higher variability (due to the lower number of wavelet coefficients on which the WV is estimated at these scales), it could be considered reasonable to ignore this approximation depending on the setting. Nevertheless, with the procedure proposed in Appendix \ref{app:IminusH}, we can still perform this approximation in a computationally efficient manner.
\end{Remark}

\begin{figure}
    \centering
    \includegraphics[width=0.7\linewidth]{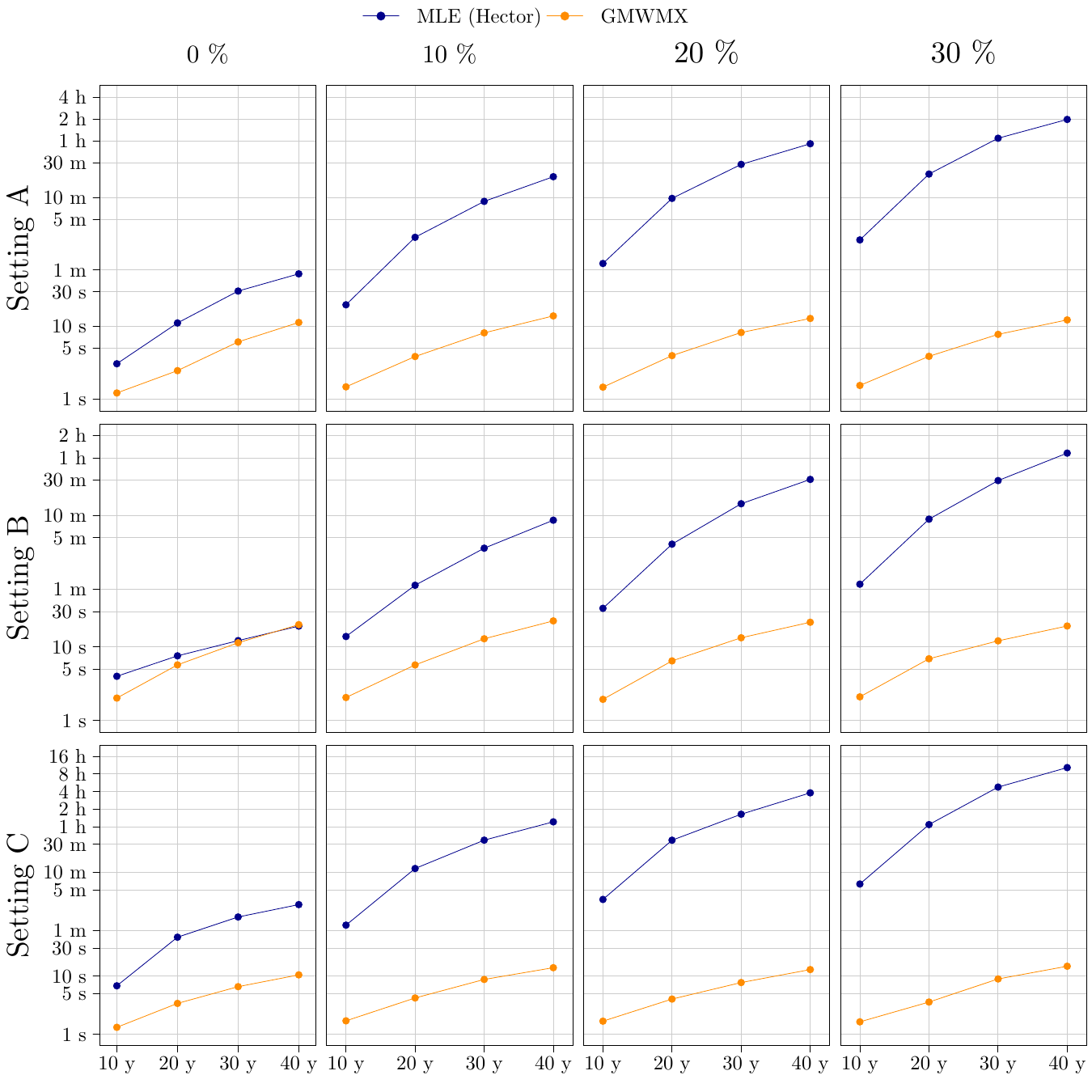}
    \caption{Mean running time for the GMWMX (orange line) and the MLE (blue line) for the different settings, different proportion of missing observations and different sample sizes.}
    \label{fig:computation_times}
\end{figure}

The final and equally important aspect of the proposed framework is its computational efficiency. Figure \ref{fig:computation_times} reports the mean running times for all sample sizes and proportions of missing observations in all of the considered settings (i.e. \textbf{Setting A} to \textbf{Setting C}). As \texttt{Hector} and the GMWMX can be implemented for parallel computation in various ways and run on different processing units, we report the computation time for single-thread execution. It can be observed that, already in the absence of missing observations, the GMWMX scales better than the MLE estimated via Hector: while the GMWMX takes 10 seconds on average for the largest sample size, the MLE can take more than 1 minute. This difference in scalability becomes even more evident as the proportion of missing observations increases, with the GMWMX never going beyond 30 seconds mean running time while the MLE (Hector) is consistently above 1 hour and even goes beyond 8 hours in the worst case (see \textbf{Setting C}). This confirms that the proposed GMWMX framework appears to generally achieve acceptable statistical performance compared to the MLE, especially with respect to the desired inferential quantities for which it actually compares favorably in the considered settings, and does so with significant reduction in computational times highlighting important advantages in terms of scalability compared to the efficient implementation of the MLE. These computational advantages are highlighted also in the following section where it can be seen how also the inferential conclusions of the GMWMX remain comparable to the MLE.

\section{Case Study}
\label{sec:case_study}

To illustrate the performance of the proposed GMWMX framework as well as the extent of analyses that can be performed seamlessly through its computational advantages, we consider a network of $508$ GNSS stations located in the North American region. We downloaded daily position time series processed by the Geodesy Advancing Geosciences and EarthScope (GAGE) GPS Analysis Center Coordinator (ACC) at MIT \citep{herring2016plate} computed within the International Terrestrial Reference Frame 2014 (\citealp{altamimi2016itrf2014}) and made available by the NSF GAGE Facility. For each station, we consider the first difference in the  North, East and vertical component for each daily position time series associated with each axis. We removed extreme observations in the time series using a standard procedure in the field which consists in filtering observations using a criterion based on the interquartile range of residuals from a linear trend fit (see e.g. \citealp{he2017review, langbein2004high}) and implemented in the \texttt{removeoutliers} utility within the \texttt{Hector} software. Based on the choice of GNSS stations, the data results in $1524$ times series ($508 \text{ stations} \times 3 \text{ axes}$) each of which has a minimum length of $487$, a maximum length of $7760$, with a median of $4170$ data points corresponding to roughly $11$ years of daily observations. The median proportion of missing observations was of $7\%$, with this proportion ranging from $0\%$ to $69 \%$. 

For the trajectory model, we consider the model defined in \eqref{eq:func_model}, which includes a seasonal and half-seasonal component to which we additionally include offsets in the trajectory model using the Heaviside step function as classically done in this context (see \citealp{he2017review}). Station offsets can be due to earthquakes and equipment changes and their time indices are provided by Plate Boundary Observatory (PBO)(\citealp{herring2016plate}), while we take care of estimating their amplitudes. The stochastic model we choose consists in a combination of flicker noise and white noise, which corresponds to the model considered in \textbf{Setting B} for the simulations discussed in Section \ref{sec:simulations}. This model is the most commonly used for the noise characterization of position time series in GNSS time series signal modeling (\citealp{wang2019impact}). As a way to compare conclusions from the analysis based on the GMWMX, also in this section we employ the MLE implemented through the \texttt{Hector} software. It must be noted however that the latter implementation only allows for the specification of stationary noise models because it relies on the Toeplitz structure of the error covariance matrix (which allows it to apply fast inversion techniques). For flicker noise, \texttt{Hector} uses an approximation which allows to obtain a near-Toeplitz covariance matrix for non-stationary PL noise.

Before analyzing the results of the GMWMX and MLE on this data, the main point to keep in mind for the following discussion is that, in order to perform the full estimation and inference on all signals associated to all three axes of the $508$ GNSS stations network, the GMWMX employed less than $1$ hour and $3$ minutes on a standard computer, while the MLE required $12$ hours and $26$ minutes. This is an extremely important result also in the light of the corresponding inferential conclusions delivered by these two approaches which highlight an almost perfect overlap. Hence, this is a strong indication that for these sizes of data and types of models, the GMWMX framework can deliver reliable statistical conclusions in drastically reduced computational times compared to the state-of-the-art fast MLE implementation.

\begin{figure}
    \centering
    \includegraphics[width=1\linewidth]{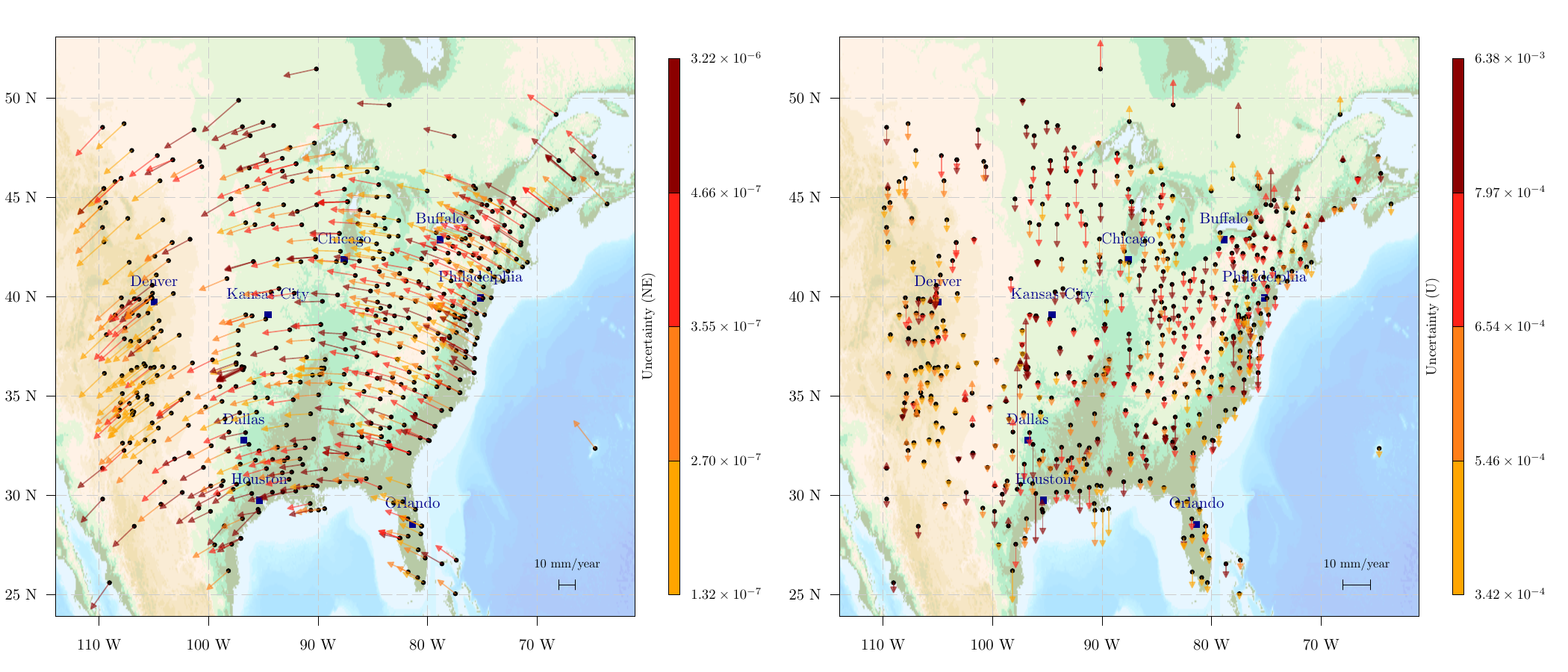}
    \caption{Estimated North-East velocity and crustal uplift for the $508$ GNSS stations of the network considered.}
    \label{fig:map_case_study}
\end{figure}

Let us now consider the results of the analyses. In Figure \ref{fig:map_case_study}, we represent the amplitude of the estimated tectonic rate and the crustal uplift estimated by the GMWMX and its estimated uncertainty. We represent the magnitude of the estimated velocity by the length of the arrow and its associated uncertainty by its color (the darker the color, the larger the uncertainty). In the case of the estimated tectonic rate of the North and East component, we compute the associated uncertainty of the 2 dimensional vector as the $\ell$-2 norm of the estimated standard error of the estimated coefficient associated with the North and East component. For the estimated crustal uplift, the uncertainty corresponds to the estimated standard deviation of the trend parameter when estimating the models corresponding to the first difference of the signal of the vertical component.

\begin{figure}
    \centering
    \includegraphics[width=0.5\linewidth]{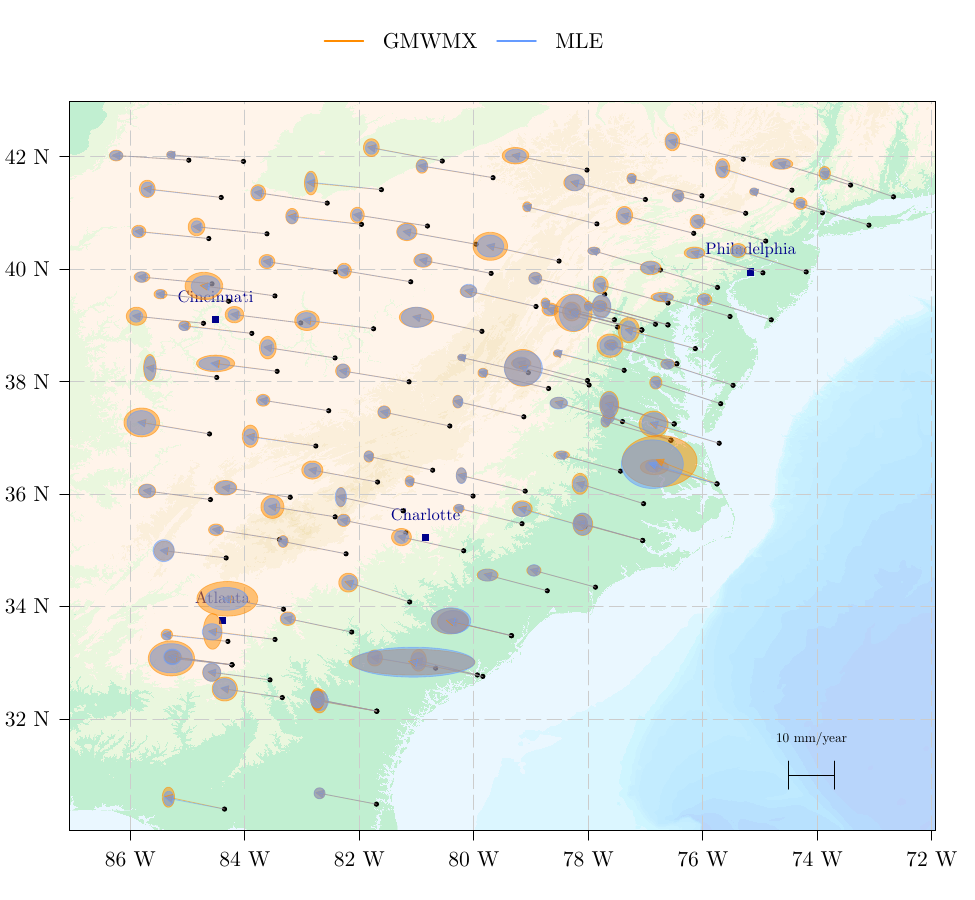}
    \caption{Estimated North-East velocity solutions and associated uncertainty for the subset of $122$ GNSS receivers illustrated using the GMWMX and \texttt{Hector} software (MLE).}
    \label{fig:map_case_stuy_vs_mle}
\end{figure}

We then focus on a specific region of the map to allow a more precise representation of the difference between the estimated trend and its associated standard deviation computed using the GMWMX and the MLE. We represent only the stations that fall within a defined geographic range, specifically those with longitudes between 85 degrees and 72 degrees West and latitudes between 30 and 42 degrees North. This subset of stations considered in the network results in $122$ stations. We compare our results to the results obtained with the MLE that estimated the same trajectory and stochastic model in Figure \ref{fig:map_case_stuy_vs_mle}. We note that the velocity estimates obtained from GMWMX and MLE are almost identical, as the arrows overlap for all stations. Similarly, the estimated standard deviations of the parameters are nearly the same, with most of the ellipses for GMWMX and MLE coinciding in Figure \ref{fig:map_case_stuy_vs_mle}.

\begin{figure}[h!]
    \centering
    \includegraphics[width=0.5\linewidth]{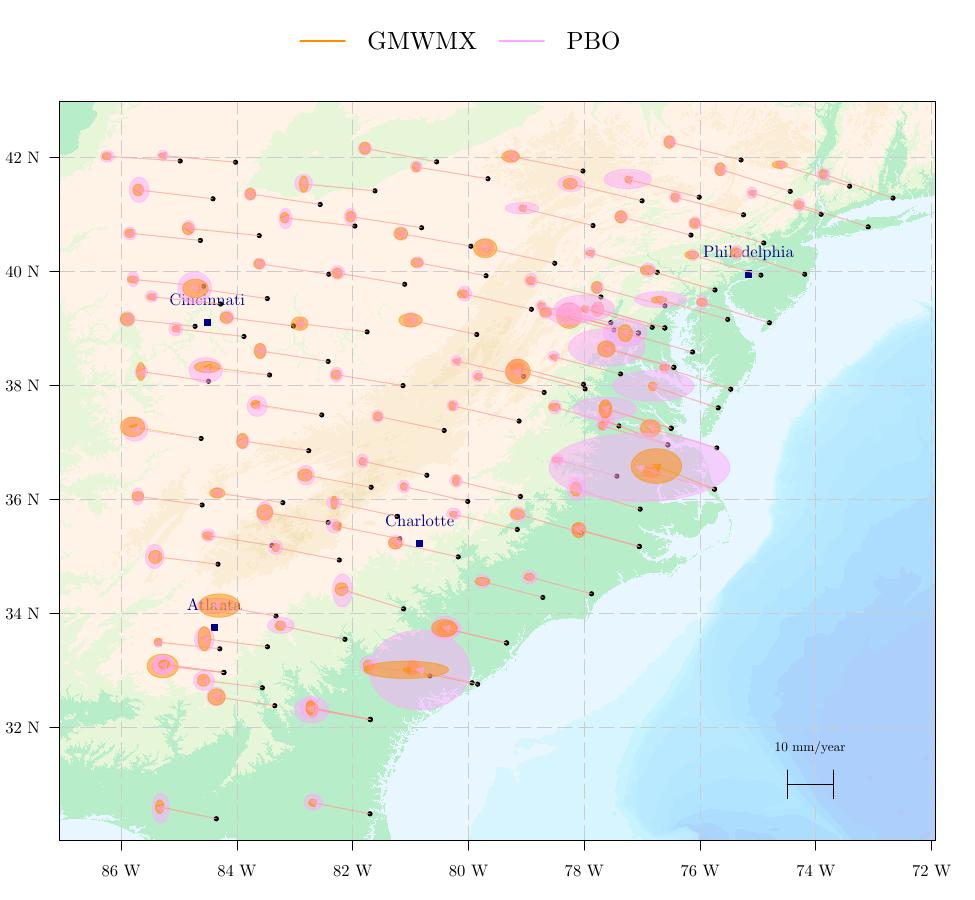}
    \caption{ Estimated North-East velocity solutions and associated uncertainty for the subset of $122$ GNSS receivers illustrated using the GMWMX and comparing with PBO solutions.}
   \label{fig:map_case_study_vs_gmwmx_vs_pbo}
\end{figure}

As a reference, we also compare our results with the velocity solutions provided by the PBO and made available at GAGE data archive. This comparison is represented in Figure \ref{fig:map_case_study_vs_gmwmx_vs_pbo}. As for the previous comparison with the MLE, we can observe that we have almost identical velocity estimates. However, we observe that for several stations, the estimated standard deviations between the GMWMX and PBO solutions differ. This difference can be attributed to the fact that both GMWMX and MLE jointly estimate stochastic noise alongside the trajectory model, whereas the PBO solution uses a faster statistical approach which relies on a Kalman filter based on a first-order Gauss–Markov noise model \citep{floyd2020fast}.

\newpage

\bibliographystyle{chicago.bst}
\bibliography{biblio}

\begin{thebibliography}{}

\bibitem[\protect\citeauthoryear{Altamimi, Rebischung, M{\'e}tivier, and Collilieux}{Altamimi et~al.}{2016}]{altamimi2016itrf2014}
Altamimi, Z., P.~Rebischung, L.~M{\'e}tivier, and X.~Collilieux (2016).
\newblock {ITRF2014: A New Release of the International Terrestrial Reference Frame Modeling Nonlinear Station Motions}.
\newblock {\em Journal of Geophysical Research: Solid Earth\/}~{\em 121\/}(8), 6109--6131.

\bibitem[\protect\citeauthoryear{Amiri-Simkooei}{Amiri-Simkooei}{2009}]{amiri2009noise}
Amiri-Simkooei, A. (2009).
\newblock {Noise in Multivariate GPS Position Time-series}.
\newblock {\em Journal of Geodesy\/}~{\em 83}, 175--187.

\bibitem[\protect\citeauthoryear{Amiri-Simkooei, Tiberius, and Teunissen}{Amiri-Simkooei et~al.}{2007}]{amiri2007assessment}
Amiri-Simkooei, A.~R., C.~C. Tiberius, and P.~J. Teunissen (2007).
\newblock {Assessment of Noise in GPS Coordinate Time Series: Methodology and Results}.
\newblock {\em Journal of Geophysical Research: Solid Earth\/}~{\em 112\/}(B7).

\bibitem[\protect\citeauthoryear{Banerjee, Dunson, and Tokdar}{Banerjee et~al.}{2013}]{banerjee2013efficient}
Banerjee, A., D.~B. Dunson, and S.~T. Tokdar (2013).
\newblock {Efficient Gaussian Process Regression for Large Datasets}.
\newblock {\em Biometrika\/}~{\em 100\/}(1), 75--89.

\bibitem[\protect\citeauthoryear{Banna, Merlev{\`e}de, and Youssef}{Banna et~al.}{2016}]{banna2016bernstein}
Banna, M., F.~Merlev{\`e}de, and P.~Youssef (2016).
\newblock {Bernstein-type Inequality for a Class of Dependent Random Matrices}.
\newblock {\em Random Matrices: Theory and Applications\/}~{\em 5\/}(02), 1650006.

\bibitem[\protect\citeauthoryear{Bao, Chang, Zhang, Chen, and Zhang}{Bao et~al.}{2021}]{bao2021filling}
Bao, Z., G.~Chang, L.~Zhang, G.~Chen, and S.~Zhang (2021).
\newblock {Filling Missing Values of Multi-station GNSS Coordinate Time Series Based on Matrix Completion}.
\newblock {\em Measurement\/}~{\em 183}, 109862.

\bibitem[\protect\citeauthoryear{Beran, Feng, Ghosh, and Kulik}{Beran et~al.}{2013}]{beran2013long}
Beran, J., Y.~Feng, S.~Ghosh, and R.~Kulik (2013).
\newblock {\em {Long-memory Processes}}.
\newblock Springer.

\bibitem[\protect\citeauthoryear{Bock, Nikolaidis, de~Jonge, and Bevis}{Bock et~al.}{2000}]{bock2000instantaneous}
Bock, Y., R.~M. Nikolaidis, P.~J. de~Jonge, and M.~Bevis (2000).
\newblock {Instantaneous Geodetic Positioning at Medium Distances with the Global Positioning System}.
\newblock {\em Journal of Geophysical Research: Solid Earth\/}~{\em 105\/}(B12), 28223--28253.

\bibitem[\protect\citeauthoryear{Bos, Fernandes, Williams, and Bastos}{Bos et~al.}{2013}]{bos2013fast}
Bos, M., R.~Fernandes, S.~Williams, and L.~Bastos (2013).
\newblock {Fast Error Analysis of Continuous GNSS Observations with Missing Data}.
\newblock {\em Journal of Geodesy\/}~{\em 87\/}(4), 351--360.

\bibitem[\protect\citeauthoryear{Bos, Fernandes, Williams, and Bastos}{Bos et~al.}{2008}]{Bos2008}
Bos, M.~S., R.~M.~S. Fernandes, S.~D.~P. Williams, and L.~Bastos (2008).
\newblock {Fast Error Analysis of Continuous GPS Observations}.
\newblock {\em Journal of Geodesy\/}~{\em 82\/}(3), 157--166.

\bibitem[\protect\citeauthoryear{Bradley}{Bradley}{2005}]{bradley2005basic}
Bradley, R.~C. (2005).
\newblock {Basic Properties of Strong Mixing Conditions. A Survey and Some Open Questions}.

\bibitem[\protect\citeauthoryear{Calais}{Calais}{1999}]{calais1999continuous}
Calais, E. (1999).
\newblock {Continuous GPS Measurements Across the Western Alps, 1996--1998}.
\newblock {\em Geophysical Journal International\/}~{\em 138\/}(1), 221--230.

\bibitem[\protect\citeauthoryear{Chib}{Chib}{2001}]{chib2001markov}
Chib, S. (2001).
\newblock {Markov Chain Monte Carlo Methods: Computation and Inference}.
\newblock {\em Handbook of econometrics\/}~{\em 5}, 3569--3649.

\bibitem[\protect\citeauthoryear{Cucci, Voirol, Kermarrec, Montillet, and Guerrier}{Cucci et~al.}{2023}]{cucci2023generalized}
Cucci, D.~A., L.~Voirol, G.~Kermarrec, J.-P. Montillet, and S.~Guerrier (2023).
\newblock {The Generalized Method of Wavelet Moments with eXogenous inputs: A Fast Approach for the Analysis of GNSS Position Time Series}.
\newblock {\em Journal of Geodesy\/}~{\em 97\/}(2), 14.

\bibitem[\protect\citeauthoryear{Floyd and Herring}{Floyd and Herring}{2020}]{floyd2020fast}
Floyd, M.~A. and T.~A. Herring (2020).
\newblock {Fast Statistical Approaches to Geodetic Time Series Analysis}.
\newblock {\em Geodetic time series analysis in earth Sciences\/}, 157--183.

\bibitem[\protect\citeauthoryear{Fuentes}{Fuentes}{2002}]{fuentes2002spectral}
Fuentes, M. (2002).
\newblock {Spectral Methods for Nonstationary Spatial Processes}.
\newblock {\em Biometrika\/}~{\em 89\/}(1), 197--210.

\bibitem[\protect\citeauthoryear{Gardner}{Gardner}{1978}]{gardner1978mathematical}
Gardner, M. (1978).
\newblock {Mathematical games: White and Brown Music, Fractal Curves and one-over-f Fluctuations}.
\newblock {\em Scientific American\/}~{\em 238\/}(4), 16--32.

\bibitem[\protect\citeauthoryear{Gerber, de~Jong, Schaepman, Schaepman-Strub, and Furrer}{Gerber et~al.}{2018}]{gerber2018predicting}
Gerber, F., R.~de~Jong, M.~E. Schaepman, G.~Schaepman-Strub, and R.~Furrer (2018).
\newblock {Predicting Missing Values in Spatio-Temporal Remote Sensing Data}.
\newblock {\em IEEE Transactions on Geoscience and Remote Sensing\/}~{\em 56\/}(5), 2841--2853.

\bibitem[\protect\citeauthoryear{Gibbs and MacKay}{Gibbs and MacKay}{2000}]{gibbs2000variational}
Gibbs, M.~N. and D.~J. MacKay (2000).
\newblock {Variational Gaussian Process Classifiers}.
\newblock {\em IEEE Transactions on Neural Networks\/}~{\em 11\/}(6), 1458--1464.

\bibitem[\protect\citeauthoryear{Gramacy}{Gramacy}{2020}]{gramacy2020surrogates}
Gramacy, R.~B. (2020).
\newblock {\em Surrogates: Gaussian Process Modeling, Design, and Optimization for the Applied Sciences}.
\newblock Chapman and Hall/CRC.

\bibitem[\protect\citeauthoryear{Guerrier, Molinari, and Stebler}{Guerrier et~al.}{2016}]{guerrier2016theoretical}
Guerrier, S., R.~Molinari, and Y.~Stebler (2016).
\newblock {Theoretical Limitations of Allan Variance-based Regression for Time Series Model Estimation}.
\newblock {\em IEEE Signal Processing Letters\/}~{\em 23\/}(5), 597--601.

\bibitem[\protect\citeauthoryear{Guerrier, Molinari, Victoria-Feser, and Xu}{Guerrier et~al.}{2021}]{guerrier2021robust}
Guerrier, S., R.~Molinari, M.-P. Victoria-Feser, and H.~Xu (2021).
\newblock {Robust Two-Step Wavelet-Based Inference for Time Series Models}.
\newblock {\em Journal of the American Statistical Association\/}~{\em 0\/}(0), 1--18.

\bibitem[\protect\citeauthoryear{Guerrier, Skaloud, Stebler, and Victoria-Feser}{Guerrier et~al.}{2013}]{guerrier2013wavelet}
Guerrier, S., J.~Skaloud, Y.~Stebler, and M.-P. Victoria-Feser (2013).
\newblock {Wavelet-Variance-based Estimation for Composite Stochastic Processes}.
\newblock {\em Journal of the American Statistical Association\/}~{\em 108\/}(503), 1021--1030.

\bibitem[\protect\citeauthoryear{Guinness}{Guinness}{2019}]{guinness2019spectral}
Guinness, J. (2019).
\newblock {Spectral Density Estimation for Random Fields via Periodic Embeddings}.
\newblock {\em Biometrika\/}~{\em 106\/}(2), 267--286.

\bibitem[\protect\citeauthoryear{Hansen}{Hansen}{1982}]{hansen1982large}
Hansen, L.~P. (1982).
\newblock {Large Sample Properties of Generalized Method of Moments Estimators}.
\newblock {\em Econometrica: Journal of the econometric society\/}, 1029--1054.

\bibitem[\protect\citeauthoryear{He, Montillet, Fernandes, Bos, Yu, Hua, and Jiang}{He et~al.}{2017}]{he2017review}
He, X., J.-P. Montillet, R.~Fernandes, M.~Bos, K.~Yu, X.~Hua, and W.~Jiang (2017).
\newblock {Review of Current GPS Methodologies for Producing Accurate Time Series and their Error Sources}.
\newblock {\em Journal of Geodynamics\/}~{\em 106}, 12--29.

\bibitem[\protect\citeauthoryear{Heaton, Datta, Finley, Furrer, Guinness, Guhaniyogi, Gerber, Gramacy, Hammerling, Katzfuss, et~al.}{Heaton et~al.}{2019}]{heaton2019case}
Heaton, M.~J., A.~Datta, A.~O. Finley, R.~Furrer, J.~Guinness, R.~Guhaniyogi, F.~Gerber, R.~B. Gramacy, D.~Hammerling, M.~Katzfuss, et~al. (2019).
\newblock {A Case Study Competition Among Methods for Analyzing Large Spatial Data}.
\newblock {\em Journal of Agricultural, Biological and Environmental Statistics\/}~{\em 24}, 398--425.

\bibitem[\protect\citeauthoryear{Herring, Melbourne, Murray, Floyd, Szeliga, King, Phillips, Puskas, Santillan, and Wang}{Herring et~al.}{2016}]{herring2016plate}
Herring, T.~A., T.~I. Melbourne, M.~H. Murray, M.~A. Floyd, W.~M. Szeliga, R.~W. King, D.~A. Phillips, C.~M. Puskas, M.~Santillan, and L.~Wang (2016).
\newblock {Plate Boundary Observatory and Related Networks: GPS Data Analysis Methods and Geodetic Products}.
\newblock {\em Reviews of Geophysics\/}~{\em 54\/}(4), 759--808.

\bibitem[\protect\citeauthoryear{Hosking}{Hosking}{1981}]{hosking1981fractional}
Hosking, J. (1981).
\newblock {Fractional Differencing}.
\newblock {\em Biometrika\/}.

\bibitem[\protect\citeauthoryear{Kasdin}{Kasdin}{1995}]{kasdin1995discrete}
Kasdin, N.~J. (1995).
\newblock {Discrete Simulation of Colored Noise and Stochastic Processes and 1/f Power Law Noise Generation}.
\newblock In {\em Proceedings of the IEEE}, Volume~83.

\bibitem[\protect\citeauthoryear{Kermarrec, Maddanu, Klos, Proietti, and Bogusz}{Kermarrec et~al.}{2024}]{kermarrec2024modeling}
Kermarrec, G., F.~Maddanu, A.~Klos, T.~Proietti, and J.~Bogusz (2024).
\newblock {Modeling Trends and Periodic Components in Geodetic Time Series: A Unified Approach}.
\newblock {\em Journal of Geodesy\/}~{\em 98\/}(3), 17.

\bibitem[\protect\citeauthoryear{Langbein and Bock}{Langbein and Bock}{2004}]{langbein2004high}
Langbein, J. and Y.~Bock (2004).
\newblock {High-Rate Real-time GPS Network at Parkfield: Utility for Detecting Fault Slip and Seismic Displacements}.
\newblock {\em Geophysical Research Letters\/}~{\em 31\/}(15).

\bibitem[\protect\citeauthoryear{L{\'a}zaro-Gredilla and Titsias}{L{\'a}zaro-Gredilla and Titsias}{2011}]{lazaro2011variational}
L{\'a}zaro-Gredilla, M. and M.~K. Titsias (2011).
\newblock {Variational Heteroscedastic Gaussian Process Regression}.
\newblock In {\em ICML}, pp.\  841--848.

\bibitem[\protect\citeauthoryear{Lilly, Sykulski, Early, and Olhede}{Lilly et~al.}{2017}]{lilly2017fractional}
Lilly, J.~M., A.~M. Sykulski, J.~J. Early, and S.~C. Olhede (2017).
\newblock {Fractional Brownian motion, the Mat{\'e}rn process, and Stochastic Modeling of Turbulent Dispersion}.
\newblock {\em Nonlinear Processes in Geophysics\/}~{\em 24\/}(3), 481--514.

\bibitem[\protect\citeauthoryear{Maddanu and Proietti}{Maddanu and Proietti}{2023}]{maddanu2023trends}
Maddanu, F. and T.~Proietti (2023).
\newblock {Trends in Atmospheric Ethane}.
\newblock {\em Climatic Change\/}~{\em 176\/}(5), 53.

\bibitem[\protect\citeauthoryear{Mandelbrot and Van~Ness}{Mandelbrot and Van~Ness}{1968}]{mandelbrot1968fractional}
Mandelbrot, B.~B. and J.~W. Van~Ness (1968).
\newblock {Fractional Brownian Motions, Fractional Noises and Applications}.
\newblock {\em SIAM review\/}~{\em 10\/}(4), 422--437.

\bibitem[\protect\citeauthoryear{Montillet, Finsterle, Kermarrec, Sikonja, Haberreiter, Schmutz, and Dudok~de Wit}{Montillet et~al.}{2022}]{montillet2022data}
Montillet, J.-P., W.~Finsterle, G.~Kermarrec, R.~Sikonja, M.~Haberreiter, W.~Schmutz, and T.~Dudok~de Wit (2022).
\newblock {Data Fusion of Total Solar Irradiance Composite Time Series Using 41 Years of Satellite Measurements}.
\newblock {\em Journal of Geophysical Research: Atmospheres\/}~{\em 127\/}(13), e2021JD036146.

\bibitem[\protect\citeauthoryear{Montillet, Kermarrec, Forootan, Haberreiter, He, Finsterle, Fernandes, and Shum}{Montillet et~al.}{2024}]{montillet2024big}
Montillet, J.-P., G.~Kermarrec, E.~Forootan, M.~Haberreiter, X.~He, W.~Finsterle, R.~Fernandes, and C.~Shum (2024).
\newblock {How Big Data Can Help to Monitor the Environment and to Mitigate Risks due to Climate Change: A Review}.
\newblock {\em IEEE Geoscience and Remote Sensing Magazine\/}.

\bibitem[\protect\citeauthoryear{Mudelsee}{Mudelsee}{2019}]{mudelsee2019trend}
Mudelsee, M. (2019).
\newblock {Trend Analysis of Climate Time Series: A Review of Methods}.
\newblock {\em Earth-science reviews\/}~{\em 190}, 310--322.

\bibitem[\protect\citeauthoryear{Muyskens, Goumiri, Priest, Schneider, Armstrong, Bernstein, and Dana}{Muyskens et~al.}{2022}]{muyskens2022star}
Muyskens, A.~L., I.~R. Goumiri, B.~W. Priest, M.~D. Schneider, R.~E. Armstrong, J.~Bernstein, and R.~Dana (2022).
\newblock {Star-Galaxy Image Separation with Computationally Efficient Gaussian Process Classification}.
\newblock {\em The Astronomical Journal\/}~{\em 163\/}(4), 148.

\bibitem[\protect\citeauthoryear{Pipiras and Taqqu}{Pipiras and Taqqu}{2000}]{pipiras2000convergence}
Pipiras, V. and M.~S. Taqqu (2000).
\newblock {Convergence of Weighted Sums of Random Variables with Long-range Dependence}.
\newblock {\em Stochastic Processes and their Applications\/}~{\em 90\/}(1), 157--174.

\bibitem[\protect\citeauthoryear{Proietti and Maddanu}{Proietti and Maddanu}{2022}]{proietti2022modelling}
Proietti, T. and F.~Maddanu (2022).
\newblock {Modelling Cycles in Climate Series: The Fractional Sinusoidal Waveform Process}.
\newblock {\em Journal of Econometrics\/}, 105299.

\bibitem[\protect\citeauthoryear{Rosenthal}{Rosenthal}{1995}]{rosenthal1995convergence}
Rosenthal, J.~S. (1995).
\newblock {Convergence Rates for Markov Chains}.
\newblock {\em Siam Review\/}~{\em 37\/}(3), 387--405.

\bibitem[\protect\citeauthoryear{Sang, Jun, and Huang}{Sang et~al.}{2011}]{sang2011covariance}
Sang, H., M.~Jun, and J.~Z. Huang (2011).
\newblock {Covariance Approximation for Large Multivariate Spatial Data Sets with an Application to Multiple climate Model Errors}.
\newblock {\em The Annals of Applied Statistics\/}, 2519--2548.

\bibitem[\protect\citeauthoryear{Schmidt, Petrovic, G{\"u}ntner, Barthelmes, W{\"u}nsch, and Kusche}{Schmidt et~al.}{2008}]{schmidt2008periodic}
Schmidt, R., S.~Petrovic, A.~G{\"u}ntner, F.~Barthelmes, J.~W{\"u}nsch, and J.~Kusche (2008).
\newblock {Periodic Components of Water Storage Changes from GRACE and Global Hydrology Models}.
\newblock {\em Journal of Geophysical Research: Solid Earth\/}~{\em 113\/}(B8).

\bibitem[\protect\citeauthoryear{Stebler, Guerrier, Skaloud, and Victoria-Feser}{Stebler et~al.}{2014}]{stebler2014generalized}
Stebler, Y., S.~Guerrier, J.~Skaloud, and M.-P. Victoria-Feser (2014).
\newblock {Generalized Method of Wavelet Moments for Inertial Navigation Filter Design}.
\newblock {\em IEEE Transactions on Aerospace and Electronic Systems\/}~{\em 50\/}(3), 2269--2283.

\bibitem[\protect\citeauthoryear{Stein, Chen, and Anitescu}{Stein et~al.}{2013}]{stein2013stochastic}
Stein, M.~L., J.~Chen, and M.~Anitescu (2013).
\newblock {Stochastic Approximation of Score Functions for Gaussian Processes}.

\bibitem[\protect\citeauthoryear{Tehranchi, Moghtased-Azar, and Safari}{Tehranchi et~al.}{2021}]{tehranchi2021fast}
Tehranchi, R., K.~Moghtased-Azar, and A.~Safari (2021).
\newblock {Fast Approximation Algorithm to Noise Components Estimation in long-term GPS Coordinate Time Series}.
\newblock {\em Journal of Geodesy\/}~{\em 95\/}(2), 1--16.

\bibitem[\protect\citeauthoryear{Tunini, Magrin, Rossi, and Zuliani}{Tunini et~al.}{2024}]{tunini2024global}
Tunini, L., A.~Magrin, G.~Rossi, and D.~Zuliani (2024).
\newblock {Global Navigation Satellite System (GNSS) Time Series and Velocities About a Slowly Convergent Margin Processed on High-Performance Computing (HPC) Clusters: Products and Robustness Evaluation}.
\newblock {\em Earth System Science Data\/}~{\em 16\/}(2), 1083--1106.

\bibitem[\protect\citeauthoryear{Wang and Herring}{Wang and Herring}{2019}]{wang2019impact}
Wang, L. and T.~Herring (2019).
\newblock {Impact of Estimating Position Offsets on the Uncertainties of GNSS Site Velocity Estimates}.
\newblock {\em Journal of Geophysical Research: Solid Earth\/}~{\em 124\/}(12), 13452--13467.

\bibitem[\protect\citeauthoryear{Williams}{Williams}{2008}]{williams2008cats}
Williams, S.~D. (2008).
\newblock {CATS: GPS Coordinate Time Series Analysis Software}.
\newblock {\em GPS solutions\/}~{\em 12\/}(2), 147--153.

\bibitem[\protect\citeauthoryear{Williams, Bock, Fang, Jamason, Nikolaidis, Prawirodirdjo, Miller, and Johnson}{Williams et~al.}{2004}]{williams2004error}
Williams, S.~D., Y.~Bock, P.~Fang, P.~Jamason, R.~M. Nikolaidis, L.~Prawirodirdjo, M.~Miller, and D.~J. Johnson (2004).
\newblock {Error Analysis of Continuous GPS Position Time Series}.
\newblock {\em Journal of Geophysical Research: Solid Earth\/}~{\em 109\/}(B3).

\bibitem[\protect\citeauthoryear{Wu}{Wu}{2005}]{wu2005nonlinear}
Wu, W.~B. (2005).
\newblock {Nonlinear System Theory: Another Look at Dependence}.
\newblock {\em Proceedings of the National Academy of Sciences\/}~{\em 102\/}(40), 14150--14154.

\bibitem[\protect\citeauthoryear{Wu}{Wu}{2007}]{wu2007m}
Wu, W.~B. (2007).
\newblock {M-estimation of Linear Models with Dependent Errors}.
\newblock {\em The Annals of Statistics\/}~{\em 35\/}(2), 495--521.

\bibitem[\protect\citeauthoryear{Xu, Dubey, and Yu}{Xu et~al.}{2021}]{xu2021online}
Xu, H., P.~Dubey, and Y.~Yu (2021).
\newblock {Online Network Change Point Detection with Missing Values and Temporal Dependence}.
\newblock {\em arXiv preprint arXiv:2110.06450\/}.

\bibitem[\protect\citeauthoryear{Xu, Guerrier, Molinari, and Zhang}{Xu et~al.}{2017}]{xu2017study}
Xu, H., S.~Guerrier, R.~Molinari, and Y.~Zhang (2017).
\newblock {A Study of the Allan Variance for Constant-mean Nonstationary Processes}.
\newblock {\em IEEE Signal Processing Letters\/}~{\em 24\/}(8), 1257--1260.

\bibitem[\protect\citeauthoryear{Zhang, Bock, Johnson, Fang, Williams, Genrich, Wdowinski, and Behr}{Zhang et~al.}{1997}]{zhang1997southern}
Zhang, J., Y.~Bock, H.~Johnson, P.~Fang, S.~Williams, J.~Genrich, S.~Wdowinski, and J.~Behr (1997).
\newblock {Southern California Permanent GPS Geodetic Array: Error Analysis of Daily Position Estimates and Site Velocities}.
\newblock {\em Journal of Geophysical Research: Solid Earth\/}~{\em 102\/}(B8), 18035--18055.

\bibitem[\protect\citeauthoryear{Zhang}{Zhang}{2008}]{zhang2008allan}
Zhang, N.~F. (2008).
\newblock {Allan Variance of Time Series Models for Measurement Data}.
\newblock {\em Metrologia\/}~{\em 45\/}(5), 549.

\end{thebibliography}

\newpage


\setcounter{page}{1}
\section{Appendices}

\subsection{Auxiliary Results}

In this appendix we list and state some auxiliary results that will be used for the proofs given in the following appendices. For this reason, we also denote the operator norm as $\|\cdot\|_{\op}$.

\begin{Lemma}[Theorem 1 in \cite{banna2016bernstein}]\label{lemma:matrix_bernstein}
Let $\{\bm{M}_i\}_{i \in \mathbb{N}^*}$ is a family of self-adjoint random matrices of size $d$. Assume that there exists a constant $c > 0$ such that for any $\ell \geq 1$, $\beta_{\bm{M}}(\ell) \leq \exp(1-c\,\ell)$, and there exists a positive constant $D$ such that for any $i \in \mathbb{N}^*$,
\begin{align*}
    \mathbb{E}[\bm{M}_i] = \bm{0} \quad \text{and} \quad \|\bm{M}_i\|_{\mathrm{op}} \leq D \quad \text{almost surely}.
\end{align*}
Then there exists an absolute constant $C$ such that for any $x > 0$ and any integer $n \geq 2$,
\[
\mathbb{P}\left(\left\|\sum_{i=1}^n \bm{M}_i\right\|_{\mathrm{op}} \geq x\right) \leq d\exp\left(-\frac{C\,x^2}{\iota^2\,n + c^{-1}\,D^2 + x\,D\,\gamma(c,n)}\right),
\]
where
\[
\iota^2 = \sup_{\mathcal{K} \subseteq\{1, \dots, n\}}\frac{1}{|\mathcal{K}|}\left\|\mathbb{E}\left(\sum_{i \in \mathcal{K}}\bm{M}_i\right)^2\right\|_{\mathrm{op}}
\]
and
\[
\gamma(c,n) = \frac{\log n}{\log 2}\max\left\{2, \frac{32\log n}{c\log 2}\right\}.
\]
\end{Lemma}

\begin{Lemma}[Weyl’s inequality]\label{lemma:weyl_ineq}
For symmetric matrices $\bm{A}, \bm{B} \in \mathbb R^{r \times r}$, we have that
    $$\max_{i = 1}^r|\lambda_{i}(\bm{A}) - \lambda_i(\bm{B})| \leq \|\bm{A} - \bm{B} \|_{\op}.$$
\end{Lemma}

The next theorem is from \cite{pipiras2000convergence}, which is an invariance principle for  weighted sums of long-range dependent process. To state the theorem, we need the following definitions.
Define the function space
\[
|\Lambda|^{d} = \left\{ f : \int_{\mathbb{R}} \int_{\mathbb{R}} \| f(u) \| \| f(v) \| |u - v|^{2d - 1} \, \mathrm{d}u \, \mathrm{d}v < \infty \right\},
\]
for $d \in (0,1/2)$. Define the norm $\|f\|_{|\Lambda|^d}$ on $|\Lambda|^d$ by
\[
\|f\|^2_{|\Lambda|^{d}} = \int_{\mathbb{R}} \int_{\mathbb{R}} \| f(u) \| \| f(v) \| |u - v|^{2d - 1} \, \mathrm{d}u \, \mathrm{d}v.
\]
For $k \in \mathbb N \cup \{\infty\}$, define the approximation
\[
f^{+}_{n,k} = \sum_{i=0}^{k} f\left( \frac{i}{n} \right) \mathbbm{1}_{[i/n, (i+1)/n)},
\quad
f^{-}_{n,k} = \sum_{i=-k}^{-1} f\left( \frac{i}{n} \right) \mathbbm{1}_{[i/n, (i+1)/n)},
\]
\[
f^{+}_{n} = f^{+}_{n,\infty},
\quad
f^{-}_{n} = f^{-}_{n,\infty},
\quad
f_{n} = f^{+}_{n} + f^{-}_{n}.
\]
Moreover, define the following two-sided partial sum process of $\{e_i\}_{i \in \mathbb Z}$ as
\[
B^{d}_n(u) =
\begin{cases}
\frac{1}{n^{d + 1/2}} \sum_{i=1}^{\lfloor nu \rfloor} e_i, & u \geq 0, \\[10pt]
-\frac{1}{n^{d + 1/2}} \sum_{i=\lfloor nu \rfloor + 1}^{0} e_i, & u < 0,
\end{cases}
\]

\begin{Theorem}[Theorem 3.1 in \cite{pipiras2000convergence}]\label{thm:clt_long-memory}
    Let $d \in (0, 1/2)$. Let $f, f^{\pm}_n, f^{\pm}_{n,k}, f_n$ be deterministic functions defined above. Suppose that the following conditions are satisfied:
\begin{enumerate}
    \item[(i)] $f, f^{\pm}_n \in |\Lambda|^{d}$, $\|f^{\pm}_n - f^{\pm}_{n,k}\|_{|\Lambda|^{d}} \to 0$, as $k \to \infty$, $\|f - f_n\|_{|\Lambda|^{d}} \to 0$, as $n \to \infty$,
    \item[(ii)] $\{e_i\}_{i \in \mathbb{Z}}$ is a stationary process with $\mathbb E(e_i) = 0$ and $\var(e_i)<\infty$, such that $|E(e_0 e_\ell)| \leq c |\ell|^{2d - 1}$, $\ell \in \mathbb{N}$, and is such that the sequence of processes $B_n^{d}$ converges to $B^{d}$ in the sense of the finite-dimensional distributions.
\end{enumerate}
Then
\[
\frac{1}{n^{d + 1/2}} \sum_{i = -\infty}^{\infty} f\left(\frac{i}{n}\right) e_i \xrightarrow{\mathcal{D}} \int_{\mathbb{R}} f(u) \, \mathrm{d}B^{d}(u),
\]
where $\{B^d(u)\}_{u \in \mathbb R}$ is a two-sided standard fractional Brownian motion.
\end{Theorem}

\subsection{Proof of Lemma \ref{lemma:beta_cov}}
\label{app:lemma_beta_cov}

Before we proceed to the proof of Lemma \ref{lemma:beta_cov}, we firstly state and prove additional (auxiliary) lemmas that will be needed to deliver the proof of the lemma of interest of this appendix.

The first auxiliary lemma concerns the properties of the missing observation process $\bm{Z}$. For example, in \cite{xu2021online} it was shown that the missingness pattern of \eqref{eq:markov_missinigness} is geometrically ergodic, i.e.~the $\beta$-mixing coefficients satisfying $\beta_Z(\ell) \leq \exp(1-c\ell)$ for some constant $c > 0$ only depending on $\rho$. In the context of this work, we note that since $\{Z_i\}$ is stationary and using the notation from \eqref{eq:markov_missinigness}, we have
\[
\mathbb{E}[Z_i] = \rho\mathbb{E}[Z_{i-1}] + (1 - \rho)\mu(\bm{\bm{\vartheta}}),
\]
which leads to
\[
\mu(\bm{\bm{\vartheta}}) = \mathbb E[Z_i].
\]
Note also, when $\rho = 0$, \eqref{eq:markov_missinigness} reduces to a model of i.i.d. Bernoulli random variables. With these definitions we have the following auxiliary lemma.

\begin{Lemma}\label{lemma:ergodic}
    Let the missingness sequence $\bm{Z}$ be a stationary Markov chain satisfying \eqref{eq:markov_missinigness} with $\rho \in [0,1)$ and $\mu(\bm{\bm{\vartheta}}) \in (0,1]$. Then $\bm{Z}$ is a geometric ergodic Markov chain, and its $\beta$-mixing coefficients satisfy $\beta_{Z}(\ell) \leq \rho^{\ell} = \exp(1-c\ell)$, for some constant $c > 0$ only depending on $\rho$. 
\end{Lemma}
\begin{proof}
    Note that $\bm{Z} = \{Z_i\}$ is a stationary Markov chain with state space $\mathcal{X} = \{0, 1\}$. By Theorem 3.7 in \cite{bradley2005basic}, it suffices to verify Doeblin’s condition \citep[e.g.][]{rosenthal1995convergence}, which holds on the transition probability $\mathbb P(\cdot, \cdot)$. Indeed, if there exists an $\epsilon > 0$ and a probability measure $\eta(\cdot)$, such that for all $x \in \mathcal{X}$ and measurable subsets $A \subseteq \mathcal{X}$, then we have that
    $\mathbb P(x, A) \geq \epsilon\eta(A)$. With this, we notice that we can write \eqref{eq:markov_missinigness} equivalently as
    \[
        Z_i = (1-U)\cdot Z_{i-1} + U\cdot B,
    \]
    where $U \sim \mathrm{Bernoulli}(1-\rho)$, $B \sim \mathrm{Bernoulli}(\mu(\bm{\bm{\vartheta}}))$ and $\{Z_i\}$ are mutually independent. We have that for any $A \subseteq \mathcal{X}$
    \begin{align*}
        \mathbb P(x, A) =& \mathbb P(Z_i \in A| Z_{i-1} = x)\\
        =& \mathbb P(Z_i \in A, U = 1| Z_{i-1} = x)\\
        &+ \mathbb P(Z_i \in A, U = 0| Z_{i-1} = x)\\
        \geq& \mathbb P(Z_i \in A, U = 1| Z_{i-1} = x)\\
        =& \mathbb P(B \in A, U = 1)\\
        =& (1-\rho)\mathbb P(B \in A).
    \end{align*}
    Thus, we have $\epsilon = 1-\rho > 0$, which concludes the proof.
\end{proof}

The following auxiliary lemma on the other hand provides a probabilistic guarantee on the structure of the design matrix under missingness (ensuring that the estimator $\hat{\bm{\beta}}$ can be computed with high probability in these settings).

\begin{Lemma}\label{lemma:restricted_eigen}
    Suppose Assumption~\ref{ass:missing} holds. Then with probability $1 - c\,n^{-3}$
    \begin{align*}
    \lambda_{\min}(\widetilde{\bm{X}}^T \widetilde{\bm{X}}) \geq C_{\tilde{x}}\,\mu(\bm{\vartheta})\,\lambda_{\min}(\bm{X}^T \bm{X}),
\end{align*}
where $c > 0$ and $C_{\tilde{x}} \in (0,1)$ are absolute constants.
\end{Lemma}
\begin{proof}
    It follows from Assumption~\ref{ass:missing} that $\bm{Z}$ is a strictly stationary, finite state Markov chain. Then, Theorem 3.7 in \cite{bradley2005basic} implies that $\{Z_i\}_{i = 1, \hdots, n}$ is a geometric ergodic sequence, i.e.~its $\beta$-mixing coefficients with lag $k$ satisfy the condition $\beta_Z(k) \leq \exp(1-c\,k)$. We apply Lemma~\ref{lemma:matrix_bernstein} with $\bm{M}_i = (Z_i - \mu(\bm{\vartheta}))\bm{X}_{i \cdot}\bm{X}_{i \cdot}^T$. With $\|\cdot\|_{\op}$ representing the operator norm, we have that
$$\max_i\|(Z_i - \mu(\bm{\vartheta}))\bm{X}_{i \cdot}\bm{X}_{i \cdot}^T\|_{\op} \leq \max_i \{\|\bm{X}_{i \cdot}\bm{X}_{i \cdot}^T\|_{\op}\} = \max_i\{\|\bm{X}_{i \cdot}\|_2^2\},$$
and
$$\iota^2 \leq C_{z}\max_i\{\|\bm{X}_{i \cdot}\bm{X}_{i \cdot}^T\|_{\op}^2\} = C_{z}\max_i\{\|\bm{X}_{i \cdot}\|_2^4\},$$
where $C_{z} > 0$ is a constant only depending on the Markov model of $\bm{Z}$. By Lemma~\ref{lemma:matrix_bernstein}, we have that with probability at least $1 - p\,n^{-3}$
\begin{align}\label{eq:event}
    \|\widetilde{\bm{X}}^T \widetilde{\bm{X}} -\mu(\bm{\vartheta}) \bm{X}^T \bm{X}\|_{\op} \leq  C_{\op}\max_i\{\|\bm{X}_{i \cdot}\|_{2}^2\}\,n^{1/2}\,\log n,
\end{align}
for some constant $C_{\op} > 0$.
By Lemma~\ref{lemma:weyl_ineq}, we have that
\begin{align*}
    \lambda_{\min}(\widetilde{\bm{X}}^T \widetilde{\bm{X}}) \geq & \,\mu(\bm{\vartheta})\,\lambda_{\min}(\bm{X}^T \bm{X}) - \|\widetilde{\bm{X}}^T \widetilde{\bm{X}} - \mu(\bm{\vartheta}) \bm{X}^T \bm{X}\|_{\op}\\
    \geq & \,\mu(\bm{\vartheta})\,\lambda_{\min}(\bm{X}^T \bm{X}) - C_{\op}\,\max_i\|\bm{X}_{i \cdot}\|_{2}^2\,n^{1/2}\,\log n\\
    \geq & \, (1 - C_{\op}\,C^{-1}_{x}\,\alpha_n^{-1})\,\mu(\bm{\vartheta})\,\lambda_{\min}(\bm{X}^T \bm{X}),
\end{align*}
where the last inequality follows from Assumption \ref{ass:missing}.
\end{proof}

With this additional lemma we can now proceed to the proof of \Cref{lemma:beta_cov}. Now we assume that $\widetilde{\bm{X}}^T \widetilde{\bm{X}}$ is non-singular, which holds with high probability due to \Cref{lemma:restricted_eigen}. Based on this, the least-squares estimator satisfies the following:
\begin{align}\label{eq:beta}
    \hat{\bm{\beta}} - \bm{\beta}_0 =& (\widetilde{\bm{X}}^T \widetilde{\bm{X}})^{-1}\widetilde{\bm{X}}^T\widetilde{\bm{Y}} - \bm{\beta}_0 = (\widetilde{\bm{X}}^T \widetilde{\bm{X}})^{-1}\widetilde{\bm{X}}^T\widetilde{\bm{\varepsilon}}\nonumber\\
    =& \left\{(\widetilde{\bm{X}}^T \widetilde{\bm{X}})^{-1} - \mu(\bm{\vartheta})^{-1}( \bm{X}^T\bm{X})^{-1}\right\} \bm{X}^T \widetilde{\bm{\varepsilon}} + \mu(\bm{\vartheta})^{-1}(\bm{X}^T\bm{X})^{-1}\bm{X}^T \widetilde{\bm{\varepsilon}} \nonumber\\ 
    =& (1+o_p(1)) \mu(\bm{\vartheta})^{-1}( \bm{X}^T \bm{X})^{-1} \bm{X}^T \widetilde{\bm{\varepsilon}},
\end{align}
where the last equality follows from the following fact. Since $\|(\bm{A}^{-1} - \bm{B}^{-1}) \bm{B}\|_{\op} \leq \lambda_{\min}^{-1}(\bm{A})\|\bm{A} - \bm{B}\|_{\op}$, we have that
\begin{align}\label{eq:rev_ratio}
    &\left\|\left\{(\widetilde{\bm{X}}^T \widetilde{\bm{X}})^{-1} - \mu(\bm{\vartheta})^{-1}(\bm{X}^T \bm{X})^{-1}\right\}\mu(\bm{\vartheta}) \bm{X}^T\bm{X}\right\|_{\op}\nonumber\\
    \leq& \lambda^{-1}_{\min}(\widetilde{\bm{X}}^T \widetilde{\bm{X}})\left\|\widetilde{\bm{X}}^T \widetilde{\bm{X}} - \mu(\bm{\vartheta}) \bm{X}^T \bm{X}\right\|_{\op}\nonumber\\
    =& O_p\left((1 - C_{\op}C^{-1}_{\mathrm{gap}}\alpha_n^{-1})^{-1}\mu(\bm{\vartheta})^{-1}\lambda^{-1}_{\min}(\bm{X}^T \bm{X})\,C_{\op}\,\max_i\{\|\bm{X}_{i \cdot}\|_{2}\}\, n^{1/2}\, \log n\right)\nonumber\\
    =& O_p\left(C_{\op}C^{-1}_{\mathrm{gap}}\alpha^{-1}_n(1 - C_{\op}C^{-1}_{\mathrm{gap}}\alpha_n^{-1})^{-1}\right)\nonumber\\
    =& o_p(1),
\end{align}
where the first equality follows from \eqref{eq:event} in the proof of \Cref{lemma:restricted_eigen}, and the second equality follows from \Cref{ass:missing}. 
Note that $\hat{\bm{\beta}}$ is unbiased,
and $\hat{\bm{\beta}} - \bm{\beta}_0 - \mu(\bm{\vartheta})^{-1}(\bm{X}^T\bm{X})^{-1} \bm{X}^T \widetilde{\bm{\varepsilon}} = o_p(1)\,\mu(\bm{\vartheta})^{-1}(\bm{X}^T \bm{X})^{-1} \bm{X}^T \widetilde{\bm{\varepsilon}} \overset{p}{\to} 0$, which implies that
\begin{align*}
    &\var\left(\mu(\bm{\vartheta})^{-1}(\bm{X}^T \bm{X} )^{-1}\bm{X}^T \widetilde{\bm{\varepsilon}} \right)\\
    =&
    \mu(\bm{\vartheta})^{-2}\mathbb E\left[(\bm{X}^T \bm{X})^{-1}\widetilde{\bm{X}}^T \bm{\Sigma}(\bm{\gamma})  \widetilde{\bm{X}} ( \bm{X}^T \bm{X})^{-1} \right]\\
    =&  \mu(\bm{\vartheta})^{-2}\mathbb E\left[(\bm{X}^T\bm{X})^{-1}(\bm{Z} \otimes \bm{1}_p^T \odot \bm{X})^T \bm{\Sigma}(\bm{\gamma})  (\bm{Z} \otimes \bm{1}_p^T \odot \bm{X}) ( \bm{X}^T \bm{X})^{-1} \right]\\
    =& \mu(\bm{\vartheta})^{-2}(\bm{X}^T \bm{X})^{-1} \bm{X}^T \left\{\mathbb{E}[\bm{Z}\bm{Z}^T] \odot \bm{\Sigma}(\bm{\gamma})\right\}  \bm{X} ( \bm{X}^T \bm{X})^{-1}\\
    =& \mu(\bm{\vartheta})^{-2}(\bm{X}^T \bm{X})^{-1} \bm{X}^T \left\{(\bm{\Lambda}(\bm{\vartheta}) + \mu(\bm{\vartheta})^2\bm{1}\bm{1}^T) \odot \bm{\Sigma}(\bm{\gamma})\right\} \bm{X} ( \bm{X}^T \bm{X})^{-1}.
\end{align*}

\subsection{Approximation of Theoretical WV using Residuals}
\label{app:IminusH}

Consider the covariance function of the residual (estimated error) process:
$$\mathbb{V}(\hat{\boldsymbol{\varepsilon}}) = 
\boldsymbol{\Sigma}_{\boldsymbol{\hat{\varepsilon}}}(\bm{\gamma})
= \left(\boldsymbol{I}-\boldsymbol{P} \right) \boldsymbol{\Sigma}(\boldsymbol{\gamma}) \left(\boldsymbol{I}-\boldsymbol{P} \right) \odot \left\{\boldsymbol{\Lambda}(\boldsymbol{\vartheta}) + \mu(\boldsymbol{\vartheta})^2 \mathbf{1} \mathbf{1}^T \right\},$$
which would be the true covariance structure of the residuals used to compute $\hat{\boldsymbol{\nu}}$ used in step 5 of the Algorithm \ref{algo:gmwmx}. A large-sample modification (under certain conditions) of the above expression would consist in the approximation:
$$\left(\boldsymbol{I}-\boldsymbol{P}\right)~\boldsymbol{\Sigma}(\boldsymbol{\gamma})~\left(\boldsymbol{I}-\boldsymbol{P} \right)~\approx~\boldsymbol{\Sigma}(\boldsymbol{\gamma}).$$
However, if possible, we would prefer to use the true covariance structure of the residuals $\boldsymbol{\Sigma}_{\boldsymbol{\hat{\varepsilon}}}(\bm{\gamma})$ on which  the theoretical WV should be based. A problem with this though is that, within the GMWM, the matrix $\boldsymbol{\Sigma}_{\boldsymbol{\hat{\varepsilon}}}(\bm{\gamma})$ should be computed at each evaluation of the GMWM objective function which requires a multiplicaiton of two $n \times n$ matrices (entailing a computational complexity of order $\mathcal{O}(n^3)$) as well as the Hadamard product of two $n \times n $ matrices (which has a computational complexity of order $\mathcal{O}(n^2)$).  

To overcome this computational bottleneck, we rely on the result of Lemma 1 of \cite{xu2017study}, which states that the theoretical WV of a (zero-mean) non-stationary process is a linear function in $L_J$ of the averages taken over the diagonal, sub- and super-diagonal elements of the process covariance matrix $\bm{\Sigma}$. Considering that $\left\{\boldsymbol{\Lambda}(\hat{\boldsymbol{\vartheta}}) + \mu(\hat{\boldsymbol{\vartheta}})^2 \mathbf{1} \mathbf{1}^T \right\} $ is a Toeplitz matrix due to the stationarity of $\boldsymbol{Z}$, the strategy that we propose to efficiently compute $\boldsymbol{\nu}\left( \boldsymbol{\Sigma}_{\boldsymbol{\hat{\epsilon}}}(\bm{\gamma}, \boldsymbol{\vartheta}) \right)$ is built on different considerations. We start by considering the case where the error process $\bm{\varepsilon}$ is stationary. Based on this, using the notation $\boldsymbol{R} = \boldsymbol{I}- \boldsymbol{P}$, the first consideration is that $\boldsymbol{R}\boldsymbol{\Sigma}(\boldsymbol{\gamma})$ is a close approximation to $\boldsymbol{R}\boldsymbol{\Sigma}(\boldsymbol{\gamma})\boldsymbol{R}$. This is intuitively justified because $\boldsymbol{H}\overset{p}{\rightarrow}\boldsymbol{I} $ and therefore $\boldsymbol{R}\overset{p}{\rightarrow} \bm{0}_{n \times n}$. Hence, we will first approximate $\boldsymbol{R}\boldsymbol{\Sigma}(\boldsymbol{\gamma})\boldsymbol{R}$ by $\boldsymbol{R}\boldsymbol{\Sigma}(\boldsymbol{\gamma})$. Hence, we are interested in efficiently computing the average over the diagonal and super/sub-diagonals of the matrix $\boldsymbol{R}\boldsymbol{\Sigma}(\boldsymbol{\gamma})$ where $\boldsymbol{R}$ is idempotent and symmetric and $\boldsymbol{\Sigma}(\boldsymbol{\gamma})$ is symmetric and Toeplitz (assuming a stationarity of the error process $\bm{\varepsilon}$). Using $\rho_{\boldsymbol{\varepsilon}}(k) = \cov(\varepsilon_t, \varepsilon_{t+k})$, with $k=0, \ldots , n-1$, to denote the autocovariance function of $\varepsilon$, without lack of generality, we define the sum of elements over the diagonal and over the $l^{th}$ super/sub-diagonals of $\boldsymbol{R} \boldsymbol{\Sigma}(\boldsymbol{\gamma})$ as

$$S_l=\sum_{i=1}^{n-l}(\boldsymbol{R} \boldsymbol{\Sigma}(\boldsymbol{\gamma}))_{i, i+l} = \sum_{k=1}^n \sum_{i=1}^{n-l} R_{i k} \rho_{\boldsymbol{\varepsilon}}(k-i-l) \quad \text{ for } l=0, \ldots , n-1 .$$

Following this definition, we can underline that  that the WV of the process $\hat{\bm{\varepsilon}}$ is a function of the vector $\left[\frac{1}{n}S_0, \frac{1}{n-1}S_1,  \ldots S_{n-1}\right]$. However, to avoid computing all the terms in $S_l$ within the evaluation of the objective function, it can be shown that 
\begin{equation*}
\label{eq::decomposition_s_l}
    S_l = \sum_{j=1}^{m-1} \left[\gamma_{\boldsymbol{\varepsilon}}(-l-j) \sum_{i=1}^{m-j} R_{i+j, i}\right]
+
\sum_{j=0}^l \left[ \gamma_{\boldsymbol{\varepsilon}}(-j) \sum_{i=1}^m R_{i, i+l-j}\right]
+
\sum_{j=1}^{m-1} \left[ \gamma_{\boldsymbol{\varepsilon}}(j) \sum_{i=1}^{m-j} R_{i, i+l+j}\right] \quad \text{ where } m = n-l,
\end{equation*}
which allows to pre-compute the quantities that depend on $\boldsymbol{R}$ outside of the optimization function and to simply index them in the optimization operation.

 Nevertheless, we still need to compute all these quantities on $\boldsymbol{R}$, which can be become computationally demanding as $n$ increases. Hence we propose the following computational trick. Defining the vector $\boldsymbol{S} = [S_0, S_1, \ldots, S_{n-1}]$ and the vector of autocovariance of $\boldsymbol{\bm{\varepsilon}}$, $\boldsymbol{\rho}_{\boldsymbol{\varepsilon}} = [\rho_{\boldsymbol{\varepsilon}}(0), \rho_{\boldsymbol{\varepsilon}}(1),\ldots \rho_{\boldsymbol{\varepsilon}}(n-1)]$, we propose to approximate the vector $\boldsymbol{S}$ by only computing some specific entries $i$ between $0$ and $n-1$  of the vector $\boldsymbol{S}$ and computing the difference between $\boldsymbol{S}_i$ and $\boldsymbol{\rho}_{\boldsymbol{\varepsilon}_i}$. After this we can then correct the vector $\boldsymbol{\rho}_{\boldsymbol{\varepsilon}}$ by performing a linear interpolation using the errors between $\boldsymbol{S}_i$ and $\boldsymbol{\rho}_{\boldsymbol{\varepsilon}_i}$. Our empirical results suggest that the number of points on which to evaluate 
$\boldsymbol{S}$, $n^{\star} = a + b\log_2(n)$ with $a=1$ and $b=3$ seems to provide a good approximation in the resulting WV. Finally, if we denote the latter approximation as $\boldsymbol{S}^\star$, we can then define $B_l = \sum_{i=1}^{n-l}\left(\left[\boldsymbol{R} \boldsymbol{\Sigma}(\boldsymbol{\gamma})\right] \odot \left\{\boldsymbol{\Lambda}({\boldsymbol{\vartheta}}) + \mu({\boldsymbol{\vartheta}})^2 \mathbf{1} \mathbf{1}^T \right\} \right)_{i, i+l}$ (with $\boldsymbol{B} = [B_0, B_1, \ldots, B_{n-1}]$). Hence, we can approximate the vector $\boldsymbol{B}$ with $\boldsymbol{B}^\star = \boldsymbol{S}^\star \odot \left( \boldsymbol{\Lambda}(\boldsymbol{\vartheta})_1 + \mu(\boldsymbol{\vartheta})^2 \mathbf{1} \mathbf{1}^T \right)$ where $\boldsymbol{\Lambda}(\boldsymbol{\vartheta})_1$ represents the first column of the covariance matrix of $Z$ (i.e. the autocovariance vector of the stationary process $Z$). With this final approximation we can then rescale the vector $\boldsymbol{B}^\star $ to obtain an approximation of the average over the diagonal and super/sub-diagonals of $\left(\left[\boldsymbol{R} \boldsymbol{\Sigma}(\boldsymbol{\gamma})\right] \odot \left\{\boldsymbol{\Lambda}({\boldsymbol{\vartheta}}) + \mu({\boldsymbol{\vartheta}})^2 \mathbf{1} \mathbf{1}^T \right\} \right)$ by computing $\boldsymbol{B}^{\star \star } = [\frac{1}{n}B_0, \frac{1}{n-1}B_1, \frac{1}{n-2}B_3, \ldots B_{n-1}]$. We can then apply the results of \cite{xu2017study} to obtain the desired approximation of the WV when computed on the residual process (i.e. estimated error process).

It is important to note that when the stochastic process $\boldsymbol{\varepsilon}$ is stationary, the approximation is an approximation due to 1) considering $\boldsymbol{R}\boldsymbol{\Sigma}(\boldsymbol{\gamma})$ instead of $\boldsymbol{R}\boldsymbol{\Sigma}(\boldsymbol{\gamma})\boldsymbol{R}$ and 2) because we approximate $\boldsymbol{S}$ by $\boldsymbol{S}^\star$. When the process  $\boldsymbol{\varepsilon}$ is not stationary (e.g. the stochastic model considered in Setting B of the Section \ref{sec:simulations}), we introduce an additional approximation step. Indeed the result obtained in \ref{eq::decomposition_s_l} is only valid in case $\boldsymbol{\Sigma}(\boldsymbol{\gamma})$ is a Toeplitz matrix which allows to work directly with the autocovariance function. However, in the case of a non-stationary process, the autocovariance function is not clearly defined. Therefore, we consider the same approximation strategy but we consider the average over the diagonal and super/sub-diagonals as a function of the vector $\bm{\rho}_{\boldsymbol{\varepsilon}}$. For the power-law noises considered in this work, the average of the diagonal and super/sub-diagonals can be efficiently computed using the strategy described below.

A power-law noise is defined by its power spectrum $P$ where $P(f) = \frac{P_0}{f^{\alpha}}$ where $f$ is the frequency, $P_0$ is a constant and $\alpha$ is the spectral index. This type of noise is parametrized by its variance $\sigma^2_{\text{PL}}$ and its spectral index $\alpha_{\text{PL}}$. A power-law noise, say $\boldsymbol{e}$, with a  spectral index between $-1$ and $1$ can be generated by convolving it with an independent and i.i.d. noise $w_k$ \citep{hosking1981fractional}:

$$e_k=\sum_{i=0}^{\infty} h_i w_{k-i},$$

where, following \cite{kasdin1995discrete}, we have:
$$
\begin{aligned}
h_0 & =1 \\
h_i & =\left(\frac{\alpha}{2}+i-1\right) \frac{h_{i-1}}{i} .
\end{aligned}
$$

For a spectral index value $\alpha_{\text{PL}} < 1$ and assuming an infinite sequence of zero-mean white noise $\boldsymbol{w}$, with variance $\sigma^2_{\text{PL}}$, the noise is stationary with covariance function \citep{hosking1981fractional}:

$$
\begin{aligned}
& \rho(0)=\sigma_{\text{PL}}^2 \frac{\Gamma(1-\alpha)}{\left(\Gamma\left(1-\frac{\alpha}{2}\right)\right)^2} \\
& \rho(k)=\frac{\frac{\alpha}{2}+k-1}{-\frac{\alpha}{2}+k} \rho(k-1) \text { for } k>0 ,
\end{aligned}
$$
where $\Gamma(\cdot)$ is the gamma function. Assuming that $w_i= 0$  for $i < 0$, the covariance between $e_k$ and $e_l$ ($l>k$) is given by:

$$
\operatorname{cov}\left(e_k, e_l\right)=\sigma_{p l}^2 \sum_{i=0}^k h_i h_{i+(l-k)}.
$$

Similarly, this can be written in a matrix form as:

$$
\mathbb{V}[\boldsymbol{e}] = \sigma^2_{\text{PL}}\boldsymbol{U}^T \boldsymbol{U} ,
$$
where
$$
\mathbf{U}=\left(\begin{array}{cccc}
h_0 & h_1 & \ldots & h_N \\
0 & h_0 & & h_{N-1} \\
\vdots & & \ddots & \vdots \\
0 & 0 & \ldots & h_0
\end{array}\right) .
$$

Hence, we can easily obtain the average over the entries of the $k^{\text{th}}$ super-diagonal ($k=0$ being equal to $\frac{1}{n}\tr(\mathbb{V}[\boldsymbol{e}])$) defined as: 

$$ 
d_k = \frac{1}{n-k}\sum_{i=1}^{n-k}\mathbb{V}[\boldsymbol{e}]_{i, i+k} = \frac{1}{n-k} \left(\sum_{i=1}^{n-k} (n-i-k+1) h_{i-1}h_{i-1+k}\right).
$$

\subsection{Beta-Mixing Missingness Process for Simulation}
\label{app:beta_mixing}

In this appendix we briefly show that the Markov chain used for the simulations in Section \ref{sec:simulations} is a special case of \eqref{eq:markov_missinigness}. We first see that
\[
P(Z_1 = 1) = \mu(\bm{\vartheta}), P(Z_1 = 0) = 1- \mu(\bm{\vartheta}),
\]
\[
P(Z_1 = 1, Z_2 = 1) + P(Z_1 = 0, Z_2= 0) = (1-p_1)\mu(\bm{\vartheta}) + (1-p_2)(1- \mu(\bm{\vartheta})) = \rho,
\]
\[
P(Z_1 = 0, Z_2 = 1) + P(Z_1 = 1, Z_2 = 0) = p_2(1-\mu(\bm{\vartheta})) + p_1 \mu(\bm{\vartheta}) = 1-\rho.
\]
By stationarity we have
\[
P(Z_2 = 1) = P(Z_1 = 0, Z_2 = 1) + P(Z_1 = 1, Z_2 = 1) = p_2(1-\mu(\bm{\vartheta})) + (1-p_1) \mu(\bm{\vartheta}) = p_2 + (1-p_1- p_2)\mu(\bm{\vartheta}) = \mu(\bm{\vartheta}),
\]
which leads to
\[
p_2 = (p_1+ p_2)\mu(\bm{\vartheta}).
\]
By stationarity again, we have
\[
P(Z_2 = 0) = (1-p_2)(1- \mu(\bm{\vartheta})) + 
p_1 \mu(\bm{\vartheta}) = 1-p_2 + (p_1 + p_2 -1) \mu(\bm{\vartheta}) = 1 - \mu(\bm{\vartheta}),\]
which also leads to
\[
p_2 = (p_1+ p_2)\mu(\bm{\vartheta}).
\]
Thus, we have 
\begin{align*}
    \begin{cases}
        p_2 = (p_1+ p_2)\mu(\bm{\vartheta}), \,\,\, \text{and}\\
        (1-p_1)\mu(\bm{\vartheta}) + (1-p_2)(1- \mu(\bm{\vartheta})) = \rho,
    \end{cases}
\end{align*}
where
\[
\mu(\bm{\vartheta}) = \frac{p_2}{p_1 + p_2},
\]
and
\[
\rho = \frac{p_2(1-p_1)}{p_1 + p_2} + \frac{(1-p_2)^2}{p_1 + p_2}.
\]

\subsection{Proof of Theorem \ref{theorem:short_memory}}
\label{proof:theorem_short_memory}
We recall that
\[
\widetilde{\bm{\Pi}}_n = \widetilde{\bm{X}}^T \widetilde{\bm{X}}
\]
and
\[
\hat{\bm{\beta}} - \bm{\beta}_0 = (\widetilde{\bm{X}}^T \widetilde{\bm{X}})^{-1}\widetilde{\bm{X}}^T \widetilde{\bm{Y}} - \bm{\beta}_0 = (\widetilde{\bm{X}}^T \widetilde{\bm{X}})^{-1}\widetilde{\bm{X}}^T \widetilde{\bm{\varepsilon}}.
\]
Then it follows that
\[
\widetilde{\bm{\Pi}}_n^{1/2}(\hat{\bm{\beta}} - \bm{\beta}_0) = \widetilde{\bm{\Pi}}_n^{-1/2} \widetilde{\bm{X}}^T \widetilde{\bm{\varepsilon}} = \sum_{i = 1}^n\widetilde{\varepsilon}_{i} \widetilde{\bm{Q}}_{i \cdot},
\]

from which we can also conclude that $\bigg\| \sum_{i = 1}^n\widetilde\varepsilon_{i} \widetilde{\bm{Q}}_{i \cdot} \bigg\|_2 = \, \bigg\| \sum_{i = 1}^n \varepsilon_{i} \widetilde{\bm{Q}}_{i \cdot} \bigg\|_2$ since $Z_i\widetilde{\bm{Q}}_{i \cdot} = \widetilde{\bm{Q}}_{i \cdot}$. For the next steps of the proof, assuming $i = \{1, \hdots, n\}$ is an ordered sequence (e.g. a time series), we define the projection operator as $\mathcal{P}_{i}(\cdot) = \mathbb{E}[\cdot | \mathcal{F}_i] - \mathbb{E}[\cdot | \mathcal{F}_{i-1}]$ which represents a measure of how much the conditional expectation of a process can change once the immediately previous information is removed. Based on this, straightforward computations show that $\varepsilon_i = \sum_{l = 0}^\infty \mathcal{P}_{i-l}(\varepsilon_i)$ which allows us to do the following:

It follows that
    \begin{align*}
        \, \bigg\| \sum_{i = 1}^n \varepsilon_{i} \widetilde{\bm{Q}}_{i} \bigg\|_2 =& \bigg\| \sum_{i = 1}^n \sum_{l = 0}^{\infty}\mathcal{P}_{i-l}(\varepsilon_{i}) \widetilde{\bm{Q}}_{i \cdot} \bigg\|_2 \leq \sum_{l = 0}^{\infty}\bigg\| \sum_{i = 1}^n \mathcal{P}_{i-l}(\varepsilon_{i}) \widetilde{\bm{Q}}_{i \cdot} \bigg\|_2\\
        =& \sum_{l = 0}^{\infty}\sqrt{\mathbb E\bigg[\bigg| \sum_{i = 1}^n \mathcal{P}_{i-l}(\varepsilon_{i}) \widetilde{\bm{Q}}_{i \cdot} \bigg|_2^2\bigg]} = \sum_{l = 0}^{\infty}\sqrt{\mathbb E\bigg[\sum_{j = 1}^p\bigg( \sum_{i = 1}^n \mathcal{P}_{i-l}(\varepsilon_{i}) \widetilde{Q}_{ij} \bigg)^2\bigg]}\\
        =& \sum_{l = 0}^{\infty}\sqrt{\mathbb E \bigg[\sum_{j = 1}^p\mathbb E\bigg[\bigg( \sum_{i = 1}^n \mathcal{P}_{i-l}(\varepsilon_{i}) \widetilde Q_{ij} \bigg)^2\bigg| \bm{Z} \bigg]\bigg]} = \sum_{l = 0}^{\infty}\sqrt{\mathbb E [(\mathcal{P}_{0}(\varepsilon_{l}))^2]\mathbb E \bigg[\sum_{j = 1}^p\sum_{i = 1}^n  \widetilde Q_{ij}^2\bigg]}\\
        =& p^{1/2}\sum_{l = 0}^{\infty}\big\|\mathcal{P}_{0}(\epsilon_{k}))\big\|_2 \leq p^{1/2}\Delta_{0,2} < \infty,
    \end{align*}
where (i) the second equality follows from the definition of the projection operator $\mathcal{P}_i(\cdot)$; (ii) the first inequality follows from the triangle inequality; (iii) the fifth equality follows from the orthogonality of the projection operator, the stationarity of $\{\varepsilon_{i}\}$ and the independence between $\{\varepsilon_i\}$ and $\bm{Z}$; and finally (iv) the seventh equality follows from the fact that $\sum_{i = 1}^n \widetilde{\bm{Q}}_{i \cdot}^T \widetilde{\bm{Q}}_{i \cdot} = p$. Thus, by Markov's inequality, we have that
    \begin{align*}
        \sum_{i = 1}^n\widetilde\varepsilon_{i}\widetilde{\bm{Q}}_{i \cdot} = O_p(1).
    \end{align*}

The proof of the limiting distribution of $\sum_{i = 1}^n\widetilde\varepsilon_{i}\widetilde{\bm{Q}}_{i \cdot}$ follows that of Lemma~2 in \cite{wu2007m}, which relies on the requirement that, if conditioning on the random missingness vector $\bm{Z}$, we have
\begin{align}\label{eq:rescaling_condition}
    \max_{i \leq n}(\widetilde{\bm{X}}_{i \cdot}^T \widetilde{\bm{\Pi}}_n^{-1} \widetilde{\bm{X}}_{i \cdot})^{1/2} = o(1).
\end{align}

To link the above requirement to \Cref{ass:X},
we will apply \Cref{lemma:matrix_bernstein} to obtain a high probability bound on $\|\widetilde{\bm{\Pi}}_n - \mu(\bm{\vartheta}) \bm{\Pi}_n\|_{\op}$ and a lower bound of $\lambda_{\min}(\widetilde{\bm{\Pi}}_n)$ using \Cref{lemma:weyl_ineq}. 

It follows from Theorem 3.7 in \cite{bradley2005basic} that the strictly stationary, finite state Markov chain $\bm{Z} = \{Z_i\}$ is geometrically ergodic, i.e.~$\beta_Z(\ell) \leq \exp(1-c\ell)$ (see Lemma \ref{lemma:ergodic}). We apply \Cref{lemma:matrix_bernstein} with $\bm{M}_i = (Z_i - \mu(\bm{\vartheta}))\bm{X}_{i \cdot}\bm{X}_{i \cdot}^T$. We have that
$$\max_i\|(Z_i - \mu(\bm{\vartheta}))\bm{X}_{i \cdot}\bm{X}_{i \cdot}^T\|_{\op} \leq \max_i\|\bm{X}_{i \cdot}\bm{X}_{i \cdot}^T\|_{\op},$$
and
$$\iota^2 \leq C\max_i\|\bm{X}_{i \cdot}\bm{X}_{i \cdot}^T\|_{\op}^2,$$
where $C > 0$ is a constant only depending on the Markov model of $Z_i$. By \Cref{lemma:matrix_bernstein}, we have that with probability at least $1 - p\,n^{-3}$
\begin{align}\label{eq:event}
    \|\widetilde{\bm{\Pi}}_n - \mu(\bm{\vartheta})\bm{\Pi}_n\|_{\op} \leq C_{\op}\max_i\|\bm{X}_{i \cdot}\bm{X}_{i \cdot}^T\|_{\op}n^{1/2}\log n,
\end{align}
for some sufficient large constant $C_{\op} > 0$. The following proof proceeds by conditioning on the high probability event in \eqref{eq:event}. By \Cref{lemma:weyl_ineq}, we have that
\begin{align*}
    \lambda_{\min}(\widetilde{\bm{\Sigma}}_n) \geq \mu(\bm{\vartheta})\lambda_{\min}(\bm{\Pi}_n) - \|\widetilde{\bm{\Pi}}_n - \mu(\bm{\vartheta})\bm{\Pi}_n\|_{\op} \geq \mu(\bm{\vartheta})\lambda_{\min}(\bm{\Pi}_n) - C_{\op}\max_i\|\bm{X}_{i \cdot}\bm{X}_{i \cdot}^T\|_{\op}n^{1/2}\log n = C^{\prime}\mu(\bm{\vartheta})\lambda_{\min}(\bm{\Pi}_n),
\end{align*}
where the equality follows from \Cref{ass:missing}, i.e.~$\mu(\bm{\bm{\vartheta}})\,\lambda_{\min}(\bm{X}^T \bm{X}) \geq C_{\mathrm{gap}} \, \alpha_n \, \max_i\{\|\bm{X}_{i \cdot}\|_{2}^2\}n^{1/2} \, \log (n)$, and $C^{\prime} \in (0,1)$ is an absolute constant. Consequently, we have that
$$\max_{i \leq n}(\widetilde{\bm{X}}_{i \cdot}^T \widetilde{\bm{\Pi}}_n^{-1} \widetilde{\bm{X}}_{i \cdot})^{1/2} \leq \max_{i \leq n}(\bm{X}_{i \cdot}^T \widetilde{\bm{\Pi}}_n^{-1} \bm{X}_{i \cdot})^{1/2} \leq \sqrt{\frac{\max_{i \leq n}|\bm{X}_{i \cdot}|_2^2}{C^{\prime}\mu(\bm{\vartheta})\lambda_{\min}(\bm{\Pi}_n)}}  = o(1).$$

The proof of \eqref{eq:apply_CLT} then follows that of Lemma 2 in \cite{wu2007m}.

\subsection{Proof of Theorem \ref{theorem:long_memory}}
\label{proof:theorem_long_memory}

    Recall from \eqref{eq:beta} that
    the least squares estimator satisfies the following
\begin{align*}
    \hat{\bm{\beta}} - \bm{\beta}_0 
    =& (1+o_p(1)) \, \mu(\bm{\vartheta})^{-1}( \bm{X}^T \bm{X} )^{-1} \bm{X}^T\widetilde{\bm{\varepsilon}}.
\end{align*}
Then
\begin{align*}
    \bm{D}_n^{1/2}(\hat{\bm{\beta}} - \bm{\beta}_0) 
    =& o_p(1) \mu(\bm{\vartheta})^{-1}(\bm{D}_n^{-1/2} \bm{X}^T \bm{X} \bm{D}_n^{-1/2})^{-1} \bm{D}_n^{-1/2}\bm{X}^T \widetilde{\bm{\varepsilon}}\\
    &+ \mu(\bm{\vartheta})^{-1}(\bm{D}_n^{-1/2} \bm{X}^T  \bm{D}_n^{-1/2}\bm{X})^{-1}\bm{D}_n^{-1/2}\bm{X}^T \widetilde{\bm{\varepsilon}}\\
    =& o_p(1) \mu(\bm{\vartheta})^{-1}\bm{C}_n^{-1} \bm{D}_n^{-1/2}\bm{X}^T \widetilde{\bm{\varepsilon}} + \mu(\bm{\vartheta})^{-1}\bm{C}_n^{-1}\bm{D}_n^{-1/2}\bm{X}^T \widetilde{\bm{\varepsilon}}\\
    =& \underbrace{o_p(1) \mu(\bm{\vartheta})^{-1} \bm{C}_n^{-1} \bm{G}_n \widetilde{\bm{\varepsilon}}}_{I} + \underbrace{\mu(\bm{\vartheta})^{-1} \bm{C}_n^{-1}\bm{G}_n (\widetilde{\bm{\varepsilon}} - \bm{\varepsilon})}_{II} + \underbrace{\mu(\bm{\vartheta})^{-1} \bm{C}_n^{-1}\bm{G}_n  \bm{\varepsilon}}_{III},
\end{align*}
where we let $\bm{C}_n = \bm{D}_n^{-1/2}\bm{X}^T\bm{X} \bm{D}_n^{-1/2} \in \mathbb R^{p \times p}$ and $\bm{G}_n = \bm{D}_n^{-1/2}\bm{X}^T \in \mathbb R^{p \times n}$. Note that the $j^{th}$ column of $\bm{G}_n$ satisfies that
\begin{align*}
    \bm{G}_{n_{\cdot j}} = \bm{D}_n^{-1/2}\bm{X}_{\cdot j} = \begin{bmatrix}
\frac{X_{j,1}}{\sqrt{\sum^n_{l=1}X_{l,1}^2}} \\
\vdots \\
\frac{X_{j,p}}{\sqrt{\sum^n_{l=1}X_{l,p}^2}},
\end{bmatrix}
\end{align*}
and the $(i,j)^{th}$ entry of $\bm{G}_n$ is
$$G_{n_{(i, j)}} = \frac{X_{j,i}}{\sqrt{\sum_{l = 1}^nX_{l,i}^2}} = n^{-1/2}\frac{X_{j,i}}{\sqrt{\frac{1}{n}\sum_{l = 1}^nX_{l,i}^2}}.$$
For $i = 1, \dots, p$, define the function $G_{n_i}: \mathbb R \mapsto \mathbb R$, such that, for $j = 1, \dots, n$, we have
$$G_{n_i}\left(\frac{j}{n}\right) = \frac{X_{j,i}}{\sqrt{\frac{1}{n}\sum_{l = 1}^nX_{l,i}^2}},$$
which can be represented by the piece-wise constant function
$$G_{n_i}(u) = \begin{cases}
    G_{n_i}\left(\frac{1}{n}\right), & \frac{1}{n} \leq u < \frac{2}{n}\\
    \cdots \\
    G_{n_i}\left(\frac{n-1}{n}\right), & \frac{n-1}{n} \leq u < 1\\
    G_{n_i}\left(1\right), & u = 1\\
    0 \quad &\text{otherwise}
\end{cases}.$$
Define also $G_i(u) = \lim_{n \to \infty}G_{n_i}(u)$ for any $u \in \mathbb R$ whose existence can be verified case by case. For instance, 
Example~7.1 in \cite{beran2013long} considers the polynomial regression model. In our case, the $(i,j)^{th}$ entry of $\bm{C}_n$ is given by
\begin{align*}
    C_{n_{i,j}} = \frac{\sum_{l = 1}^nX_{l,j}X_{l,i}}{\sqrt{\sum_{l = 1}^nX_{l,j}^2}\sqrt{\sum_{l = 1}^nX_{l,i}^2}},
\end{align*}
based on which it can be shown that also $\bm{C}_n$ converges to a nondegenerate $\bm{C} \in \mathbb R^{p \times p}$ as $n \to \infty$.

It is straightforward to see that term $I$ is dominated by term $III$, leaving us to show that $|II|_2 = o_p(1)$. Note that $II$ is a $p$-dimensional vector with fixed $p$, so it suffices to show that for each $j \in \{1, \dots, p\}$, the $j^{th}$ entry $[\bm{G}_n (\widetilde{\bm{\varepsilon}} - \bm{\varepsilon})]_j = o_p(1)$. Note that
\begin{align*}
    [\bm{G}_n (\widetilde{\bm{\varepsilon}} - \bm{\varepsilon})]_j = \sum_{i = 1}^n G_{n_{i, j}} (Z_i - \mu(\bm{\vartheta}))\varepsilon_i,
\end{align*}
and the lag-$k$ autocovariance of the process $\{(Z_i - \mu(\bm{\vartheta}))\varepsilon_i\}$ is
\begin{align*}
    \cov\left[(Z_i - \mu(\bm{\vartheta}))\varepsilon_i, (Z_{i+k} - \mu(\bm{\vartheta}))\varepsilon_{i+k}\right] =& \mathbb E\left[ (Z_i - \mu(\bm{\vartheta}))(Z_{i+k} - \mu(\bm{\vartheta}))\,\varepsilon_i \varepsilon_{i+k} \right]\\
    =& \mathbb E\left[ (Z_i - \mu(\bm{\vartheta}))(Z_{i+k} - \mu(\bm{\vartheta}))\right] \mathbb E\left(\varepsilon_i \varepsilon_{i+k} \right)\\
    \leq& \, \sigma_\varepsilon^2 \cov\left[Z_i - \mu(\bm{\vartheta}), Z_{i+k} - \mu(\bm{\vartheta})\right],
\end{align*}
where $\sigma_{\varepsilon}^2 = \mathbb{V}[\varepsilon_i] < \infty$ is the marginal variance of $\bm{\varepsilon}$. Thus, we have for each $j \in \{1, \dots, p\}$,
\begin{align*}
    \mathbb{V}[\bm{G}_n (\widetilde{\bm{\varepsilon}} - \bm{\varepsilon})]_i =&  \mathbb E\left(\left\{ \sum_{i = 1}^n G_{n_{i,j}} (Z_i - \mu(\bm{\vartheta}))\varepsilon_i\right\}^2 \right)\\
    \leq& \max_i \frac{X_{i,j}^2}{\sum_{l = 1}^nX_{l,j}^2}\mathbb E\left(\left\{ \sum_{i = 1}^n (Z_i - \mu(\bm{\vartheta}))\varepsilon_i\right\}^2 \right)\\
    \leq& \max_i \frac{X_{i,j}^2}{\sum_{l = 1}^nX_{l,j}^2}\sigma_{\varepsilon}^2\sum_{k = -\infty}^{\infty} \cov\left(Z_i - \mu(\bm{\vartheta}), Z_{i+k} - \mu(\bm{\vartheta})\right)\\
    =& \max_i \frac{X_{i,j}^2}{\sum_{l = 1}^nX_{l,j}^2}\,\sigma_\varepsilon^2\,\sigma_{Z, \infty}^2\\
    =& \, o(1),
\end{align*}
where $\sigma_{z, \infty}^2 < \infty$ is the long-run variance of $\{Z_i\}$, and the last line follows from \Cref{ass:X}. 

We then consider term $III$, which dominates the other two terms and thus provides the limiting distribution. For this purpose, we consider the long-memory regime of $\{\varepsilon_i\}$, i.e.~$0 < d < 1/2$, for which we verify the condition of \Cref{thm:clt_long-memory}. Hereinafter we define $G_{n_i}(u) = f_n(u)$ and $G_i(u) = f(u)$, for $j = 1, \dots, p$. Since $G_{n_j}(u) < \infty$ and the support of $G_{n_j}(u)$ is $(0, 1]$ by definition, Condition (i) of  \Cref{thm:clt_long-memory} is satisfied. For Condition (ii) of  \Cref{thm:clt_long-memory}, note that $B_n^d(u)$ is a partial sum process of a stationary process with long-range dependence. The convergence to a standard fractional Brownian motion is satisfied by many examples, such as Gaussian processes \cite[see Theorem~4.2 in][]{beran2013long} and the linear processes \cite[see Theorem~4.6 in][]{beran2013long}. Hence, after proper rescaling, by \Cref{thm:clt_long-memory} we have
\[
n^{-d} \mu(\bm{\vartheta})^{-1}\bm{C}_n^{-1}\bm{G}_n  \bm{\varepsilon} \xrightarrow{\mathcal{D}} \mu(\bm{\vartheta})^{-1}\bm{C}^{-1}\int_{0}^1 G(u) \, \mathrm{d}B^{d}(u),
\]
where $\{B^d(u)\}_{u \geq 0}$ is a standard fractional Brownian motion.  Therefore we have that
\begin{align*}
    n^{-d}\bm{D}_n^{1/2}(\hat{\bm{\beta}} - \bm{\beta}_0) \xrightarrow{\mathcal{D}} \mu(\bm{\vartheta})^{-1}\bm{C}^{-1}\int_{0}^1 G(u) \, \mathrm{d}B^{d}(u),
\end{align*}
which concludes the proof.

\FloatBarrier
\subsection{Extended Simulation Results}
\label{app:extended_simu_results}

In this appendix we give all simulation results that were not directly discussed in Section \ref{sec:simulations}. 

\subsubsection{Setting A}

\begin{figure}[h]
    \centering
    \includegraphics[width=1\linewidth]{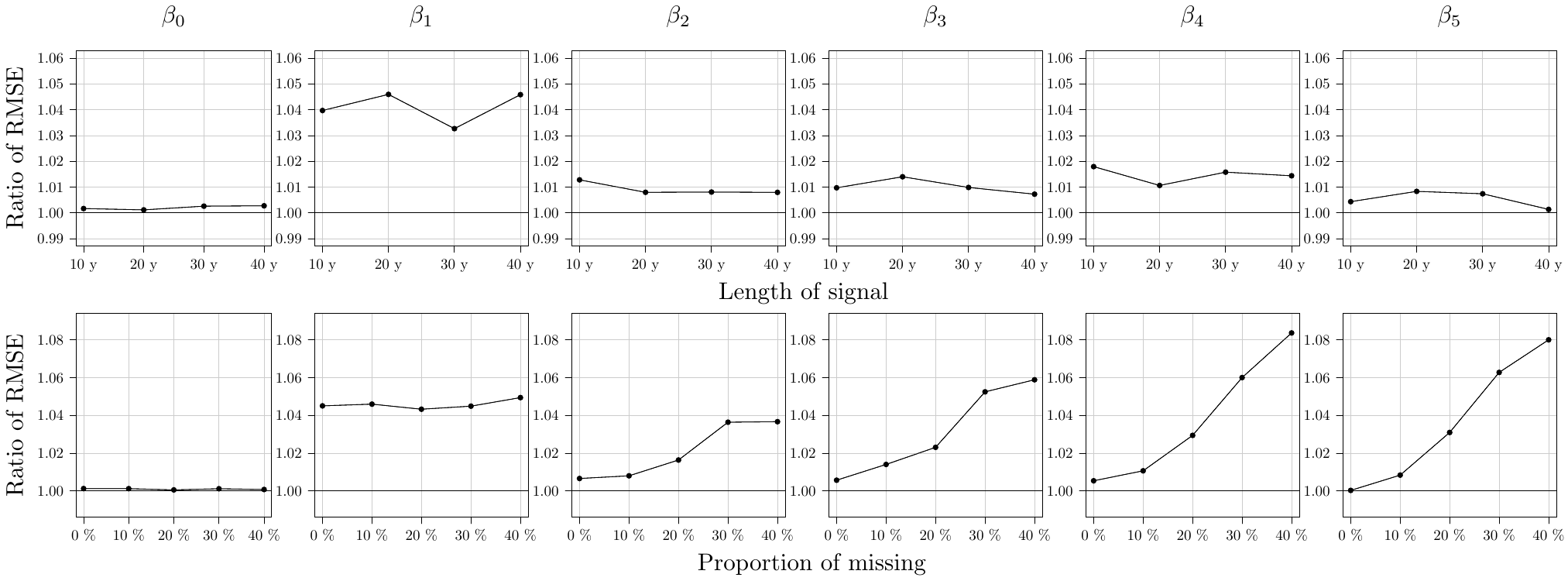}
    \caption{Ratio of RMSE for parameters ${\boldsymbol{\beta}}$ for Setting A1 and A2.}
    \label{fig:rmse_all_beta_a1_a2}
\end{figure}

\begin{figure}
    \centering
    \includegraphics[width=1\linewidth]{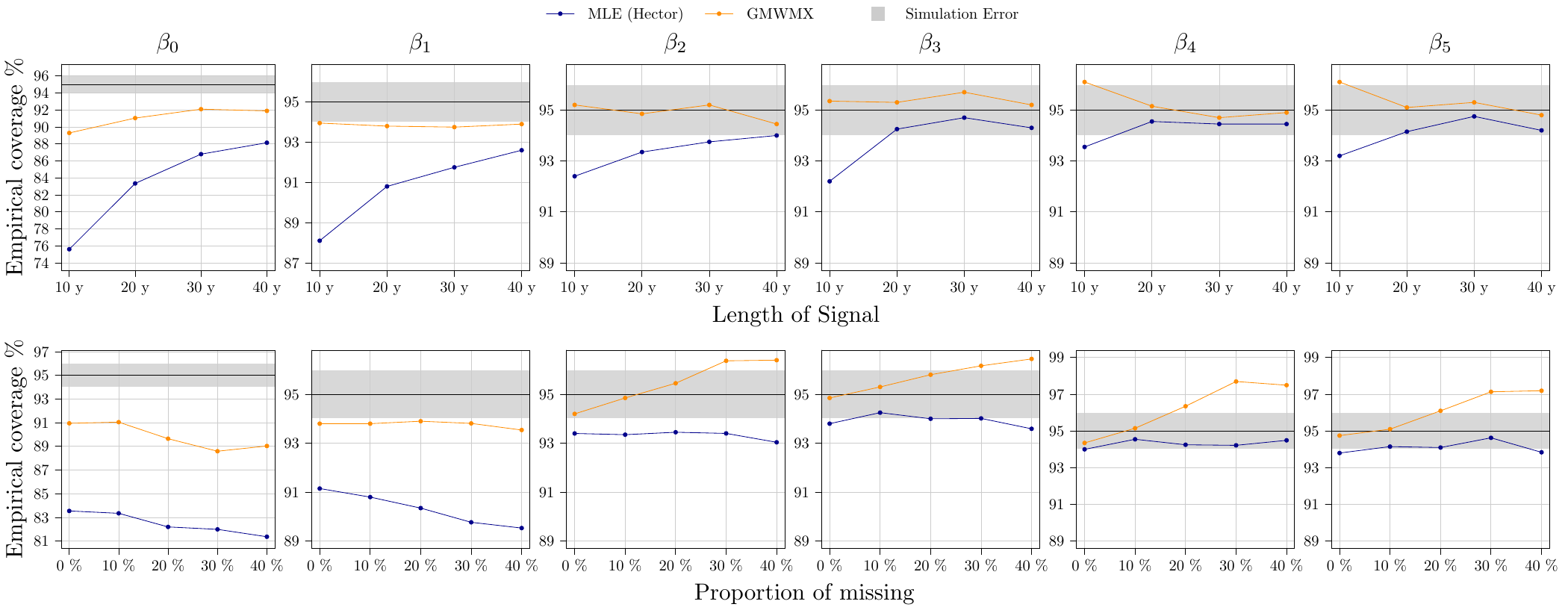}
    \caption{Empirical coverage for parameters $\boldsymbol{\beta}$ for Setting A1 and A2.}
    \label{fig:emp_coverage_all_beta_a1_a2}
\end{figure}

\begin{figure}
    \centering
    \includegraphics[width=.5\linewidth]{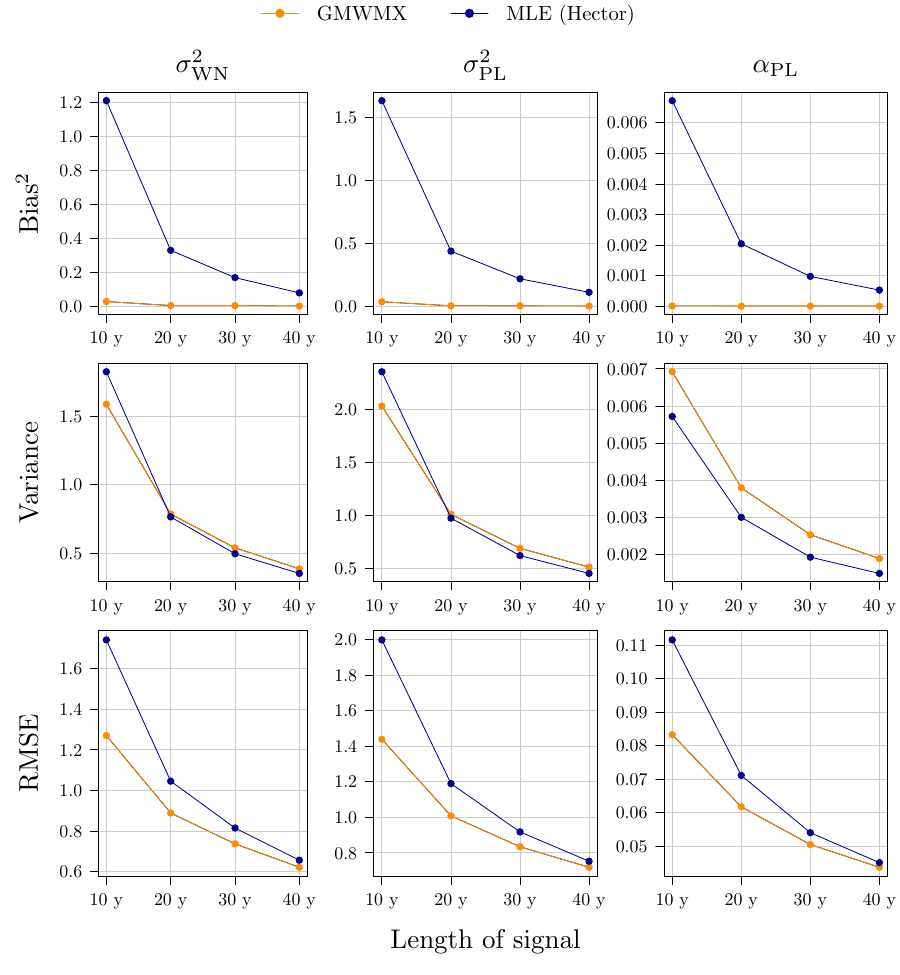}
    \caption{Bias$^2$, variance and RMSE for parameters ${\boldsymbol{\gamma}}$ for Setting A1.}
    \label{fig:bias_variance_rmse_gamma_a1}
\end{figure}

\begin{figure}
    \centering
    \includegraphics[width=.55\linewidth]{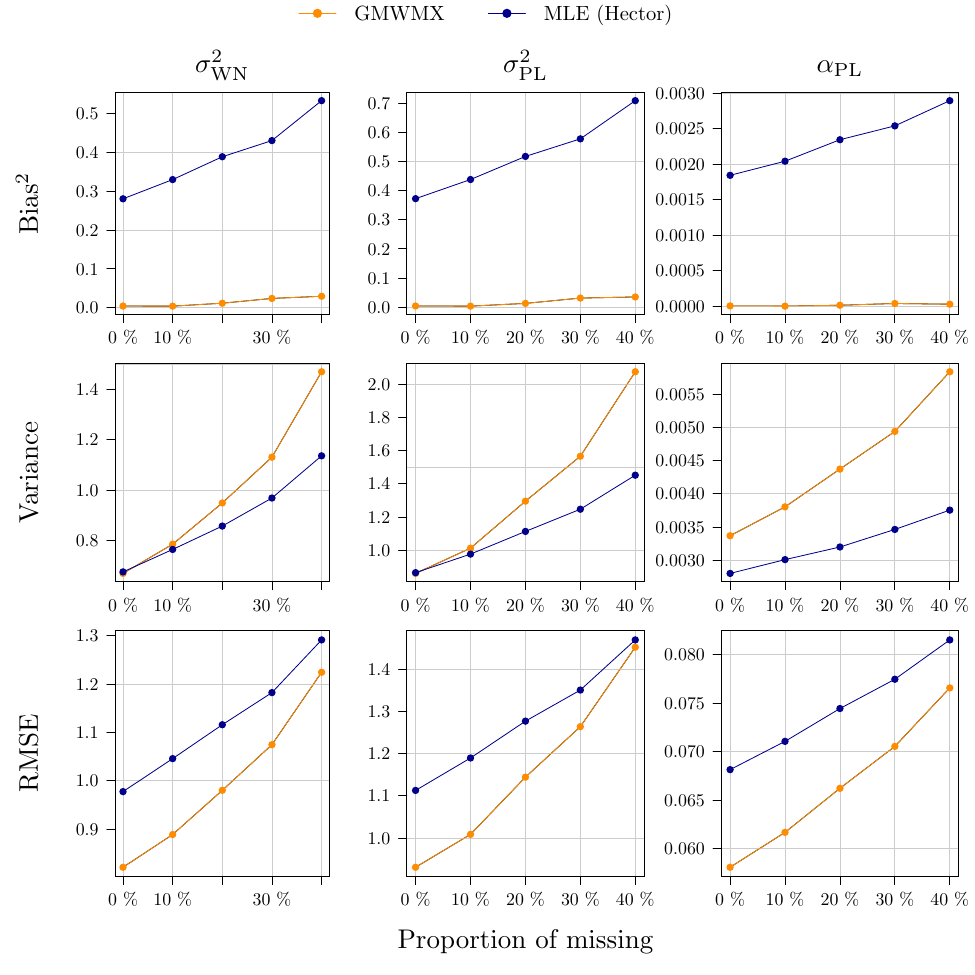}
    \caption{Bias$^2$, variance and RMSE for parameters ${\boldsymbol{\gamma}}$ for Setting A2.}
    \label{fig:bias_variance_rmse_gamma_a2}
\end{figure}

To start, we provide the results (in all sample size and missingness scenarios) for all elements of the parameter vector $\bm{\beta}$ in \textbf{Setting A}. More precisely, the ratios of RMSE for the point estimates of $\boldsymbol{\beta}$ obtained with the GMWMX and \texttt{Hector} are presented in Figure \ref{fig:rmse_all_beta_a1_a2}. The empirical coverages for parameters $\boldsymbol{\beta}$ computed using  the GMWMX and \texttt{Hector} are presented in Figure \ref{fig:emp_coverage_all_beta_a1_a2}. We present the squared bias, the variance and the RMSE of the point estimates of $\bm{\gamma}$ obtained with the GMWMX and \texttt{Hector} for Setting A1 and A2 in Figure~\ref{fig:bias_variance_rmse_gamma_a1} and Figure~\ref{fig:bias_variance_rmse_gamma_a2}, respectively.

\FloatBarrier
\subsubsection{Setting B}

\begin{figure}[h]
    \centering
    \includegraphics[width=0.45\linewidth]{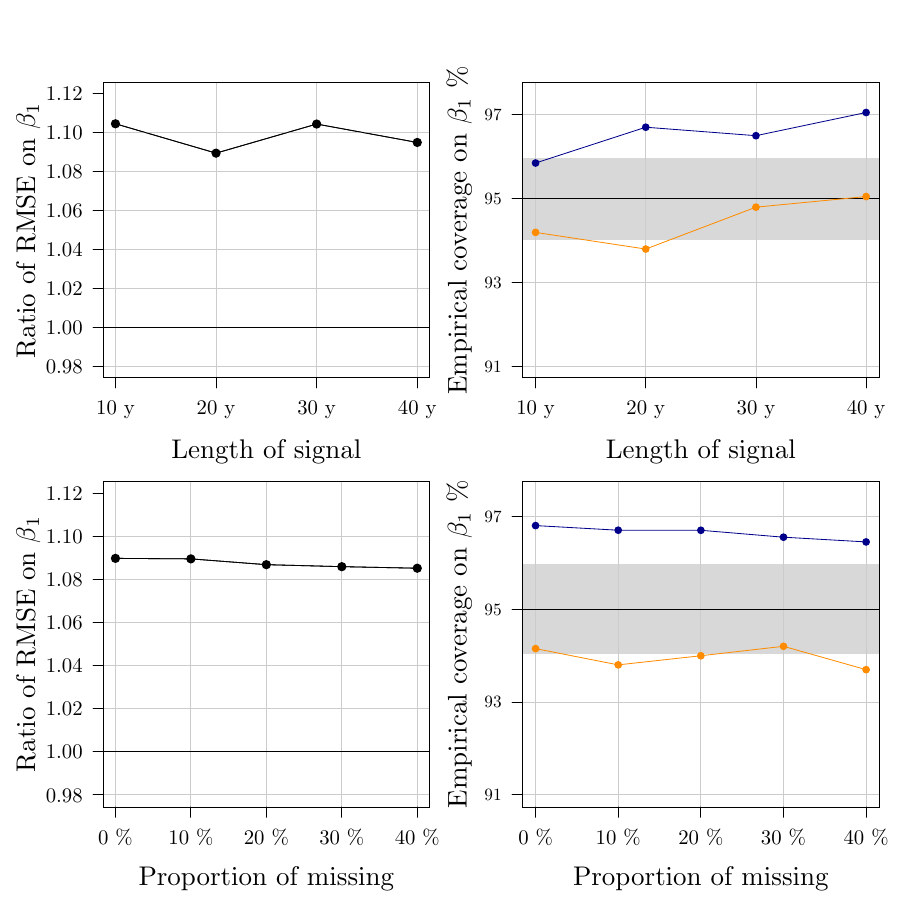}
    \caption{Ratio of RMSE and empirical coverage for parameter $\beta_1$ for Setting B1 and B2.}
    \label{fig:ratio_beta_b}
\end{figure}

\begin{figure}
    \centering
    \includegraphics[width=0.4\linewidth]{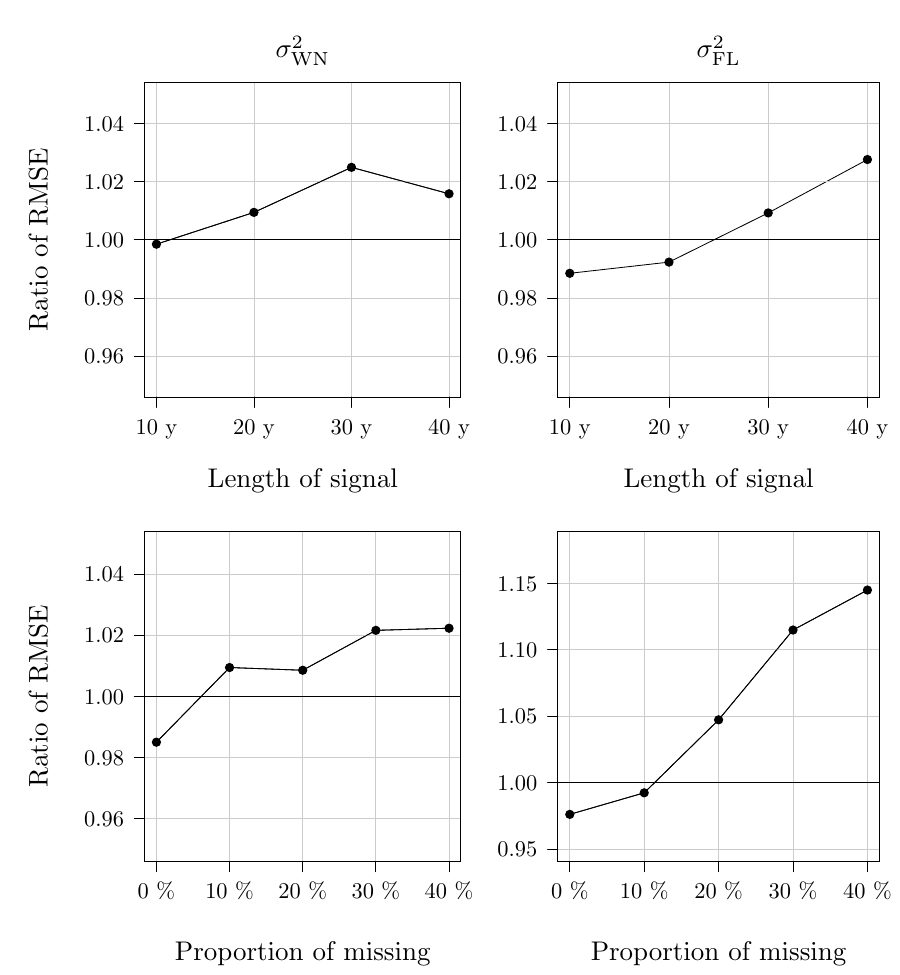}
    \caption{Ratio of RMSE for parameter $\boldsymbol{\gamma}$ for Setting B1 and B2.}
    \label{fig:ratio_gamma_b}
\end{figure}

\begin{figure}
    \centering
    \includegraphics[width=1\linewidth]{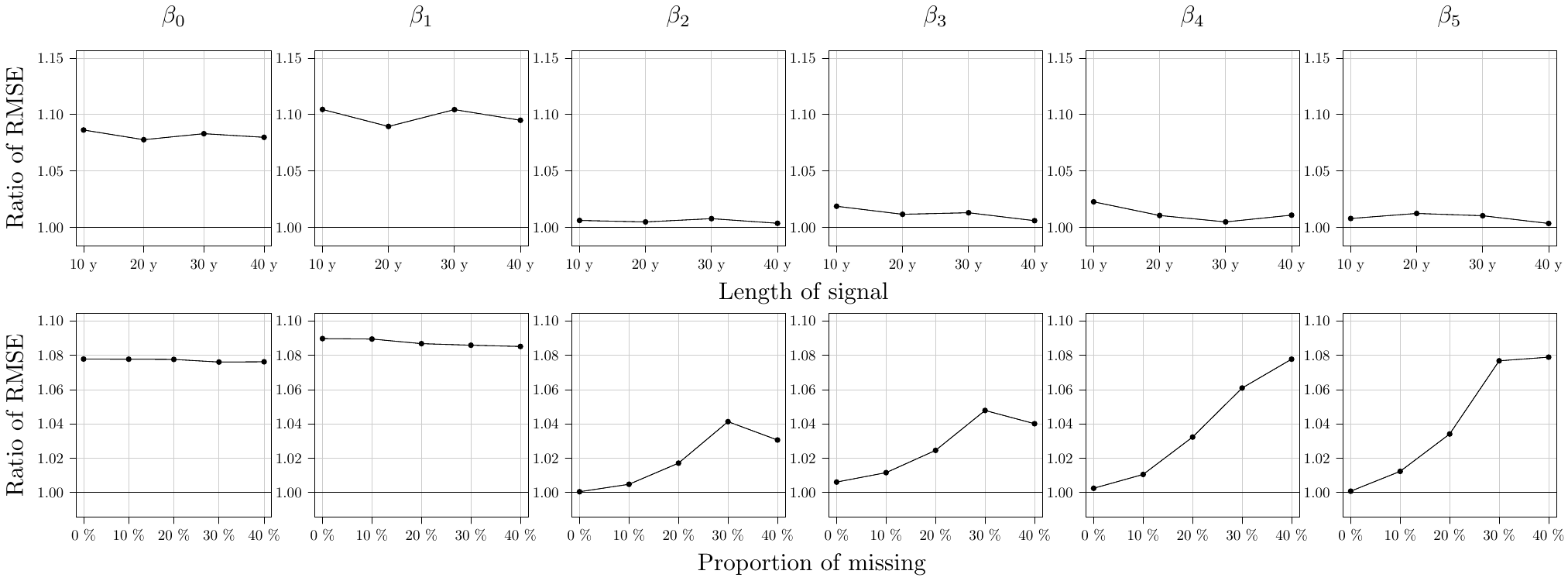}
    \caption{Ratio of RMSE for parameters ${\boldsymbol{\beta}}$ for Setting B1 and B2.}
    \label{fig:rmse_all_beta_setting_b1_b2}
\end{figure}

\begin{figure}
    \centering
    \includegraphics[width=0.5\linewidth]{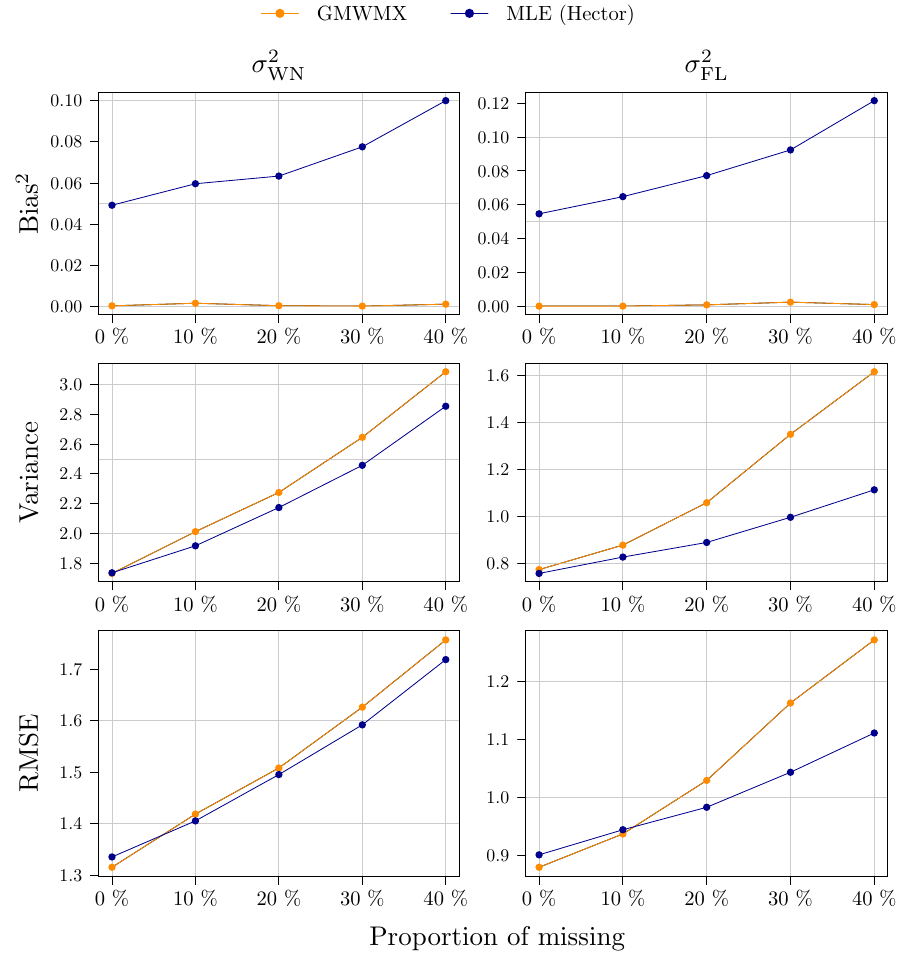}
    \caption{Bias$^2$, variance and RMSE for parameters $\boldsymbol{\gamma}$  for Setting B2.}
    \label{fig:bias_variance_rmse_gamma_setting_b2}
\end{figure}

For \textbf{Setting B}, which considers the WN plus the flicker noise, we fix the variance of the WN to $\sigma^2_{\mathrm{WN}} = 50$ and the variance of the flicker noise to  $\sigma^2_{\mathrm{FL}} = 10$. We consider the same ranges of sample sizes and proportions of missing observations as for \textbf{Setting A} where, similarly to the latter, we first focus on the RMSE ratios for the parameter $\beta_1$ and stochastic parameters (Figures \ref{fig:ratio_beta_b} and \ref{fig:ratio_gamma_b}) then present the complete results afterwards (Figures \ref{fig:rmse_all_beta_setting_b1_b2} to \ref{fig:bias_variance_rmse_gamma_setting_b2}).

\begin{figure}
    \centering
    \includegraphics[width=1\linewidth]{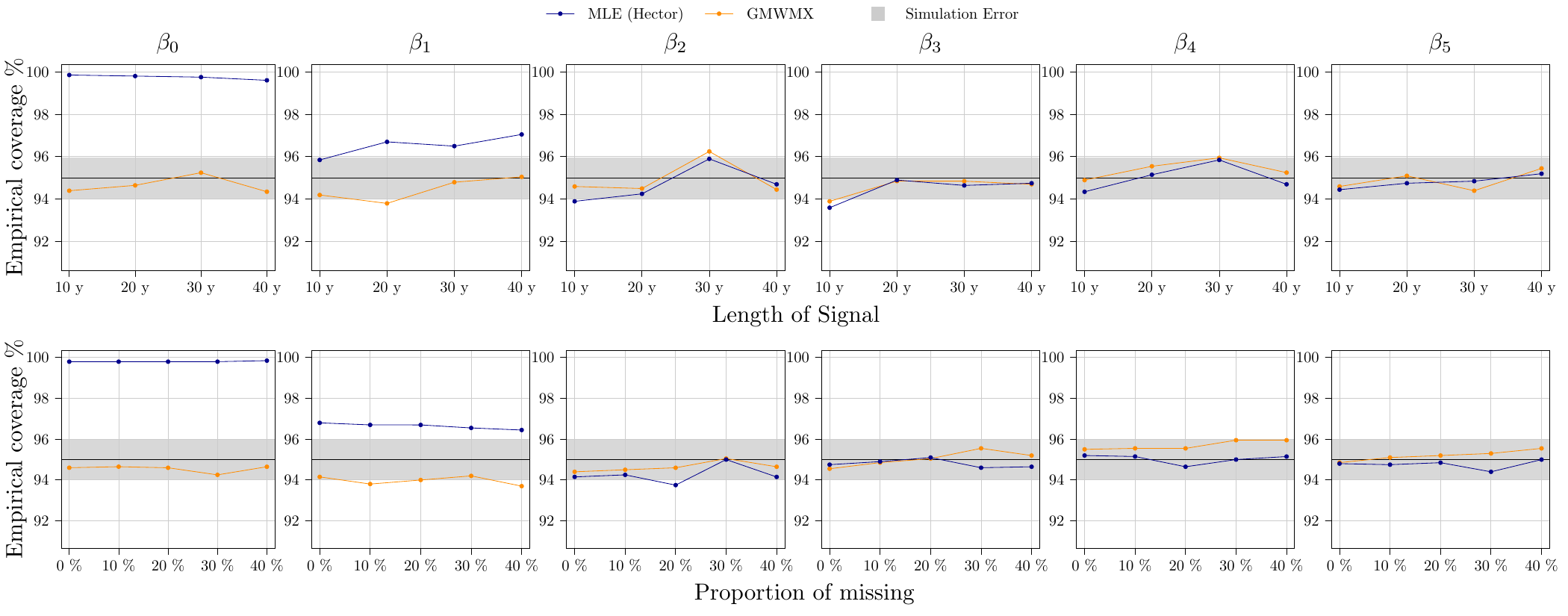}
    \caption{Empirical coverage for parameters $\boldsymbol{\beta}$ for Setting B1 and B2.}
    \label{fig:enter-label}
\end{figure}

\begin{figure}
    \centering
    \includegraphics[width=0.5\linewidth]{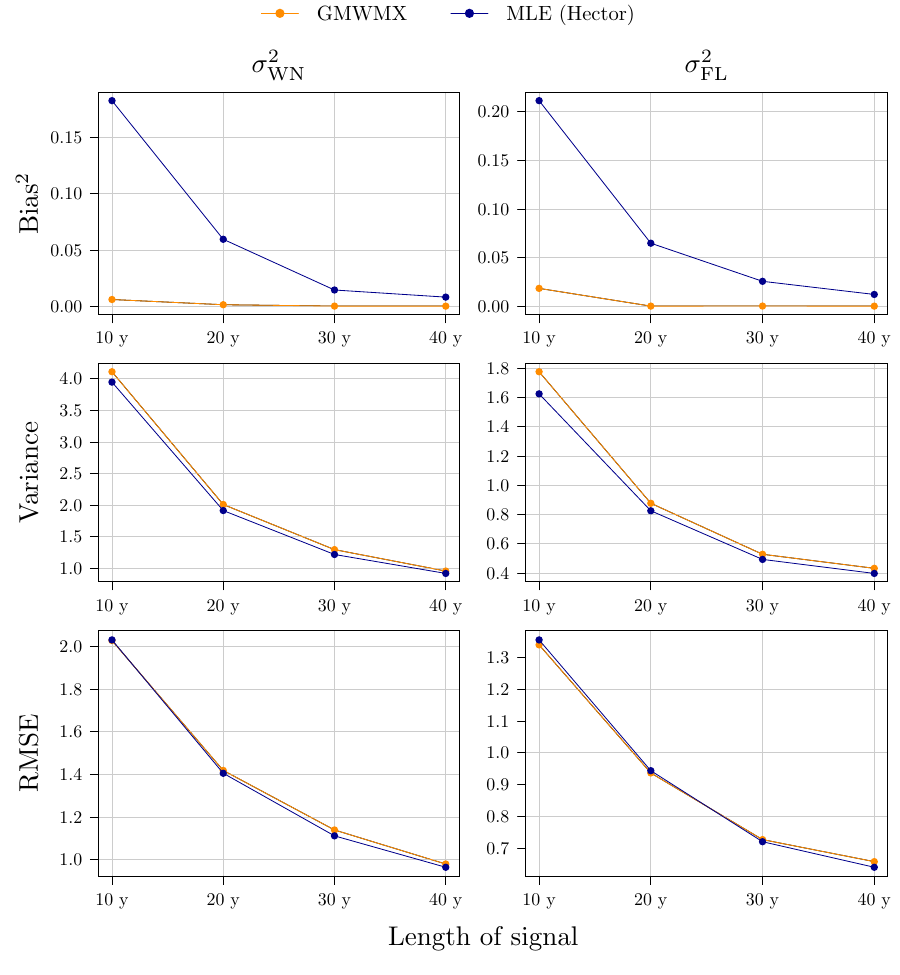}
    \caption{Bias$^2$, variance and RMSE for parameters $\boldsymbol{\gamma}$  for Setting B1.}
    \label{fig:enter-label}
\end{figure}

\FloatBarrier
\subsubsection{Setting C}

\begin{figure}[h]
    \centering
    \includegraphics[width=0.5\linewidth]{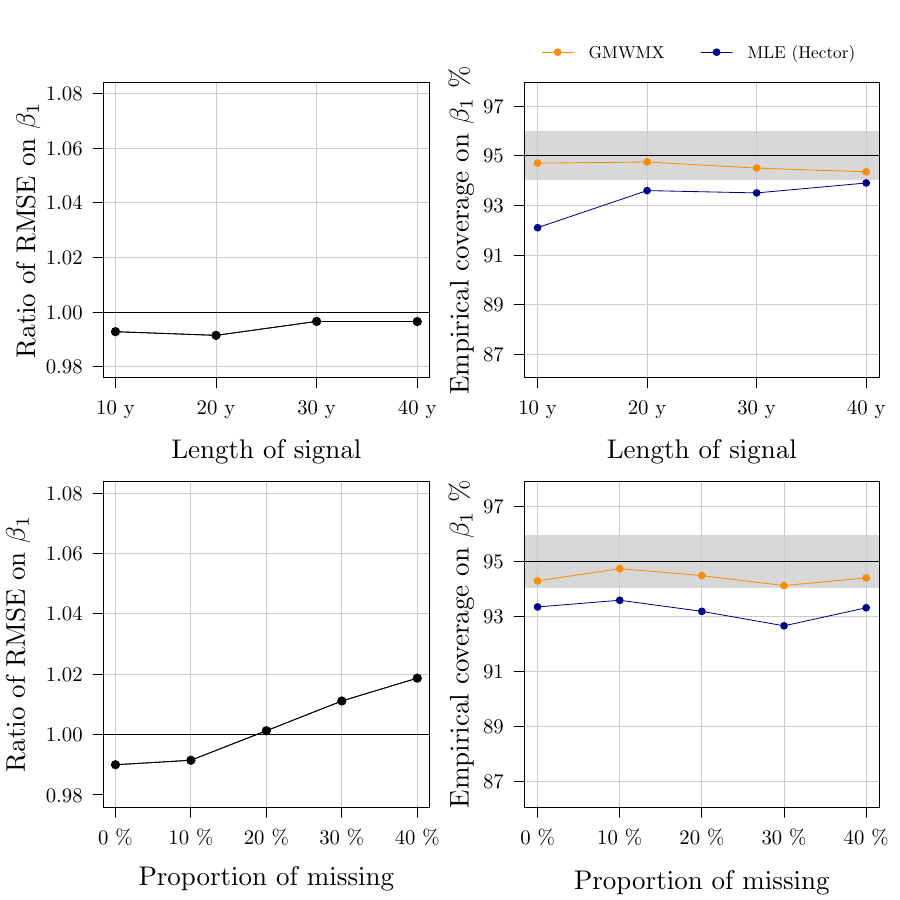}
    \caption{Ratio of RMSE and empirical coverage for parameter $\beta_1$ for Setting C1 and C2.}
    \label{fig:ratio_beta_c}
\end{figure}

\begin{figure}
    \centering
    \includegraphics[width=0.7\linewidth]{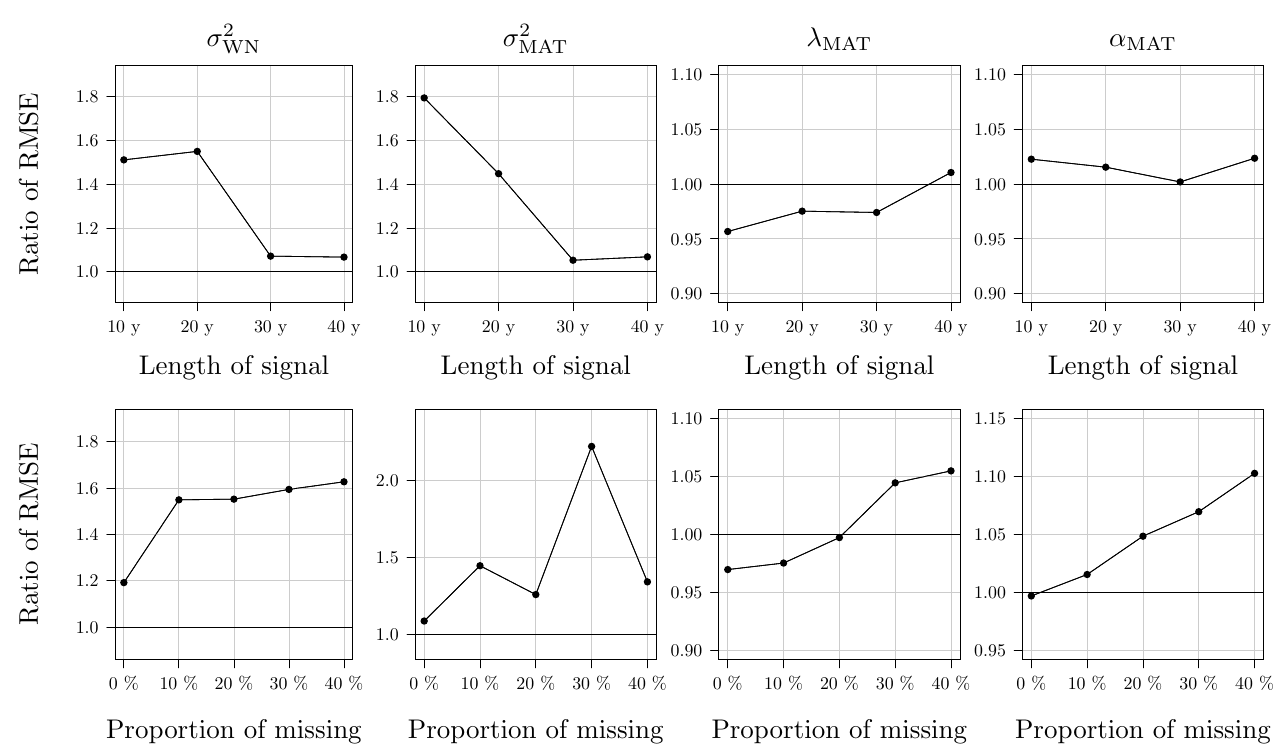}
    \caption{Ratio of RMSE of the stochastic parameters for Setting C1 and C2.}
    \label{fig:ratio_gamma_c}
\end{figure}

\begin{figure}
    \centering
    \includegraphics[width=1\linewidth]{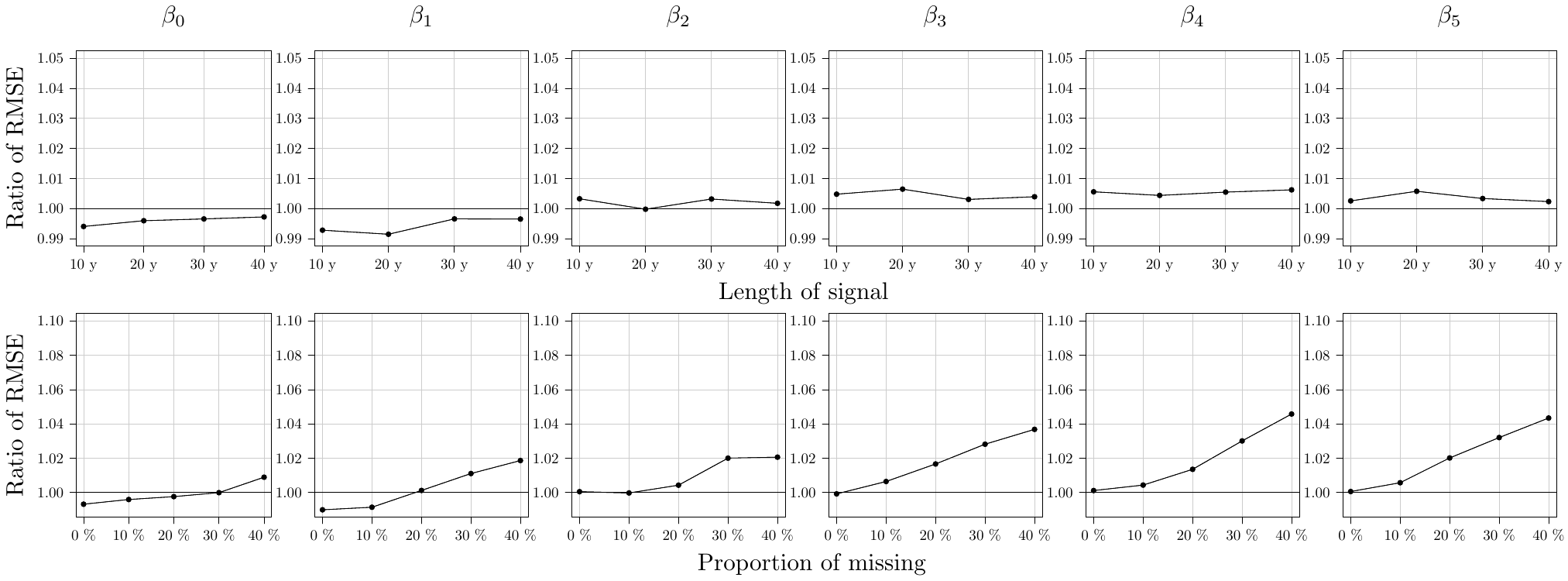}
    \caption{Ratio of RMSE for parameters $\boldsymbol{\beta}$ for Setting C1 and C2.}
    \label{fig:ratio_rmse_all_beta_seting_C1C2}
\end{figure}

\begin{figure}
    \centering
    \includegraphics[width=1\linewidth]{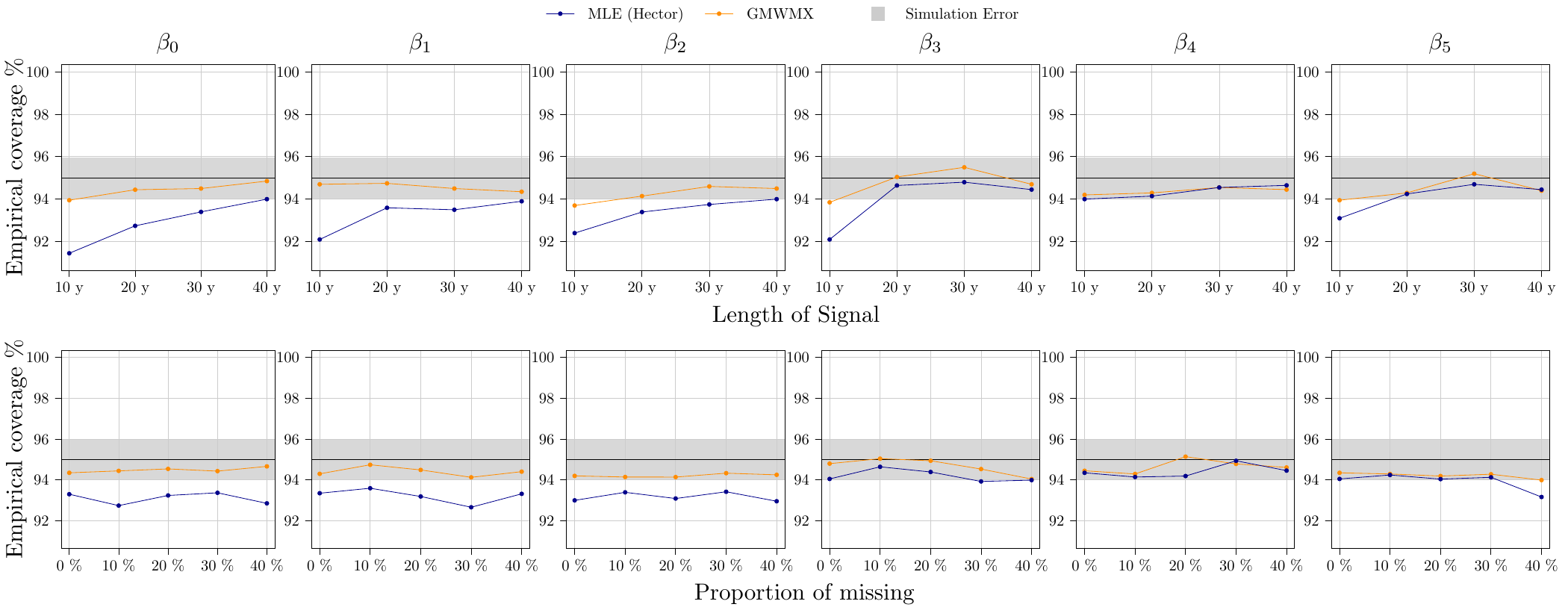}
    \caption{Empirical coverage for parameters $\boldsymbol{\beta}$ for Setting C1 and C2.}
    \label{fig:emp_coverage_all_beta_seting_C1C2}
\end{figure}

\begin{figure}
    \centering
    \includegraphics[width=0.75\linewidth]{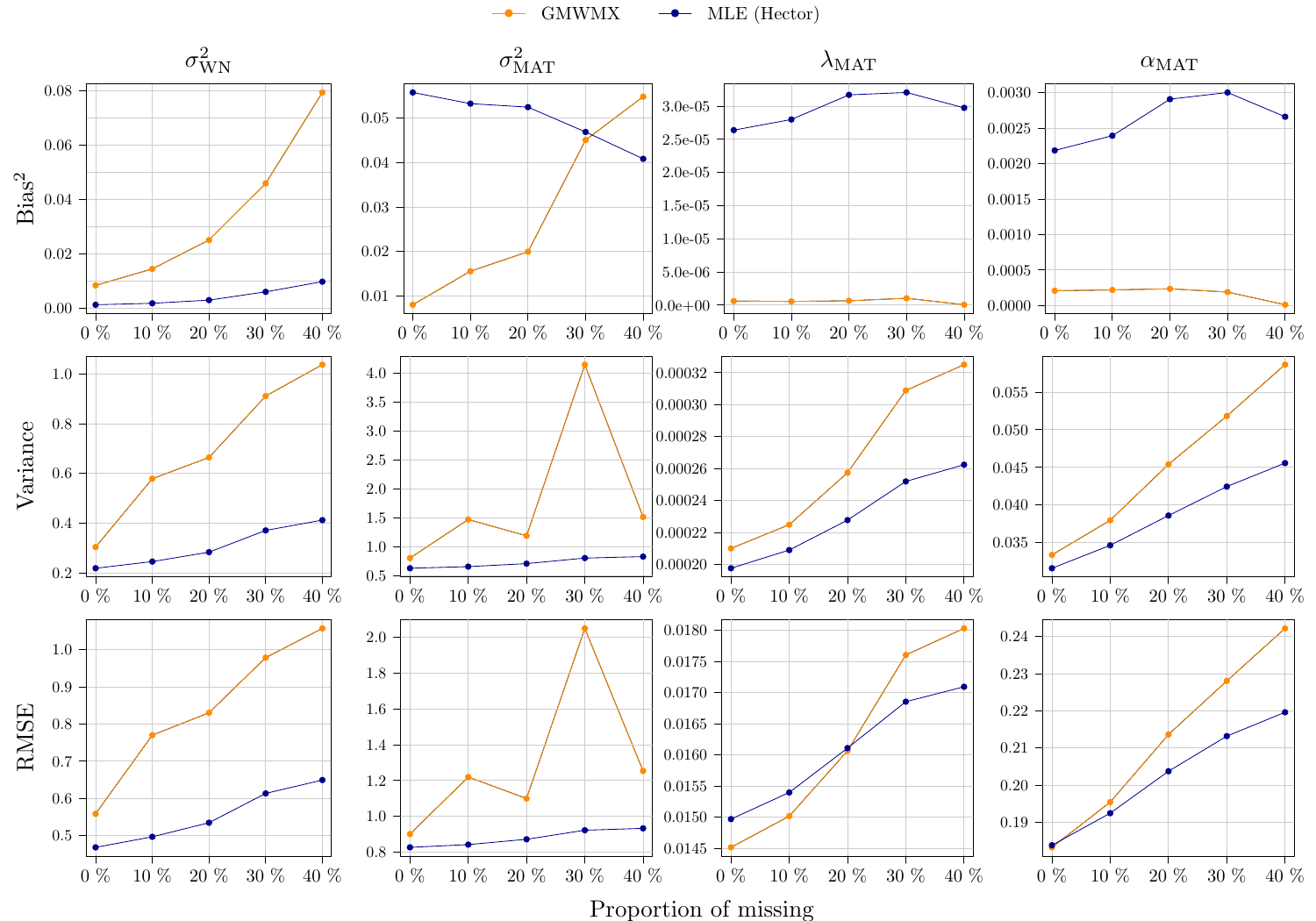}
    \caption{Bias$^2$, variance and RMSE for parameters $\boldsymbol{\gamma}$  for Setting C2.}
    \label{fig:bias_variance_rmse_stcoh_param_c2}
\end{figure}

In \textbf{Setting C}, we consider a combination of WN and Matérn noise (see \cite{lilly2017fractional} for a discussion on the Matérn process) which is a stationary process also often considered when modelling noise in GNSS observations \cite{bos2013fast}. We fix values of the process such that $\sigma^2_{\text{WN}}=20$ (innovation variance of WN), $\sigma^2_{\text{MAT}}=8$ (innovation variance of Matérn) and additional parameters of the Matérn process $\lambda_{\text{MAT}}=0.05$ and  $\alpha_{\text{MAT}}=1.1$. For this model we provide the RMSE ratios for the parameter $\beta_1$ (Figure \ref{fig:ratio_beta_c}) and for the stochastic parameters (Figure \ref{fig:ratio_gamma_c}) and the complete results afterwards (Figures \ref{fig:ratio_rmse_all_beta_seting_C1C2} to \ref{fig:bias_variance_rmse_stcoh_param_c2})

\begin{figure}
    \centering
    \includegraphics[width=0.75\linewidth]{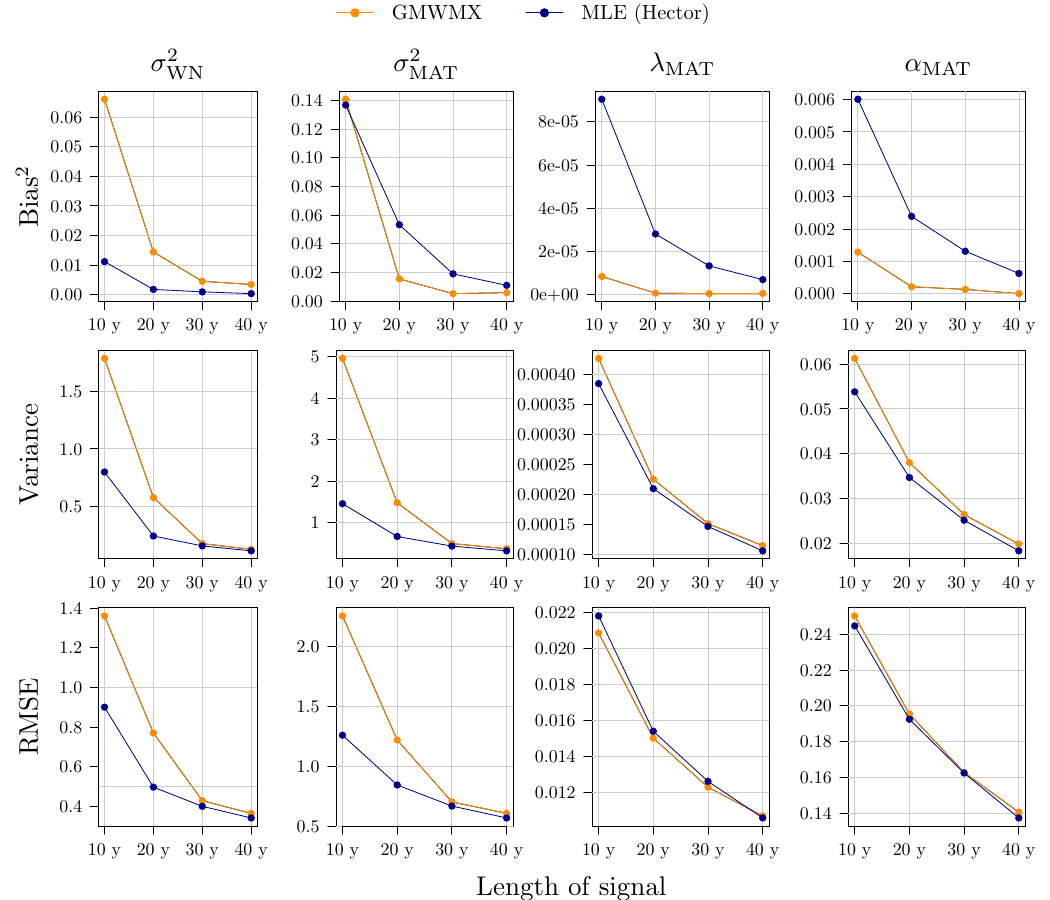}
    \caption{Bias$^2$, variance and RMSE for parameters $\boldsymbol{\gamma}$  for Setting C1.}
    \label{fig:enter-label}
\end{figure}

\FloatBarrier
\subsection{Computation of WV Covariance Matrix}
\label{app:compute_v}

Consider the error process $\bm{\varepsilon}$ to be a stationary process with  autocovariance function $\rho(k) = \cov(\varepsilon_i, \varepsilon_{i+k})$. We also define the autocovariance of the wavelet coefficients at scale $j$ as 
$$
f_j(h) = \cov\left(W_{j, t}, W_{j, t+h}\right) .
$$
We now construct the autocovariance of the wavelet coefficients issued from the \textit{Haar} wavelet filter at scale $j=1$, $f_1(k), k = 1, \ldots n$ using the following relationship:

\begin{equation*}
f_1(h)  =\frac{1}{2} \rho(k)-\frac{1}{4} \rho(k-1)-\frac{1}{4} \rho(k+1).
\end{equation*}

From this expression, we can now recursively compute the autocovariance of the wavelet coefficients for scales $2,\ldots, J$ using the following formula:

  \begin{equation*}
        f_{j+1}(k)= \frac{3}{2} f_j(k)+f_j\left(k+\frac{2^j}{2}\right)+f_j\left(k-\frac{2^j}{2}\right)+\frac{1}{4} f_j\left(k+2^j\right)+\frac{1}{4} f_j\left(k-2^j\right) .
    \end{equation*}

Using these expression, the variance of the estimated WV is given by:

\begin{equation*}
     \var\left(\hat{\nu}_j\right) =  \frac{2}{M_j^2} \sum_{i=-M_{j}+1}^{M_j-1}\left(M_j-|i|\right) f_j(i)^2,
\end{equation*}

while the covariance is given by

\begin{equation*}
  \begin{aligned}
        \cov\left(\hat{\nu}_j, \hat{\nu}_{j+l}\right) & =\cov\left(\frac{1}{M_j} W_j^T W_j, \frac{1}{M_{j+l}} W_{j+l}^T W_{j0+l}\right)\\
        & = \frac{2}{M_j M_{j+l}} \sum_{t=1}^{M_{j+l}} \sum_{m=1}^{M_j} C_{t, m}^2,
  \end{aligned}
\end{equation*}

where the summation above is as follows:

\begin{equation*}
    \sum_{t=1}^{M_{j+k}} \sum_{m=1}^{M_j} C_{t, m}^2 =\sum_{i=-M_{j}+1}^{M_{j +l}-1}\left[M_{j + l} \mathbbm{1}\left\{2^j-2^{j+l} \leqslant t \leqslant 0\right\}+\left(M_{j+l}-|i|\right) \mathbbm{1}\{i>0\}+\left(M_{j+l}-\left|i-2^j+2^{j+l}\right|\right) \mathbbm{1}\left\{i<2^j-2^{j+l}\right\}\right] \delta_i,
\end{equation*}

where
\begin{equation*}
    \delta_i  =\left[\sum_{p=0}^{2^l-2}\left\{\frac{p+1}{2^l} f_j\left(i+p  2^{j-1}\right)+\frac{2^l-1-p}{2^l} f_j\left(i+p 2^{j-1}+2^{j+l-1}\right)\right\}+f_j\left(i+2^{j+l-1}-2^{j-1}\right)\right]^2.
\end{equation*}

When the wavelet coefficients are not stationary, using $\boldsymbol{\Sigma}$ and $\boldsymbol{\Sigma}_{W}$ to represent the covariance matrix of the error process and of the wavelet coefficients respectively, then we obtain the variance of the WV using the following procedure:

\begin{enumerate}
    \item Compute $\mathbb{V}[\boldsymbol{W}_1]$, where $\boldsymbol{W}_1 \in \mathbb{R}^{M_1}$ is the vector containing the wavelet coefficients at scale $j=1$, as follows:
$$
\begin{aligned}
&\var\left(\boldsymbol{W}_1\right)_{k, l}=\sum_{\substack{i \in\{k, k+1\} \\ j \in\{l, l+1\}}} A_{k i} A_{l j}\left(\boldsymbol{\Sigma}\right)_{i j}\\
&\text { where } \quad A_{k i}=\left\{\begin{array}{cc}
-\frac{1}{2} & \text { if } i=k \\
\frac{1}{2} & \text { if } i=k+1
\end{array} \quad A_{l j}= \begin{cases}-\frac{1}{2} & \text { if } j=l \\
\frac{1}{2} & \text { if }j=l+1\end{cases}\right.
\end{aligned}
$$
    
    \item  Compute all the variances the wavelet coefficients at scale $j=2, \ldots, J$ as follows:

$$
\begin{aligned}
\var\left(\boldsymbol{W}_{j+1}\right)_{k l}= & \sum_{i=1}^{M_j} \sum_{j=1}^{M_j} B_{k i} \{\mathbb{V}\left[\boldsymbol{W}_j\right]\}_{i j}  B_{l j} \\
= 
& \sum_{\substack{
i \in\left\{k, k+\frac{L_j}{2}, k+L_j\right\}
\\
j \in\left\{l, l+\frac{L_j}{2}, 1 +L_j\right\}
}}B_{k i} B_{l j}\left[\mathbb{V}\left[\boldsymbol{W}_{j}\right]\right]_{i j}
\end{aligned}
$$
where
$$
B_{k i}=\left\{\begin{array}{ll}
\frac{1}{2} & \text { if } i=k \\
1 & \text { if }i=k+\frac{L_j}{2} \\
\frac{1}{2} &\text { if } i=k+L_j
\end{array} \quad B_{l j}=\left\{\begin{array}{lll}
\frac{1}{2} & \text { if } j=l \\
1 & \text { if } j=l+\frac{L_j}{2} \\
\frac{1}{2} & \text { if } j=l+L_j
\end{array}\right.\right.
$$

    \item Compute $C_{j, j+k} =\cov\left(\boldsymbol{W}_j, \boldsymbol{W}_{j+k}\right)$ as follows:

$$
C_{l m}=\sum_{p=0}^{2^{k+1}-2} C_p \left(\Sigma_{W_j}\right)_{l+p 2^{j-1}, m } \quad \begin{array}{r}
\text { with } l=1 \cdots M_{j+k} \\
m=1 \cdots M_j
\end{array}
$$

where

$$
C_p= \begin{cases}\frac{p+1}{2^k} & \text { if } 0 \leqslant p \leqslant 2^k-1 \\ \frac{2^{k+1}-1-p}{2^k} & \text { if } 2^k \leqslant p \leqslant 2^{k+1}-2\end{cases}
$$

    \item Compute
$$
\cov\left(\hat{\nu}_j, \hat{\nu}_{j+k}\right)=\frac{2}{M_j M_{j+k}}  \sum_{l=1}^{M_{j+k}} \sum_{m=1}^{M_j}\left[\cov\left(\boldsymbol{W}_j, \boldsymbol{W}_{j+k}\right)\right]_{l m}^2
$$

where we rewrite the above equation as

$$
\cov\left(\hat{\nu}_j, \hat{\nu}_{j+k}\right)=\frac{2}{M_j M_{j+k}} \sum_{h=-M_{j+1}}^{M_{j+k}-1} C_h \tilde{f}_{j, j+k}(h)^2
$$

where

$$
C_h= \begin{cases}M_{j+k} & \text { if } h=M_{j+k}-M_j, \cdots, 0 \\ M_{j+k}-h & \text { if } h=1, \ldots,  M_{j+k}-1 \\ M_{j+k}-\left|h-M_{j+k}+M_j\right| & \text { if } h=1-M_j, \ldots, M_{j+k}-M_j-1\end{cases}
$$

and $\tilde{f}_{j, j+k}(h)$ is defined as:

$$
\tilde{f}_{j, j+k}(h) = \begin{cases}
    \sqrt{\frac{1}{M_{j+k}} \sum_{l=1}^{M_{j+k}} \cov\left(W_{j+k, l}, W_{j, l-h}\right)^2} & \text{ if } h= M_{j+k} - M_j , \ldots, 0 \\
  \sqrt{\frac{1}{M_{j+k}-h} \sum_{l=h+1}^{M_{j+k}} \cov\left(W_{j+k, l}, W_{j, l-h}\right)^2} & \text{ if } h= 1, \ldots, M_{j+k}-1\\
   \sqrt{\frac{1}{M_{j+k}-|h - M_{j+k}+M_j|} \sum_{l=1}^{M_{j+k} - |h - M_{j+k}+M_j|} \cov\left(W_{j+k, l}, W_{j, l-h}\right)^2} & \text{ if } h= 1 - M_j, \ldots, M_{j+k} -M_j -1
\end{cases}
$$
\end{enumerate}

Rewriting with the function as $\tilde{f}_{j, j+k}(h) $ allows to construct a fast approximation of $\var(\hat{\nu})$ in the non-stationary setting.

\end{document}